%% file: main.tex
\pdfoutput=1

\documentclass[camera,letterpaper,nomarginnotes,nonarrowgutter]{jpaper}

\input{packages}
\input{macros}

\ifcameraready
\else
    \fancypagestyle{firstpage}
    {
        \fancyhead{}
        \fancyhead[C]{\textcolor{red}{CONFIDENTIAL DRAFT -- DO NOT DISTRIBUTE -- TO APPEAR IN HPCA'24} \\ \textcolor{MidnightBlue}{\emph{Version \versionnum~---~\today, \ampmtime}}  \hrule }
    }
\fi

\makeatletter
\def\bstctlcite{\@ifnextchar[{\@bstctlcite}{\@bstctlcite[@auxout]}}
\def\@bstctlcite[#1]#2{\@bsphack
  \@for\@citeb:=#2\do{%
    \edef\@citeb{\expandafter\@firstofone\@citeb}%
    \if@filesw\immediate\write\csname #1\endcsname{\string\citation{\@citeb}}\fi}%
  \@esphack}
\makeatother

\begin{document}
\bstctlcite{IEEEexample:BSTcontrol}

%%%%%%%%%%%%%%%%%%%%%%%%%%%%%%%%%%%%%%%%
%%%%%%%%%%%%%% -- UPDATE -- %%%%%%%%%%%%%%%
%\newcommand{\hpcasubmissionnumber}{132}

\title{\scalebox{0.95}{\prop: An End-to-End Processing-\gfcr{U}sing-DRAM System}
\scalebox{0.95}{ for \gfcri{High-\gfcrii{Throughput\gfcriii{,}}} Energy-Efficient \gfcriii{and Programmer-Transparent}}
\scalebox{0.95}{
\gfcri{Multiple-Instruction Multiple-Data} Processing}\vspace{10pt}}
%%%%%%%%%%%%%%%%%%%%%%%%%%%%%%%%%%%%%%%%

%%%%%%%%%%%%%%%%%%%%%%%%%%%%%%%%%%%%%%%%
%%%%%%%% -- ONLY FOR CAMERA READY -- %%%%%%%%
%\def\hpcacameraready{} % Uncomment to build camera-ready version
%\newcommand{\hpcapubid}{0000--0000/00\$00.00}
\author{
Geraldo F. Oliveira$^\dagger$~\qquad 
Ataberk Olgun$^\dagger$~\qquad 
Abdullah Giray Ya\u{g}l{\i}k\c{c}{\i}$^\dagger$~\qquad
F. Nisa Bostanc\i$^\dagger$\\
Juan Gómez-Luna$^\dagger$~\qquad 
Saugata Ghose$^\ddagger$~\qquad 
Onur Mutlu$^\dagger$\vspace{10pt}
\\
$^\dagger$~\emph{ETH Zürich} \qquad \qquad $^\ddagger$~\emph{Univ. of Illinois Urbana-Champaign}
\vspace{5pt}
}
%%%%% -- ARTEFACT EVALUATION RESULTS -- %%%%%%
% Uncomment the following based on the badges that were awarded to this paper
%\def\aeopen{}           % The artifact is publically available
%\def\aereviewed{}     % The artefact has been reviewed
%\def\aereproduced{} % The results have been reproduced
%%%%%%%%%%%%%%%%%%%%%%%%%%%%%%%%%%%%%%%%

%\input{hpca-template}

\ifcameraready
 \thispagestyle{plain}
\else
  \thispagestyle{firstpage}
\fi
\pagestyle{plain}

\maketitle

\input{sections/00_abstract}

%%%%%% -- PAPER CONTENT STARTS-- %%%%%%%%

\input{sections/01_introduction}
\input{sections/02_background}

\input{sections/03_motivation}

\input{sections/04_key_ideas}

\input{sections/05_design}

\input{sections/06_evaluation}

\input{sections/07_related_work}
\input{sections/08_conclusion}

%%%%%%% -- PAPER CONTENT ENDS -- %%%%%%%%

\section*{\gfcr{Acknowledgments}}

\gfcr{We thank the anonymous reviewers of MICRO 2023 and HPCA 2024 for their encouraging feedback. 
We thank the SAFARI Research Group members for providing a stimulating intellectual environment. We acknowledge
the generous gifts from our industrial partners\omi{; including} Google, Huawei, Intel, Microsoft\omi{.} This work is supported in part by the Semiconductor Research Corporation\gfcrii{,} the ETH Future Computing Laboratory\gfcrii{, the \omii{AI Chip Center for Emerging Smart Systems (ACCESS)}}, and the UIUC/ZJU HYBRID Center.}

%%%%%%%%% -- BIB STYLE AND FILE -- %%%%%%%%
\balance 

{
  \bstctlcite{IEEEexample:BSTcontrol}
  \let\OLDthebibliography\thebibliography
  \renewcommand\thebibliography[1]{
    \OLDthebibliography{#1}
    \setlength{\parskip}{0pt}
    \setlength{\itemsep}{0pt}
  }
  \bibliographystyle{IEEEtran}
  \bibliography{refs}
}
%%%%%%%%%%%%%%%%%%%%%%%%%%%%%%%%%%%%

\end{document}

%% file: packages.tex
\PassOptionsToPackage{hyphens}{url}
\usepackage[bookmarks=true,breaklinks=true,colorlinks,linkcolor=black,citecolor=black,urlcolor=blue]{hyperref}

\usepackage[noadjust]{cite}
\usepackage{amsmath,amssymb,amsfonts}
\usepackage{algorithmic}
\usepackage{graphicx}
\usepackage{textcomp}
\usepackage[hyphens]{url}
\usepackage{fancyhdr}

\usepackage{mathptmx} % This is Times font
\usepackage{setspace}

\usepackage{balance}

\usepackage[acronym,nonumberlist,nowarn]{glossaries}
    \glsdisablehyper
     \loadglsentries{acronyms}

\usepackage{footnote}
\usepackage{multicol}
\usepackage{xspace}
\usepackage[normalem]{ulem}
\usepackage[skip=2pt]{caption}
\usepackage{booktabs}
\usepackage{multirow}
\usepackage{ltablex}
\usepackage{makecell}
\usepackage[binary-units=true]{siunitx}
\usepackage{xargs} 
\usepackage{color, colortbl}
\usepackage[multiple]{footmisc}
\usepackage{marginnote}
\usepackage{clipboard}
\usepackage{tikz}
\usepackage{subcaption}
\usepackage{mathtools}
\usepackage{longfbox}
\usepackage[compact]{titlesec}
\usepackage{wrapfig}
\usepackage{graphbox}
\usepackage{cuted,capt-of}
\usepackage{todonotes}
\usepackage{soul}
\usepackage[many]{tcolorbox}

\usepackage{enumitem}
\usepackage{xcolor}
\usepackage[us,12hr]{datetime}

\usepackage[en-GB, useregional=numeric]{datetime2}

\usepackage{cleveref}

\crefformat{section}{\S#2#1#3}
\crefformat{subsection}{\S#2#1#3}
\crefformat{subsubsection}{\S#2#1#3}

%% file: acronyms.tex
\newacronym{ADI}{ADI}{Alternating Direction Implicit}
\newacronym{AGU}{AGU}{Address Generation Unit}
\newacronym[longplural={Access History Tables}]{AHT}{AHT}{Access History Table}
\newacronym{AHT-R}{AHT-R}{Access History Table - Read}
\newacronym{AHT-W}{AHT-W}{Access History Table - Write-Back}
\newacronym{AIP}{AIP}{Access Interval Predictor}
\newacronym{AL}{AL}{Added Latency for column accesses}
\newacronym{ASIC}{ASIC}{Application Specific Integrated Circuit}
\newacronym{AVX}{AVX}{Advanced Vector Extensions}
\newacronym[longplural={Basic Block Vectors}]{BBV}{BBV}{Basic Block Vector}
\newacronym{BHT}{BHT}{Branch History Table}
\newacronym{HBM}{HBM}{Hybrid Bandwidth Memory}
\newacronym{DDR4}{DDR4}{Double-Data Rate 4}
\newacronym{MIMD}{MIMD}{multiple-instruction multiple-data}
\newacronym{SIMT}{SIMT}{single-instruction multiple-thread}
\newacronym{SLP}{SALP}{subarray-level parallelism}

\newacronym{NN}{NN}{Neural Network}
\newacronym{BTB}{BTB}{Branch Target Buffer}
\newacronym{CAS}{CAS}{Column Address Strobe}
\newacronym{AMAT}{AMAT}{Average Memory Access Time}
\newacronym{CCD}{CCD}{Column to Column Delay}
\newacronym{AI}{AI}{Arithmetic Intensity}
\newacronym[longplural={Columnar Database Systems}]{CDBS}{CDBS}{Columnar Database System}
\newacronym[longplural={Chip Multiprocessors}]{CMP}{CMP}{Chip Multiprocessor}
\newacronym{CMT}{CMT}{Chip Multithreading}
\newacronym[longplural={Coarse-Grain Reconfigurable Arrays}]{CGRA}{CGRA}{Coarse-Grain Reconfigurable Array}
\newacronym{C-RAM}{C-RAM}{Computational-RAM}
\newacronym{CWD}{CWD}{Column Write Delay}
\newacronym{DBI}{DBI}{Dynamic Binary Instrumentation}
\newacronym[longplural={Database Management Systems}]{DBMS}{DBMS}{Database Management System}
\newacronym{DDR}{DDR}{Double Data Rate}
\newacronym{DEWP}{DEWP}{Dead Line and Early Write-Back Predictor}
\newacronym{DRAM}{DRAM}{Dynamic Random Access Memory}
\newacronym{DSBP}{DSBP}{Dead Sub-Block Predictor}
\newacronym{DSM}{DSM}{Decomposed Storage Manage}
\newacronym{FAW}{FAW}{Four row Activation Window}
\newacronym{FP}{FP}{Floating-Point}
\newacronym{EDP}{EDP}{Energy-Delay Product}
\newacronym{FCFS}{FCFS}{First-Come First-Serve}
\newacronym{FIFO}{FIFO}{Fist In, First Out}
\newacronym{FSM}{FSM}{Finite State Machine}
\newacronym{FPGA}{FPGA}{Field-Programmable Gate Array}
\newacronym[longplural={Functional Units}]{FU}{FU}{Functional Unit}
\newacronym{GAg}{GAg}{Global Adaptive branch prediction using one Global PHT}
\newacronym{GAs}{GAs}{Global Adaptive branch prediction using per-Set PHT}
\newacronym{GCC}{GCC}{GNU Compiler Collection}
\newacronym{GEMS}{GEMS}{General Execution-driven Multiprocessor Simulator}
\newacronym[longplural={General Purpose Processors}]{GPP}{GPP}{General Purpose Processor}
\newacronym{HMC}{HMC}{Hybrid Memory Cube}
\newacronym{HIVE}{HIVE}{HMC Instruction Vector Extensions}
\newacronym{IATAC}{IATAC}{Inter-Access Time per Access Count}
\newacronym{ILP}{ILP}{Instruction Level Parallelism}
\newacronym{IPC}{IPC}{Instructions per Cycle}
\newacronym{ISA}{ISA}{Instruction Set Architecture}
\newacronym{IRAM}{IRAM}{Intelligent RAM}
\newacronym{KIPS}{KIPS}{Kilo Instructions per Second}
\newacronym{LDS}{LDS}{Linked Data Structure}
\newacronym{LFSR}{LFSR}{Linear Feedback Shift Register}
\newacronym{LFMR}{LFMR}{Last-to-First Miss-Ratio}
\newacronym{MLP}{MLP}{Memory-Level Parallelism}

\newacronym{LLC}{LLC}{last-level cache}
\newacronym{LRU}{LRU}{Least Recently Used}
\newacronym{LSU}{LSU}{Load Store Unit}
\newacronym{LTP}{LTP}{Last-Touch Predictor}
\newacronym{LvP}{LvP}{Live-time Predictor}
\newacronym{LWP}{LWP}{Last Write Predictor}
\newacronym{MARSS}{MARSS}{Micro Architectural and System Simulator}
\newacronym{McPAT}{McPAT}{Multi-core Power, Area, and Timing}
\newacronym{MIPS}{MIPS}{Microprocessor without Interlocked Pipeline Stages}
\newacronym{MOB}{MOB}{Memory Order Buffer}
\newacronym{MMU}{MMU}{memory management unit}
\newacronym{PuD}{PUD}{Processing-using-DRAM}
\newacronym{MPKI}{MPKI}{Misses per Kilo-Instruction}
\newacronym{MSHR}{MSHR}{Miss-Status Handling Registers}
\newacronym{NAS}{NAS}{Numerical Aerodynamic Simulation}
\newacronym{NDP}{NDP}{Near-Data Processing}
\newacronym{NMP}{NMP}{Near Memory Processor}
\newacronym{NoC}{NoC}{Network-on-Chip}
\newacronym{NPB}{NPB}{NAS Parallel Benchmark}
\newacronym{NUCA}{NUCA}{Non-Uniform Cache Architecture}
\newacronym{NUMA}{NUMA}{Non-Uniform Memory Access}
\newacronym{OoO}{OoO}{Out-of-Order}
\newacronym{OpenMP}{OpenMP}{Open Multi-Processing}
\newacronym{OS}{OS}{Operating System}
\newacronym{PAg}{PAg}{Per-address Adaptive branch prediction using one Global PHT}
\newacronym{PAs}{PAs}{Per-address Adaptive branch prediction using per-Set PHT}
\newacronym{PC}{PC}{Program Counter}
\newacronym[longplural={Pointer-Chasing Engines}]{PCE}{PCE}{Pointer-Chasing Engine}
\newacronym{PCM}{PCM}{Performance Counter Monitor}
\newacronym{PIM}{PIM}{processing-in-memory}
\newacronym[longplural={Pattern History Tables}]{PHT}{PHT}{Pattern History Table}
\newacronym{RAPL}{RAPL}{Running Average Power Limit}
\newacronym{RAS}{RAS}{Row Address Strobe}
\newacronym{RAT}{RAT}{Registers Alias Table}
\newacronym{RC}{RC}{Row Cycle}
\newacronym{RCD}{RCD}{RAS to CAS Delay}
\newacronym[longplural={Row-based Database Systems}]{RDBS}{RDBS}{Row-based Database System}
\newacronym{ROB}{ROB}{Reorder Buffer}
\newacronym{RRD}{RRD}{Row to Row activation Delay}
\newacronym{DNN}{DNN}{Deep Neural Network}
\newacronym{ANN}{ANN}{Artificial Neural Network}
\newacronym{PnM}{PNM}{processing-near-memory}
\newacronym{PuM}{PUM}{processing-using-memory}
\newacronym{FFD}{FFD}{first-fit-decreasing}
\newacronym{HFF}{HFF}{helper flip-flop}

\newacronym{RP}{RP}{Row Precharge}
\newacronym{RTL}{RTL}{Register Transfer Level}
\newacronym{RTP}{RTP}{Read To Precharge}
\newacronym{RVU}{RVU}{Reconfigurable Vector Unit}
\newacronym{SDP}{SDP}{Skewed Dead-Block Predictor}
\newacronym{SESC}{SESC}{Superescalar Simulator}
\newacronym{SSE}{SSE}{Streaming SIMD Extensions}
\newacronym{SFP}{SFP}{Spatial Footprint Predictor}
\newacronym{SiNUCA}{SiNUCA}{Simulator of Non-Uniform Cache Architectures}
\newacronym{SMT}{SMT}{Simultaneous Multi-Threading}
\newacronym{SIMD}{SIMD}{single-instruction multiple-data}
\newacronym{SPEC}{SPEC}{Standard Performance Evaluation Corporation}
\newacronym{SoC}{SoC}{System-on-Chip}
\newacronym{SPP}{SPP}{Spatial Pattern Predictor}
\newacronym{SRAM}{SRAM}{Static Random Access Memory}
\newacronym{SSOR}{SSOR}{Symmetric Successive Over-Relaxation}
\newacronym{SSV}{SSV}{Search Set Vector}
\newacronym{TLB}{TLB}{Translation Look-aside Buffer}
\newacronym{TLP}{TLP}{Thread Level Parallelism}
\newacronym[longplural={Through-Silicon Vias}]{TSV}{TSV}{Through-Silicon Via}
\newacronym{VWQ}{VWQ}{Virtual Write Queue}
\newacronym{WTR}{WTR}{Write Recovery time}
\newacronym{WR}{WR}{Write To Read delay time}
\newacronym{TRA}{TRA}{triple row activation}

%% file: macros.tex
\newcommand\efficiencysimdram{14.3$\times$\xspace}
\newcommand\efficiencygpu{6.8$\times$\xspace}
\newcommand\efficiencycpu{30.6$\times$\xspace}

\newcommand{\circled}[1]{\tikz[baseline=(char.base)]{\node[shape=circle,draw,inner sep=0pt,fill=black, text=white] (char) {#1};}}

\newcommand{\circledgreen}[1]{\tikz[baseline=(char.base)]{\node[shape=circle,draw,inner sep=0pt,fill=dollarbill, text=white] (char) {#1};}}

\newcommand{\circledii}[1]{\tikz[baseline=(char.base)]{\node[shape=circle,draw,inner sep=0pt,fill=gray, text=white] (char) {\itshape#1};}}

\newcommand{\circlediii}[1]{\tikz[baseline=(char.base)]{\node[shape=circle,draw,inner sep=0pt,fill=white, text=black] (char) {\itshape#1};}}

\newcommand{\paratitle}[1]{\vspace{4pt}\noindent\textbf{#1.}}

\newcommand{\tempcommand}[1]{\renewcommand{\arraystretch}{#1}}

\newcommand{\li}{(\textit{i})}
\newcommand{\lii}{(\textit{ii})}
\newcommand{\liii}{(\textit{iii})}
\newcommand{\liv}{(\textit{iv})}
\newcommand{\lv}{(\textit{v})}

\newtcolorbox{NewBox}[1]{%
  floatplacement={#1}, width=\columnwidth,
  colframe=gray!10!black,colback=orange!10!white,boxrule=1pt,arc=.2em,boxsep=-1.6mm,% any tcolorbox options here
  }

\newcommand{\prop}{MIMDRAM\xspace}

\DeclareSIUnit{\flop}{FLOP}

\newcommand{\revdel}[1]{}

\newcommand\matrange{\texttt{\lbrack}\emph{mat begin}\texttt{,}\emph{mat end}\texttt{\rbrack}\xspace}

\newcommand\uprog{\textmu{}Program\xspace}
\newcommand\uprogs{\textmu{}Programs\xspace}

\newcommand\bbop{\emph{bbop}\xspace}
\newcommand\underutilization{underutilization\xspace}

\definecolor{dollarbill}{rgb}{0.52, 0.73, 0.4}

\definecolor{ao}{rgb}{0.0, 0.5, 0.0}
\definecolor{burntorange}{rgb}{0.8, 0.33, 0.0}

\newcommandx{\unsure}[2][1=]{\todo[linecolor=red,backgroundcolor=red!25,bordercolor=red,#1, size=\tiny,fancyline]{#2}}
\newcommandx{\change}[2][1=]{\todo[linecolor=blue,backgroundcolor=blue!25,bordercolor=blue,#1,size=\tiny]{\textbf{#2}}}
\newcommandx{\feedback}[2][1=]{\todo[linecolor=yellow,backgroundcolor=yellow!25,bordercolor=yellow,#1,size=\tiny]{#2}}
\newcommandx{\improvement}[2][1=]{\todo[linecolor=Plum,backgroundcolor=Plum!25,bordercolor=Plum,#1]{#2}}
\newcommandx{\thiswillnotshow}[2][1=]{\todo[disable,#1]{#2}}
\newcommandx{\completedRevision}[2][1=]{\todo[disable,backgroundcolor=red,#1]{#2}}
\newcommandx{\dataSource}[2][1=]{\todo[disable,backgroundcolor=red,#1]{#2}}
\newcommandx{\info}[2][1=]{\todo[linecolor=dollarbill,backgroundcolor=dollarbill!25,bordercolor=dollarbill,#1, size=\tiny,fancyline]{#2}}

\definecolor{blush}{rgb}{0.87, 0.36, 0.51}

\definecolor{MidnightBlue}{rgb}{0.1, 0.1, 0.44}
\newif\ifrevision
\revisionfalse
\ifrevision 
    \newcommand{\revA}[1]{\textcolor{red}{#1}}
    \newcommand{\revB}[1]{\textcolor{purple}{#1}}
    \newcommand{\revC}[1]{\textcolor{blush}{#1}}
    \newcommand{\revD}[1]{\textcolor{burntorange}{#1}}
    \newcommand{\revE}[1]{\textcolor{ao}{#1}}
    \newcommand{\revCommon}[1]{\textcolor{blue}{#1}}

    \let\marginpar\marginnote
    \newcommandx{\changeCM}[2][1=]{\todo[linecolor=blue,backgroundcolor=blue!25,bordercolor=blue,#1,size=\scriptsize]{\revCommon{\textbf{#2}}}}
    
    \newcommandx{\changeA}[2][1=]{\todo[linecolor=red,backgroundcolor=red!25,bordercolor=red,#1,size=\scriptsize]{\revA{\textbf{#2}}}}
    
    \newcommandx{\changeB}[2][1=]{\todo[linecolor=purple,backgroundcolor=purple!25,bordercolor=purple,#1,size=\scriptsize]{\revB{\textbf{#2}}}}
    
    \newcommandx{\changeC}[2][1=]{\todo[linecolor=blush,backgroundcolor=blush!25,bordercolor=blush,#1,size=\scriptsize]{\revC{\textbf{#2}}}}
    
    \newcommandx{\changeD}[2][1=]{\todo[linecolor=orange,backgroundcolor=orange!25,bordercolor=orange,#1,size=\scriptsize]{\revD{\textbf{#2}}}}
    
    \newcommandx{\changeE}[2][1=]{\todo[linecolor=ao,backgroundcolor=ao!25,bordercolor=ao,#1,size=\scriptsize]{\revE{\textbf{#2}}}}

\else
    \newcommand{\revA}[1]{\textcolor{black}{#1}}
    \newcommand{\revB}[1]{\textcolor{black}{#1}}
    \newcommand{\revC}[1]{\textcolor{black}{#1}}
    \newcommand{\revD}[1]{\textcolor{black}{#1}}
    \newcommand{\revE}[1]{\textcolor{black}{#1}}
    \newcommand{\revCommon}[1]{\textcolor{black}{#1}}

    \newcommandx{\changeCM}[2][1=]{\todo[disable,#1]{#2}}
    \newcommandx{\changeA}[2][1=]{\todo[disable,#1]{#2}}
    \newcommandx{\changeB}[2][1=]{\todo[disable,#1]{#2}}
    \newcommandx{\changeC}[2][1=]{\todo[disable,#1]{#2}}
    \newcommandx{\changeD}[2][1=]{\todo[disable,#1]{#2}}
    \newcommandx{\changeE}[2][1=]{\todo[disable,#1]{#2}}

\fi

\newif\ifsubmissionmicro
\submissionmicrotrue
\ifsubmissionmicro
    \newcommand{\juan}[1]{\textcolor{black}{#1}}
    \newcommand{\juani}[1]{\textcolor{black}{#1}}
    \newcommand{\sg}[1]{\textcolor{black}{#1}}
    \newcommand\jgl[1][0]{}
    \newcommand\gfbox[1][0]{}
    \newcommand{\gf}[1]{\textcolor{black}{#1}}
    \newcommand{\gfi}[1]{\textcolor{black}{#1}}

    \newcommand{\atb}[1]{\textcolor{black}{#1}}
    \newcommand{\agy}[1]{\textcolor{black}{#1}}
    \newcommand{\agycomment}[1]{}
\else
    \newcommand{\juan}[1]{\textcolor{black}{#1}}
    \newcommand{\juani}[1]{\textcolor{dollarbill}{#1}}
    \newcommand\jgl[1]{\noindent{\color{dollarbill} {\bf \fbox{JGL}} {\it#1}}}
    
    \newcommand{\sg}[1]{\textcolor{purple}{#1}}

    \newcommand{\gf}[1]{\textcolor{blue}{#1}}
    \newcommand{\gfi}[1]{\textcolor{orange}{#1}}
    \newcommand\gfbox[1]{\noindent{\color{red} {\bf \fbox{TODO:}} {\it#1}}}
    \definecolor{ao}{rgb}{0.0, 0.5, 0.0}
    \newcommand{\atb}[1]{\textcolor{ao}{#1}}
    \newcommand{\agy}[1]{\textcolor{orange}{#1}}
    \newcommand{\agycomment}[1]{\textcolor{orange}{\textbf{[agy:} #1\textbf{]}}}
\fi 

\newif\ifsubmissionhpca
\submissionhpcatrue
\ifsubmissionhpca
    \newcommand{\gfhpca}[1]{\textcolor{black}{#1}}
\else
    \newcommand{\gfhpca}[1]{\textcolor{blue}{#1}}  
\fi 

\newif\ifrevisionhpca
\revisionhpcafalse
\ifrevisionhpca 
    \newcommand{\rA}[1]{\textcolor{red}{#1}}
    \newcommand{\rB}[1]{\textcolor{purple}{#1}}
    \newcommand{\rC}[1]{\textcolor{blush}{#1}}
    \newcommand{\rD}[1]{\textcolor{burntorange}{#1}}
    \newcommand{\rE}[1]{\textcolor{ao}{#1}}
    \newcommand{\rCommon}[1]{\textcolor{blue}{#1}}

    \let\marginpar\marginnote
    \newcommandx{\changerCM}[2][1=]{\todo[linecolor=blue,backgroundcolor=blue!25,bordercolor=blue,#1,size=\scriptsize]{\revCommon{\textbf{#2}}}}
    
    \newcommandx{\changerA}[2][1=]{\todo[linecolor=red,backgroundcolor=red!25,bordercolor=red,#1,size=\scriptsize]{\revA{\textbf{#2}}}}
    
    \newcommandx{\changerB}[2][1=]{\todo[linecolor=purple,backgroundcolor=purple!25,bordercolor=purple,#1,size=\scriptsize]{\revB{\textbf{#2}}}}
    
    \newcommandx{\changerC}[2][1=]{\todo[linecolor=blush,backgroundcolor=blush!25,bordercolor=blush,#1,size=\scriptsize]{\revC{\textbf{#2}}}}
    
    \newcommandx{\changerD}[2][1=]{\todo[linecolor=orange,backgroundcolor=orange!25,bordercolor=orange,#1,size=\scriptsize]{\revD{\textbf{#2}}}}
    
    \newcommandx{\changerE}[2][1=]{\todo[linecolor=ao,backgroundcolor=ao!25,bordercolor=ao,#1,size=\scriptsize]{\revE{\textbf{#2}}}}

\else
    \newcommand{\rA}[1]{\textcolor{black}{#1}}
    \newcommand{\rB}[1]{\textcolor{black}{#1}}
    \newcommand{\rC}[1]{\textcolor{black}{#1}}
    \newcommand{\rD}[1]{\textcolor{black}{#1}}
    \newcommand{\rE}[1]{\textcolor{black}{#1}}
    \newcommand{\rCommon}[1]{\textcolor{black}{#1}}

    \newcommandx{\changerCM}[2][1=]{\todo[disable,#1]{#2}}
    \newcommandx{\changerA}[2][1=]{\todo[disable,#1]{#2}}
    \newcommandx{\changerB}[2][1=]{\todo[disable,#1]{#2}}
    \newcommandx{\changerC}[2][1=]{\todo[disable,#1]{#2}}
    \newcommandx{\changerD}[2][1=]{\todo[disable,#1]{#2}}
    \newcommandx{\changerE}[2][1=]{\todo[disable,#1]{#2}}

\fi

\newcommand\pnm{\cite{farmahini2015nda,babarinsa2015jafar,devaux2019true,ghiasi2022genstore,gomez2021benchmarkingcut,gomezluna2021benchmarking,gomez2022benchmarking,syncron,singh2020nero,skhynixpim,ke2021near,giannoula2022sparsep,shin2018mcdram,cho2020mcdram,denzler2021casper,asghari2016chameleon,IRAM_Micro_1997,C_RAM_1999,CASES_MVX,Xi_2015,sun2021abc,matam2019graphssd,gokhale1995processing,hall1999mapping,MEMSYS_MVX,lockerman2020livia,ahn2015scalable,nai2017graphpim,boroumand2018google,lazypim, top-pim, gao2016hrl, kim2018grim, drumond2017mondrian, RVU, NIM, PEI, gao2017tetris,Kim2016,gu2016leveraging, boroumand2019conda, hsieh2016transparent, cali2020genasm, NDC_ISPASS_2014,pattnaik2016scheduling,akin2015data,hsieh2016accelerating,lee2015bssync,boroumand2021mitigating,boroumand2021google,boroumand2022polynesia,boroumand2021polynesia,amiraliphd,besta2021sisa,fernandez2020natsa,singh2019napel,kwon202125,lee2021hardware,niu2022184qps,Sparse_MM_LiM,azarkhish2016logic,azarkhish2018neurostream,guo20143d,de2018design,akin2014hamlet,huang2020heterogeneous,dai2018graphh,liu2018processing,tsai:micro:2018:ams,gu2020ipim,DRAMA_CAL_2014,Asghari-Moghaddam_2016,huang2019active,kersey2017lightweight,li2019pims,kim2017grim,boroumand2017lazypim,zhuo2019graphq,zhang2018graphp,lim2017triple,smc_sim,HIVE,jang2019charon,IBM_ActiveCube,hadidi2017cairo,santos2018processing,lenjani2020fulcrum}\xspace}

\newcommand\pim{\cite{farmahini2015nda,babarinsa2015jafar,devaux2019true,ghiasi2022genstore,gomez2021benchmarkingcut,gomezluna2021benchmarking,gomez2022benchmarking,syncron,singh2020nero,skhynixpim,ke2021near,giannoula2022sparsep,shin2018mcdram,cho2020mcdram,denzler2021casper,asghari2016chameleon,IRAM_Micro_1997,C_RAM_1999,CASES_MVX,Xi_2015,sun2021abc,matam2019graphssd,gokhale1995processing,hall1999mapping,MEMSYS_MVX,lockerman2020livia,ahn2015scalable,nai2017graphpim,boroumand2018google,lazypim, top-pim, gao2016hrl, kim2018grim, drumond2017mondrian, RVU, NIM, PEI, gao2017tetris,Kim2016,gu2016leveraging, boroumand2019conda, hsieh2016transparent, cali2020genasm, NDC_ISPASS_2014,pattnaik2016scheduling,akin2015data,hsieh2016accelerating,lee2015bssync,boroumand2021mitigating,boroumand2021google,boroumand2022polynesia,boroumand2021polynesia,amiraliphd,besta2021sisa,fernandez2020natsa,singh2019napel,kwon202125,lee2021hardware,niu2022184qps,Sparse_MM_LiM,azarkhish2016logic,azarkhish2018neurostream,guo20143d,de2018design,akin2014hamlet,huang2020heterogeneous,dai2018graphh,liu2018processing,tsai:micro:2018:ams,gu2020ipim,DRAMA_CAL_2014,Asghari-Moghaddam_2016,huang2019active,kersey2017lightweight,li2019pims,kim2017grim,boroumand2017lazypim,zhuo2019graphq,zhang2018graphp,lim2017triple,smc_sim,HIVE,jang2019charon,IBM_ActiveCube,hadidi2017cairo,santos2018processing,Chi2016, Shafiee2016, seshadri2017ambit, seshadri2019dram, li2017drisa, seshadri2013rowclone, seshadri2016processing, deng2018dracc, xin2020elp2im, song2018graphr, song2017pipelayer,gao2019computedram, eckert2018neural, aga2017compute,dualitycache,seshadri2016buddy,seshadri.bookchapter17,seshadri2018rowclone,seshadri2015fast,li2016pinatubo,ferreira2021pluto,ferreira2022pluto,imani2019floatpim,he2020sparse,flashcosmos,truong2022adapting,truong2021racer,olgun2021quactrng,kim2019d,kim2018dram,bostanci2022dr,olgun2022pidram,ali2019memory,angizi2019graphide,li2018scope,subramaniyan2017parallel,zha2020hyper,fujiki2018memory,orosa2021codic,sharad2013ultra,rezaei2020nom,
gao2021parabit,choi2020flash,han2019novel,merrikh2017high,wang2018three,lue2019optimal,kim2021behemoth,wang2022memcore,han2021flash,kang2021s,lee2020neuromorphic,lee20223d,si2019dual,
simon2020blade,nag2019gencache,wang2019bit,al2020towards,kang2014energy,kim2021colonnade,jiang2020c3sram,jeloka201628,wang2023infinity,kang2015energy, imani2020dual,chang2016low, hajinazarsimdram,deng2019lacc,sutradhar2021look,sutradhar2020ppim,lenjani2020fulcrum,peng2023chopper,oliveira2022accelerating,singh2021fpga,oliveira2023dappa,oliveira2022methodologies,oliveira2022heterogeneous,shahroodi2023swordfish,chen2023simplepim,gupta2023evaluating,gomez2023evaluating,oliveira2023transpimlib,diab2023framework,mao2022genpip,singh2022accelerating}\xspace}

\newcommand\pum{\cite{Chi2016, Shafiee2016, seshadri2017ambit, seshadri2019dram, li2017drisa, seshadri2013rowclone, seshadri2016processing, deng2018dracc, xin2020elp2im, song2018graphr, song2017pipelayer,gao2019computedram, eckert2018neural, aga2017compute,dualitycache,besta2021sisa,seshadri2016buddy,seshadri.bookchapter17,seshadri2018rowclone,seshadri2015fast,li2016pinatubo,ferreira2021pluto,ferreira2022pluto,imani2019floatpim,he2020sparse,flashcosmos,truong2022adapting,truong2021racer,olgun2021quactrng,kim2019d,kim2018dram,bostanci2022dr,olgun2022pidram,ali2019memory,angizi2019graphide,li2018scope,subramaniyan2017parallel,zha2020hyper,fujiki2018memory,orosa2021codic,sharad2013ultra,rezaei2020nom,gao2021parabit,choi2020flash,han2019novel,merrikh2017high,wang2018three,lue2019optimal,kim2021behemoth,wang2022memcore,han2021flash,kang2021s,lee2020neuromorphic,lee20223d,si2019dual,
simon2020blade,nag2019gencache,wang2019bit,al2020towards,kang2014energy,kim2021colonnade,jiang2020c3sram,jeloka201628,wang2023infinity,kang2015energy,imani2020dual, chang2016low,hajinazarsimdram,deng2019lacc,sutradhar2021look,sutradhar2020ppim,peng2023chopper,shahroodi2023swordfish}\xspace}

\newcommand\drampum{\cite{angizi2019graphide,besta2021sisa,bostanci2022dr,deng2018dracc,ferreira2021pluto,ferreira2022pluto,gao2019computedram,li2017drisa,li2018scope,olgun2021quactrng,olgun2022pidram,seshadri.bookchapter17, seshadri2013rowclone,seshadri2015fast,seshadri2016buddy, seshadri2016processing, seshadri2017ambit,seshadri2018rowclone, seshadri2019dram, xin2020elp2im}\xspace}

\newcommand\srampum{\cite{aga2017compute,eckert2018neural,si2019dual,simon2020blade,nag2019gencache,wang2019bit,al2020towards,kang2014energy,kim2021colonnade,jiang2020c3sram,jeloka201628,wang2023infinity,kang2015energy}\xspace}

\newcommand\nvmpum{\cite{imani2019floatpim,li2016pinatubo,Shafiee2016,song2017pipelayer,song2018graphr,truong2021racer,truong2022adapting,sharad2013ultra,Chi2016,imani2020dual}\xspace}

\newcommand\flashpum{\cite{flashcosmos,gao2021parabit,choi2020flash,han2019novel,merrikh2017high,wang2018three,lue2019optimal,kim2021behemoth,wang2022memcore,han2021flash,kang2021s,lee2020neuromorphic,lee20223d}\xspace}

\newcommand\ambit{\cite{seshadri2017ambit,seshadri2019dram,seshadri2015fast,seshadri.bookchapter17,seshadri2016buddy,seshadri2016processing}\xspace}

\newcommand\drambackgroundshort{\cite{ ghose.sigmetrics20, seshadri2019dram, kim2012case, zhang2014half,  seshadri2017ambit, lee2015adaptive, seshadri2013rowclone, Dennard68field,
keeth2007dram,
yauglikcci2022hira,
luo2023rowpress,
keeth2001dram,oconnor2017fine}\xspace}

\newcommand\pnmshort{\cite{devaux2019true,ghiasi2022genstore,gomez2021benchmarkingcut,gomezluna2021benchmarking,gomez2022benchmarking,syncron,singh2020nero,skhynixpim,ke2021near,giannoula2022sparsep,denzler2021casper,IRAM_Micro_1997,C_RAM_1999,gokhale1995processing,hall1999mapping,ahn2015scalable,boroumand2018google,lazypim, top-pim, kim2018grim, RVU, NIM, PEI,Kim2016, boroumand2019conda, hsieh2016transparent, cali2020genasm,hsieh2016accelerating,boroumand2021mitigating,boroumand2021google,boroumand2022polynesia,boroumand2021polynesia,besta2021sisa,fernandez2020natsa,singh2019napel,lee2021hardware,kim2017grim,boroumand2017lazypim,santos2018processing,lenjani2020fulcrum}\xspace}

\titlespacing\section{0pt}{5pt plus 2pt minus 2pt}{0pt plus 2pt minus 2pt}
\titlespacing\subsection{0pt}{5pt plus 2pt minus 2pt}{0pt plus 2pt minus 2pt}
\titlespacing\subsubsection{0pt}{5pt plus 2pt minus 2pt}{0pt plus 2pt minus 2pt}

\makeatletter
\g@addto@macro{\normalsize}{%
  \setlength{\abovedisplayskip}{2pt plus 0.5pt minus 1pt}
  \setlength{\belowdisplayskip}{2pt plus 0.5pt minus 1pt}
  \setlength{\abovedisplayshortskip}{0pt}
  \setlength{\belowdisplayshortskip}{0pt}
  \setlength{\intextsep}{2pt plus 1pt minus 1pt}
  \setlength{\textfloatsep}{2pt plus 1pt minus 1pt}
  \setlength{\skip\footins}{4pt plus 1pt minus 1pt}}
\makeatother

\renewcommand*\footnoterule{}

\newif\ifcameraready
\camerareadytrue
\ifcameraready
    \newcommand{\gfcr}[1]{\textcolor{black}{#1}} 
    \newcommand{\gfcri}[1]{\textcolor{black}{#1}} 
     
    \newcommand{\gfcrii}[1]{\textcolor{black}{#1}} 
    \newcommand{\gfcriii}[1]{\textcolor{black}{#1}}

    \newcommand{\om}[1]{\textcolor{black}{#1}} 

    \newcommand{\omi}[1]{\textcolor{black}{#1}} 
    \newcommand{\omii}[1]{\textcolor{black}{#1}} 

    \newcommand{\omiii}[1]{\textcolor{black}{#1}} 
\else 
    \newcommand{\gfcr}[1]{\textcolor{black}{#1}} 
    \newcommand{\gfcri}[1]{\textcolor{black}{#1}} 
    \newcommand{\gfcrii}[1]{\textcolor{black}{#1}} 
    
    \newcommand{\gfcriii}[1]{\textcolor{red}{#1}}

    \newcommand{\om}[1]{\textcolor{black}{#1}} 

    \newcommand{\omi}[1]{\textcolor{black}{#1}} 

    \newcommand{\omii}[1]{\textcolor{red}{#1}} 

    \newcommand{\omiii}[1]{\textcolor{blue}{#1}} 

    \let\marginpar\marginnote
\fi

\newcommand{\versionnum}[0]{4.1}

%% file: sections/00_abstract.tex
\begin{abstract}

\gfi{\gls{PuD}} is a \gls{PIM} approach that \gf{uses} \om{a} DRAM \omi{array's} massive \om{internal}  parallelism to execute \om{very}-wide \gfcri{(e.g., \gfcrii{16,384--262,144-\omii{bit-}wide}) } \om{data-parallel} operations\gfcri{, in a \gls{SIMD} fashion}. However, DRAM rows' large and rigid granularity \gfcr{limit} the effectiveness and applicability
of \gfi{\gls{PuD}} in \gf{three} ways. 
First, since \gfcrii{applications \omiii{have} varying degrees of \gls{SIMD} parallelism} \om{(which is often smaller than the DRAM row granularity)}, \gfi{\gls{PuD}}  execution \om{often} leads
to \underutilization, throughput loss, and energy waste. 
Second, due to \om{the} \gfcri{high area cost of implementing} \om{interconnects} that connect \gfcri{columns in a wide DRAM row}\revdel{ \gf{in} a DRAM \gfcri{subarray}}, 
most \gfi{\gls{PuD}}  architectures are limited to the execution of \gfcr{parallel} map operations\gfcri{,} where \gfcrii{a single \omii{operation} is performed over equally-sized input and output arrays}.
\gf{Third, the \gfcri{need to feed the wide DRAM row with \omii{tens of} thousands of data \gfcrii{elements} \gfcri{combined with the} lack \gfi{of \om{adequate} compiler support for \gfi{\gls{PuD}}} \gfcrii{systems create a {programmability barrier}}, since \om{programmers} need to {manually} extract \gls{SIMD} parallelism from an application and map computation to the  \om{\gls{PuD}} \gfi{hardware}.}}

Our \textbf{goal} is to design a flexible \gls{PuD} system that overcomes the\revdel{ three} limitations caused by the large and rigid granularity of \gls{PuD}. To \om{this end}, we propose \prop, a hardware/software co-designed \gls{PuD} system that introduces new mechanisms to allocate and control {only} the \om{necessary}\revdel{ computing} resources for \om{a given} \gls{PuD} \omiii{operation}. 
The key idea of \prop is to leverage fine-grained DRAM (i.e., the ability to \omii{independently} access \omi{smaller} \gfcri{segments} of a \omi{large} DRAM row) for \gls{PuD} \omi{computation}. \changeD{\#D1}\revD{\prop \gfcrii{exploits this key idea to} enable a \gfcri{\gls{MIMD}} execution model \om{in each} DRAM \om{subarray} (and \gls{SIMD} execution within each \gfcri{DRAM \omi{row}} segment).} 
\revdel{On the hardware side, \prop does simple changes to 
\li~DRAM's row access circuitry to enable concurrent execution of \gls{PuD} operations in segments of the DRAM row; and 
\lii~to the\revdel{ local and global} DRAM I/O circuitry to allow data to move across DRAM columns \revD{(enabling native support for \gls{PuD} reduction operations)}. 
On the software side, \prop implements compiler passes to 
\li~identify and generate \gls{PuD} operations with the appropriate \gls{SIMD} granularity; and 
\lii~schedule the concurrent execution of independent \gls{PuD} operations.}
%
% We evaluate \prop's performance and energy efficiency using a wide range of real-world applications. Our evaluations show that \prop provides \add{}$\times$ and \add{}$\times$ the performance, and \add{}$\times$ and \add{}$\times$ the energy efficiency of a CPU and a state-of-the-art \gfi{\gls{PuD}}  framework (SIMDRAM), respectively, over \add{} applications while incurring an \add{}\% area overhead on top of a DRAM chip.

We evaluate \prop using \gfcrii{twelve} real-world applications \gfcrii{and 495 multi-programmed application mixes}. 
Our evaluation shows that \prop  \gfcri{\om{provides} 34$\times$ the performance, \efficiencysimdram the energy efficiency, 1.7$\times$ the throughput, and 1.3$\times$ the fairness of a state-of-the-art \gfi{\gls{PuD}}  framework\omi{, along with}} 
\efficiencycpu \gfcri{and} \efficiencygpu the energy efficiency of a high-end CPU \gfcri{and} GPU, \gfcri{respectively}. \prop \gfi{adds} \om{small} area cost \om{to} a DRAM chip (1.11\%) and CPU die (0.6\%). \omi{We hope and believe that \prop's ideas and results will help \omii{to enable} more efficient and easy-to-program \gls{PuD} systems. \omii{To this end, we open source} \prop at \url{https://github.com/CMU-SAFARI/MIMDRAM}.}

\end{abstract}

%% file: sections/01_introduction.tex
\section{Introduction}
\label{sec:introduction}
\glsresetall

\omi{\emph{Data movement} between computation units (e.g., CPUs, GPUs) and  main memory (e.g., DRAM) is a major \emph{performance} and \emph{energy bottleneck} in current computing systems~\gfcri{\cite{
mutlu2013memory,
mutlu2015research,
dean2013tail,
kanev_isca2015,
mutlu2019enabling,
mutlu2019processing,
mutlu2020intelligent,
ghose.ibmjrd19,
mutlu2020modern,
boroumand2018google, 
wang2016reducing, 
mckee2004reflections,
wilkes2001memory,
kim2012case,
wulf1995hitting,
ghose.sigmetrics20,
ahn2015scalable,
PEI,
sites1996,
deoliveira2021IEEE}}, and is expected to worsen due to the increasing data intensiveness of modern applications~\cite{sevilla2022compute,gholami2020ai_and_memory_wall}.} 
To mitigate the overheads \sg{caused by data movement}, several works propose \gls{PIM} architectures~\gfcriii{\pim}\revdel{, where computation is moved \juan{closer to}~\pnm or \omi{inside}~\pum memory \om{chips}}.
\gfcri{There are two main approaches to \gls{PIM}~\cite{ghose.ibmjrd19, mutlu2020modern}:
\li~\gls{PnM}~\gfcriii{\pnm}, where \omi{computation} logic is added \omi{near} the \omi{memory arrays (e.g., in a DRAM chip or at the logic layer of a 3D-stacked memory~\cite{HMC2, HBM,lee2016simultaneous})}; and
\lii~\gls{PuM}~\pum, where computation is performed by exploiting the analog operational properties of the memory \omi{circuitry}.}
\gfcrii{There are two main advantage\gfcrii{s} of \gls{PuM} \omi{over} \gls{PnM}. First, \gls{PuM}  \emph{fundamentally} \emph{eliminates} data movement by performing computation \emph{in situ}, while data movement still occurs between computation units and memory arrays in \gls{PnM}.} 
Second, \gls{PuM} architectures \omi{exploit}  the large internal bandwidth and parallelism available \omi{\emph{inside}} the memory arrays, while \gls{PnM} solutions are \emph{fundamentally} bottlenecked by the memory's internal data buses.

\gfcri{\gls{PuM} architectures can be implemented using different memory technologies\omi{,} including} SRAM~\gfcriii{\srampum}, DRAM~\drampum, emerging non-volatile~\gfcriii{\nvmpum} or \omi{NAND} flash~\gfcriii{\flashpum}.
\gfi{\gls{PuD}}~\drampum, \gfi{in \sg{particular}}, \gfcri{enables the execution of different \emph{bulk} operations in DRAM (i.e., \gls{PuD} operations), such as 
\li~data copy \gfcri{and initialization}~\cite{seshadri.bookchapter17, seshadri2013rowclone,seshadri2018rowclone,chang2016low},
\lii~\om{bitwise} Boolean operations~\cite{seshadri2017ambit, gao2019computedram, xin2020elp2im, besta2021sisa, li2017drisa}, 
\liii~arithmetic operations~\cite{deng2018dracc, gao2019computedram,li2017drisa,angizi2019graphide, hajinazarsimdram,li2018scope}, and 
\liv~lookup table \omi{based} operations~\gfcri{\cite{ferreira2021pluto,ferreira2022pluto,deng2019lacc,sutradhar2021look,sutradhar2020ppim}}}. 

%Based on these prior works, follow-up proposals have implemented frameworks that extend the functionalities of \gls{PuM} architectures by constructing new and more complex operations. For example, SIMDRAM~\cite{hajinazarsimdram}, which leverages a \gfi{\gls{PuD}} substrate proposed by Ambit~\ambit, designs a three-step methodology that allows the user to implement \emph{any} complex \gfi{\gls{PuD}} operation by orchestrating sequences of DRAM commands. 
\gfcrii{\gfi{\gls{PuD}} architectures commonly employ bit-serial computation~\cite{angizi2019graphide,besta2021sisa,gao2019computedram,seshadri.bookchapter17, seshadri2013rowclone,seshadri2015fast,seshadri2016buddy, seshadri2016processing, seshadri2017ambit,seshadri2018rowclone, seshadri2019dram, xin2020elp2im, hajinazarsimdram}, where they map \gfcrii{each data element} of a \gls{PuD} operation to a DRAM column.\footnote{\gfcri{We provide a detailed background on DRAM organization in \cref{sec:background:dram}.}} 
\gfcrii{\omi{A} \gls{PuD} architecture that leverages bit-serial computation effectively enables a very-wide \gls{SIMD} execution \gfcrii{model} in DRAM, with a \emph{large} and \emph{rigid} computation granularity.}
\gfcrii{The \emph{large} computation granularity stems from the fact that} bit-serial computation turns each one of the \emph{many}
DRAM columns in a DRAM subarray into a computing engine.
For example, there are \gfcrii{16,384--262,144} DRAM columns in a \gfcrii{DDR4 DRAM~\cite{micron2014ddr4} subarray}, and each can execute \gfcrii{a \emph{single} \gls{PuD} operation over \emph{multiple} data elements stored on the DRAM columns.} 
\gfcrii{The \emph{rigid} computation granularity stems from the fact that} the granularity \omii{at} which DRAM rows are accessed dictates the computation granularity of a \gls{PuD} operation. 
In commodity DRAM chips, all DRAM row accesses happen at a \emph{fixed} granularity\omi{:} the DRAM access circuitry addresses \emph{\omi{all}} DRAM columns in a DRAM row \emph{simultaneously}. As such, \emph{every} \gls{PuD} operation operates simultaneously on \emph{all} the 16,384 to 262,144 data elements (one per DRAM column) a DDR4 DRAM subarray stores, for example.}

%\gfi{\gls{PuD}} operations are executed in a \gls{SIMD} manner by storing all bits of a operand vertically in a single column of a DRAM array. 
\revdel{In this execution model, the \gfhpca{granularity where DRAM rows are opened (i.e., activated)} fixed size of the DRAM row (e.g.,  \SI{8}{\kilo\byte} in DDR4 DRAM~\cite{standard2012jesd79}) defines the \gfi{computation granularity}.}

\sg{We highlight three limitations} of state-of-the-art \gfi{\gls{PuD}} architectures that stem from the \emph{large} and \emph{rigid} granularity of \gfhpca{\gls{PuD}'s \om{very}-wide \gls{SIMD} execution model.} 
%\juan{DRAM rows}. 
First, to \revdel{make \juan{efficient} use of the %bulk \gls{SIMD} 
parallelism available inside DRAM and }sustain high \omii{\emph{\gls{SIMD} utilization}} \gfcri{(i.e., the fraction of \gls{SIMD} lanes executing a useful operation)}, \om{each} \gfi{\gls{PuD}} \om{operation needs to operate on} a \emph{large} amount of data. \om{Unfortunately, not all applications \omii{have} the required large amount of data parallelism to sustain high \gls{SIMD} utilization in \gls{PuD} architectures. Our} analysis of \gfcrii{twelve} general-purpose  applications \gfcrii{(compiled for a high-end CPU system)} shows that the degree of \gls{SIMD} parallelism an application \gfcrii{has} \emph{varies} significantly, from as low as 8 to as high as 134,217,729 \omi{data elements} per \gls{SIMD} instruction (\cref{sec:motivation}). 
Second, the large granularity of \gfi{\gls{PuD}} execution makes \sg{it \gfcri{costly} to implement} \gfcri{interconnects that connect columns in a DRAM row. \gfcri{Such \omi{interconnects} could allow for the implementation of}} \gfi{\gls{PuD}} operations that require \gfcri{shifting data across DRAM columns, which is a common computing pattern present in vector-to-scalar reduction operations, \omi{e.g.,} \texttt{sum += A[i]}}. This \sg{limits} \gfi{\gls{PuD}} operations to \emph{only} parallel map operations \gfcri{(i.e., operations where an output data of the same \omi{dimension} as the input data is produced by applying a computation \omi{independently} to each input operand)}. \gfcri{\gfi{A} prior work~\cite{li2017drisa} proposes modifications to the DRAM array to enable communication across DRAM columns\gfi{. However,} \juani{the \omi{interconnection network that this work proposes can lead} to prohibitive area \gfi{cost} \omi{in} commodity DRAM \omi{chips} \gfcri{(i.e., 21\% DRAM area overhead, see \cref{sec:eval:otherpims})}.}} 
\gfcri{Third, the need to feed a \omii{very-}wide DRAM
row with thousands of data \gfcrii{elements} combined with the
lack of adequate compilation tools that can identify very-wide \gls{SIMD}  instructions in general-purpose applications and map such instructions to equivalent \gls{PuD} operations create a \emph{programmability barrier} for \gls{PuD} architectures. In state-of-the-art \gls{PuD} architectures~\gfcrii{\cite{angizi2019graphide,besta2021sisa,bostanci2022dr,deng2018dracc,ferreira2022pluto,gao2019computedram,li2017drisa,li2018scope,olgun2021quactrng,olgun2022pidram, seshadri2013rowclone, seshadri2016processing, seshadri2017ambit, seshadri2019dram, xin2020elp2im, hajinazarsimdram}}, the programmer needs to \emph{manually} extract a \emph{large} and \emph{fixed} amount of data-parallelism from an application and map computation to the underlying \gls{PuD} hardware, which is a daunting task.}

Our \emph{goal} is to design a flexible \gfi{\gls{PuD}} substrate that overcomes the \gf{three} limitations caused by the large and rigid granularity of \gfi{\gls{PuD}} execution. To \om{this end}, we propose \prop, a hardware/software co-designed \gfi{\gls{PuD}} \gfi{system} that introduces the ability to allocate and control \emph{only} the \om{required amount of} computing resources inside the DRAM \gfcri{subarray} for \gfi{\gls{PuD}} computation. The \emph{key idea} of \prop is to leverage fine-grained DRAM for \gfi{\gls{PuD}} operations. 
Prior works on fine-grained DRAM~\gfcriii{\cite{cooper2010fine,udipi2010rethinking,zhang2014half,ha2016improving,lee2017partial,olgun2022sectored,o2021energy,oconnor2017fine}} \gfcri{leverage the hierarchical design of a DRAM subarray \gfcrii{to enable DRAM row accesses with \emph{flexible} granularity. A DRAM subarray is composed of multiple (e.g., \gfcrii{32--128}) smaller 2D arrays of \gfcrii{512--1024} DRAM rows and \gfcrii{512--1024} columns, called \emph{DRAM mats}~\cite{standard2012jesd79,seshadri2019dram,seshadri2016processing,zhang2014half}.
\gfcrii{During a row access, the DRAM access circuitry \emph{simultaneously} addresses \emph{all} columns across \emph{all} mats in a subarray, composing a \emph{large} DRAM row of size \texttt{[\#columns\_per\_mat $\times$ \#mats\_per\_subarray]}.
Fine-grained DRAM \omiii{modifies} the DRAM access circuitry to}} enable reading/writing data \omi{from/}to \gfcri{individual DRAM mats}, allowing the access of DRAM rows with a \omii{smaller} number of columns. } 

\gfcrii{Inspired by fine-grained DRAM, we propose
simple modifications to the DRAM access circuitry to enable
addressing individual DRAM mats during \gls{PuD} computation.
Fine-grained DRAM for \gls{PuD}  computation brings four main advantages. 
First, by addressing individual DRAM mats, \prop better matches the granularity of a \gls{PuD} operation to the available data-parallelism \omii{present in} the application.
Second, with fewer DRAM columns per \gls{PuD} operation, \prop makes it feasible to implement low-cost interconnects inside a DRAM subarray, allowing for data movement \emph{within} and \emph{across} DRAM mats.
Third, since the number of columns in a single DRAM mat
is on par with the number of \gls{SIMD} lanes in modern processors' \gls{SIMD} engines  (e.g., 512 \gls{SIMD} lanes in  AVX-51 \gls{SIMD} engines2~\cite{firasta2008intel}), \prop can leverage
traditional compilers to map \gls{SIMD} instructions to \gls{PuD} operations.
Fourth, \prop leverages \emph{unused} DRAM
mats to \omii{\emph{concurrently}} execute \emph{independent} \gls{PuD} operations
\emph{across} DRAM mats in a \emph{single} DRAM subarray. This enables the \gls{PuD} substrate to execute a \emph{not-so-wide} \gls{PuD} \omii{operation} in a subset of the DRAM mats and other independent \gls{PuD} operations across the remaining DRAM mats within a single DRAM subarray, in a \gls{MIMD} fashion~\cite{smith1986pipelined,flynn1966very,thornton1964parallel}.}

% \revD{This opens up a new opportunity for \gls{PuD} \omi{computation}: by leveraging the unused \gfcri{segments} of \gfcri{a} DRAM row \gfcri{across different DRAM mats}, the \gls{PuD} substrate can execute \emph{multiple} \gls{PuD} \gfcri{operations} over \emph{multiple} input data \emph{concurrently} in a single DRAM \gfcri{subarray}. \gfcri{Practically allowing \prop to execute \gls{PuD} operations} in a \gls{MIMD} fashion~\cite{smith1986pipelined,flynn1966very,thornton1964parallel}.
%can execute data-independent \gls{SIMD}operations within the DRAM array in a \gls{MIMD} fashion. This way, different portions of a DRAM array can execute \emph{multiple} \gls{PuD} \gls{SIMD} \gfhpca{instructions} over \emph{multiple} input data concurrently, improving throughput and utilization.

%, enabling a \gls{MIMD} execution model within the DRAM array, where different portions of a DRAM array execute multiple \gfi{\gls{PuD} \gls{SIMD}} operations over multiple input data concurrently. 

\prop leverages fine-grained DRAM for \gfi{\gls{PuD}} in hardware and software. On the hardware side, \prop \gf{proposes simple modifications to the DRAM \gfcri{subarray} and includes new mechanisms to the memory controller that allow the execution of \gfcri{independent} \gfi{\gls{PuD}} \gfcri{operations across the DRAM mats in a single  subarray}.  Concretely, \gfcrii{\prop includes}}  
\li~latches, isolation transistors, and selection logic in the DRAM \gfcri{subarray}'s \gfcri{access} circuitry to enable the execution of \gfcrii{independent} \gfi{\gls{PuD}} \gfcri{operations \omi{in} different DRAM mats}; 
\lii~two different low-cost \gfcri{interconnects} placed \omi{respectively} at the local and global DRAM I/O circuitry\sg{, which enable communication} across columns of a DRAM row at varying granularities \gfcr{and} \revdel{\sg{allowing for} }the execution of vector\gfcri{-to-scalar} reduction \gfcri{in DRAM} at low \omi{hardware} cost. 
\gf{\omii{In} the memory \sg{controller}, \prop includes a new control unit that orchestrates the concurrent execution of \gfcri{independent} \gfi{\gls{PuD}} \gfcri{operations} \gfcri{\omii{in} different \omii{mats} of a DRAM subarray}.} 
On the software side, \prop \gfi{implements} compiler passes to
\li~automatically vectorize code regions that can benefit from \gfi{\gls{PuD}} execution \omi{(called \gls{PuD}-friendly regions)};
\lii~\omi{for such regions,} generate \gfi{\gls{PuD}} \gfcri{operations} with the most appropriate \gls{SIMD} granularity; and
\liii~schedule the concurrent execution of \gfcri{independent} \gfi{\gls{PuD}} \gfcri{operations \omi{in} different DRAM mats}. \gf{\gfcri{W}e discuss how to integrate \prop in a real system\gfcr{, including how \prop deals with 
\li~data allocation \gfcri{\omi{within} a DRAM subarray} and 
\lii~mapping \gfcri{of \omi{a} \gls{PuD}'s operands to guarantee high utilization of the \gls{PuD} substrate\revdel{ and low data movement across DRAM mats}.}}}

We evaluate \prop's \gf{\gfcri{performance, energy efficiency, throughput, fairness}\gf{, and \gf{SIMD utilization}}} for \gfcri{twelve} real-world applications from four popular benchmark suites (i.e., Phoenix~\cite{yoo_iiswc2009}, Polybench~\cite{pouchet2012polybench}, Rodinia~\cite{che_iiswc2009}, and SPEC 2017~\cite{spec2017}) \gfcri{and 495 multi-programmed application mixes}. 
Our evaluation shows that \prop \om{provides} \gfcri{34$\times$ the performance}, \efficiencysimdram the energy efficiency, 1.7$\times$ the throughput, 1.3$\times$ the fairness, \gfcri{and 18.6$\times$ the \gls{SIMD} utilization} of \gfcri{SIMDRAM~\cite{hajinazarsimdram}} (a state-of-the-art \gfi{\gls{PuD}} framework); \gfcri{while providing} \efficiencycpu and \efficiencygpu the energy efficiency of a high-end CPU \gfcri{and GPU, respectively}. 
\gfcri{\omii{MIMDRAM}\omi{'s improvements are}  due to its ability to 
\li~maximize the utilization of the  \gls{PuD} substrate by \emph{concurrently} \omi{executing} independent \gls{PuD} operations within \omi{each} DRAM subarray, and 
\lii~allocate \emph{only} the necessary resources (i.e., \gfcrii{appropriate number of DRAM mats}) for a \gls{PuD} operation.}
\prop \gfcri{adds small} area cost \gfcri{to} a \gfcri{state-of-the-art} DRAM chip (1.11\%) and \gfcri{state-of-the-art} CPU die (0.6\%).

We make the following key contributions:
\begin{itemize}[leftmargin=3mm,itemsep=0mm,parsep=0mm,topsep=0mm]
\item To our knowledge, this is the first work to propose \juani{an end-to-end} \gfcri{processing-using-DRAM (PUD)} \omi{system} for general-purpose applications\omi{,} \gfcri{\omi{which} executes operations in a multiple-instruction multiple-data (MIMD) fashion}. \prop \omi{makes} low-cost modifications to the DRAM \gfcri{subarray design that enable the} \gfcri{\emph{concurrent}} \gfcri{execution of} \gfcri{multiple} \gfcri{independent} \gfi{\gls{PuD}} \gfcri{operations} in a single DRAM \gfcri{subarray}. 

\item We propose compiler passes that \gfcri{\omii{take} as input unmodified C/C++ applications and,} transparently \omii{to} the \gfcri{programmer}, 
\li~identify \gfcri{loops \omi{that are}} \gfcri{suitable} for \gfi{\gls{PuD}} execution,
\lii~transform the \gfcri{source} code to use \gfi{\gls{PuD}} \gfcri{operations}, and \liii~schedule \gfcri{independent \gls{PuD} operations for concurrent execution \omi{in each DRAM subarray}, maximizing \gfcri{utilization} \omi{of} the underlying \gls{PuD} architecture}. 

\item We evaluate \prop with a wide range of general-purpose applications and observe that \omi{it} provides higher energy efficiency \omi{and} system throughput than state-of-the-art \gfi{\gls{PuD}}\gfcri{, CPU, and GPU architectures}. 

\item \gfcrii{We open-source \prop  at \url{https://github.com/CMU-SAFARI/MIMDRAM}}.
\end{itemize}

%% file: sections/02_background.tex
\section{Background}
\label{sec:background}

We first briefly explain the architecture of a typical DRAM chip.\footnote{We refer the reader to \omii{various} prior works~\gfcriii{\drambackgroundshort}~for a more detailed description of the DRAM architecture.} Second, we describe prior \gls{PuD} works that \prop builds on top of. 

% \gf{We briefly explain 
% \li~the architecture of a typical DRAM chip, and 
% \lii~prior processing-using-DRAM \gfi{(PuD)} works.}

% {\color{red} \rule{\linewidth}{0.5mm} \\ \textbf{V3 STALE TEXT BEGIN} \\ \rule{\linewidth}{0.5mm}}

\subsection{DRAM Organization \& \juan{Operation}}
\label{sec:background:dram}

\begin{figure}[ht]
    \centering
    \includegraphics[width=\linewidth]{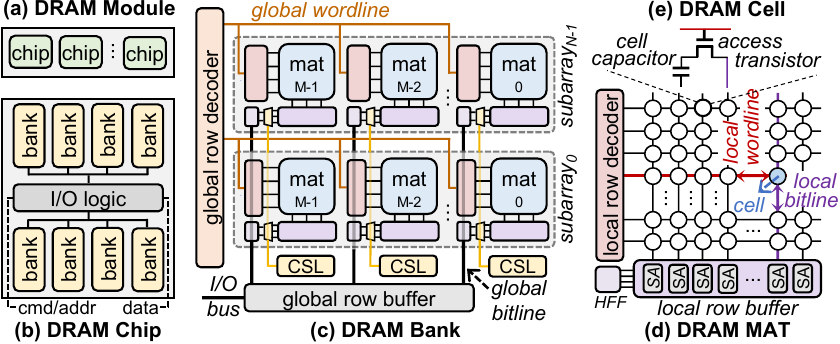}
    \caption{\gfcrii{\omii{Overview} of DRAM organization.}}
    \label{fig_subarray_dram}
\end{figure}

\paratitle{DRAM Organization} A DRAM system comprises \omi{of} a hierarchy of components, as \gfi{Fig.}~\ref{fig_subarray_dram} illustrates\revdel{, starting with a DRAM module at the highest level}. 
A \emph{DRAM module} (\gfi{Fig.}~\ref{fig_subarray_dram}a) \gfi{has} several (e.g., 8--16) DRAM chips. 
A \emph{DRAM chip} (\gfi{Fig.}~\ref{fig_subarray_dram}b) \gfi{has} multiple DRAM banks (e.g., 8--16). 
A \emph{DRAM bank} (\gfi{Fig.}~\ref{fig_subarray_dram}c) \gfi{has} multiple (e.g., 64--128) 2D arrays of DRAM cells known as \emph{DRAM mats}. Several DRAM mats (e.g., 8--16) are grouped in a \emph{DRAM subarray}. \gfcrii{In a DRAM bank, there are three global components that are used
to access the DRAM mats (as Fig.~\ref{fig_subarray_dram}c depicts):}
%three global components are used to access columns of the mats \gfi{in} a subarray:
\li~a \emph{global row decoder} that selects a row of DRAM cells \emph{across} multiple mats \omi{in a subarray},
\lii~a \emph{column select logic} (CSL) that selects portions of the DRAM row based on the column address, and
\liii~a \emph{global sense amplifier} \omi{(i.e., global row buffer)} that transfers the selected fraction of the data from the \omi{selected} DRAM row through the \emph{global \omi{bitlines}}.

A \emph{DRAM mat} (\gfi{Fig.}~\ref{fig_subarray_dram}d) \omi{consists of a 2D array of} DRAM cells \omi{organized} into multiple \emph{rows} (e.g., 512--1024) and multiple \emph{columns} (e.g., 512--1024)~\cite{kim2018solar, lee2017design, kim2002adaptive}. 
A \emph{DRAM cell} (\gfi{Fig.}~\ref{fig_subarray_dram}e) consists of an \emph{access transistor} and a \emph{storage capacitor}. 
%that encodes a single bit of data using its voltage level. 
The source nodes of the access transistors of all the DRAM cells in the same column connect the cells' storage capacitors to the same \emph{local bitline}. The gate nodes of the access transistors of all the DRAM cells in the same row connect the cells' access transistors to the same \emph{local wordline}. Each mat contains 
\li~a \emph{local row decoder} that drives the local wordlines to the appropriate voltage levels to \gfcr{open (activate)} a row, 
\lii~a row of sense amplifiers (also called a \emph{local row buffer}) that senses and latches data from the activated row, and
\liii~\emph{\glspl{HFF}} that drive a portion (e.g., \SI{4}{\bit}) of the data \gfi{in} the local row buffer to the global \omi{bitlines}. 

\paratitle{DRAM \juan{Operation}} Three major steps are involved in serving a main memory request. First, to select a DRAM row, the memory controller issues an \texttt{ACTIVATION} (\texttt{ACT}) command with the row address. On receiving this command, \omi{the} DRAM \omi{chip transfers} all the data in the row to the corresponding local row buffer. \revdel{The two-terminal sense amplifier in the local row buffer connected to each local bitline senses the voltage difference between the local bitline and a reference voltage and amplifies it to a CMOS-readable value.} %In doing so, the sense amplifier connected to the reference voltage is amplified to the \emph{opposite} (i.e., \emph{negated}) value. 
%The local row buffer maintains the sensed data for as long as the row is \emph{open} (i.e., the local and global wordlines continue to be asserted).
Second, to access a cache line from the activated row, the memory controller issues a \texttt{READ} (\texttt{RD}) command with the column address of the request. \revdel{ DRAM then transfers portions of the selected data from the local row buffer in each mat to the global row buffer (via the \glspl{HFF}) of the DRAM bank and from the DRAM bank to the memory controller over the memory channel.} Third, to enable the access of another DRAM row in the same bank, the memory controller issues a  \texttt{PRECHARGE} (\texttt{PRE}) command with the address of the currently activated \gfi{DRAM bank}. This command disconnects the local bitline by disabling the local wordline, and the local bitline voltage is restored to its quiescent state.% (e.g., typically $\frac{1}{2}V_{DD}$). 

\revdel{\paratitle{Fine-Grained DRAM Access \& Activation} 
Most commercial DRAM architectures employ a coarse-grained access and activation to exploit the temporal and spatial locality of an application's data access pattern. \gf{In this model, during an \texttt{ACT}, a single row address controls a large-size \emph{logical} DRAM row, which spreads across all the mats in a subarray, and across all DRAM chips.}
%\atb{ALTERNATIVE: In coarse-grained activation, an \texttt{ACT} command activates the local wordlines in \emph{all mats} in a subarray in every DRAM chip.}
%During DRAM access, the time taken to sense and latch a DRAM row dominates the total memory request time (the latency of an \texttt{ACT} command represents 71\% of the DRAM bank access time~\cite{Tiered-Latency_LEE}). By maintaining a large logical DRAM row containing several (e.g., 128) cache lines, DRAM can quickly service requests to consecutive cache lines (exploiting spatial locality) or repeated requests to the same cache line (exploiting spatial locality) from a previously-latched logical DRAM row. 
%However, DRAM's coarse-grained access and activation can lead to wasted energy consumption and throughput when an application does \emph{not} hold enough spatial and temporal locality.  
Many prior works~\cite{cooper2010fine,udipi2010rethinking,zhang2014half,ha2016improving,lee2017partial,olgun2022sectored} have investigated the issues caused by DRAM's coarse-grained access and activation in low-locality applications. To solve this issue, such works propose \emph{fine-grained} DRAM access and activation architectures. Their proposals are built based on the observation that DRAM's hierarchical organization \emph{already} provides a finer-grained activation granularity when considering the hardware structure of a single mat. 
%Thus, \gfi{they} focus on allowing individual control of a mat inside a DRAM array by partitioning the global wordline.
 }

\subsection{Processing-\gfcr{U}sing-DRAM}
\label{sec:background:pum}

\paratitle{In-DRAM-Row Copy} 
%RowClone~\cite{seshadri2013rowclone} is a mechanism that exploits the vast internal DRAM bandwidth to copy rows inside DRAM without CPU intervention efficiently.
RowClone~\cite{seshadri2013rowclone} enables copying a row~$A$ to a row~$B$ in the \emph{same} subarray by issuing two consecutive \texttt{ACT}
commands to these two rows, followed by a \texttt{PRE} command. This command sequence is called \texttt{AAP}~\omi{\cite{seshadri2017ambit}}.\revdel{ (\texttt{ACT}-\texttt{ACT}-\texttt{PRE})} \revdel{The first \texttt{ACT} copies the contents of the source row $A$ into the local row buffer. The second \texttt{ACT} connects the DRAM cells in the destination row~$B$ to the local bitlines. Because the sense amplifiers have already sensed and amplified the source data by the time row~$B$ is activated, the data in each cell of row~$B$ is overwritten by the data stored in the row buffer (i.e., row~$A$'s data). }%LISA~\cite{chang2016low} expands RowClone functionally to enable the execution of in-DRAM row copy operations across DRAM rows in \emph{different} subarrays of a DRAM chip

\paratitle{In-DRAM AND/OR/NOT}
Ambit~\ambit shows that simultaneously activating \emph{three} DRAM rows, via a DRAM operation called \emph{\gfi{\gls{TRA}}}, can perform \emph{in-DRAM} bitwise AND and OR operations.%When activating three rows, three cells connected to each local bitline share charge simultaneously and contribute to the perturbation of the local bitline.
\revdel{Upon sensing the perturbation \gf{of the three simultaneously activated rows}, the sense amplifier amplifies the local bitline voltage to $V_{DD}$ or 0 if at least two of the capacitors of the three DRAM cells are charged or discharged, respectively. As such, a \gfi{\gls{TRA}} results in a Boolean majority operation ($MAJ$).}
%A majority operation MAJ outputs a 1 (0) only if more than half of its inputs are 1 (0). %In terms of AND ($\cdot$) and OR (+) operations, a 3-input majority operation can be expressed as \texttt{MAJ(A, B, C) = A $\cdot$ B + A $\cdot$ C + B $\cdot$ C.} 
%
%To achieve functional completeness, Ambit implements NOT operations by exploiting the differential design of DRAM sense amplifiers. As Section~\ref{sec:background:dram} explains, the sense amplifier already generates the complement of the sensed value as part of the activation process. Therefore, Ambit simply forwards  the complement of the sensed value to a special DRAM row in the subarray that consists of DRAM cells with \emph{two} access transistors, called \emph{dual-contact cells} (DCCs). Each access transistor is connected to one side of the sense amplifier and is controlled by a separate wordline.% (\emph{d-wordline} or \emph{n-wordline}).
%By activating either the d-wordline or the n-wordline, the row of DCCs can provide the true or negated value stored in the row's cells, respectively.
%
%\rA{Ambit proposes \gfi{two} simple modifications to the DRAM array to \emph{efficiently} implement AND/OR/NOT operations. %\gfi{} internal organization of a subarray in Ambit, which has two different components than a conventional DRAM subarray. 
Ambit 
% Ambit implements MAJ by introducing a custom row decoder that can perform a TRA by simultaneously addressing three wordlines. To use this decoder, Ambit 
defines a new command called \texttt{AP} that issues a \gfi{\gls{TRA}}\revdel{ to compute a MAJ,} followed by a \texttt{PRE}\revdel{to close all three rows}.\revdel{\footnote{Although the `\texttt{A}' in \texttt{AP} refers to a \gls{TRA} operation instead of a conventional \texttt{ACT}, we use this terminology to remain consistent with the Ambit paper~\cite{seshadri2017ambit}, since an \texttt{ACT} can be internally translated to a TRA operation by the DRAM chip~\cite{seshadri2017ambit}.}} 
\rA{\changerA{\rA{\#A1}}Since TRA operations are destructive, Ambit divides DRAM rows into \emph{three groups} for \gls{PuD} computing: 
\li~the \textbf{D}ata group, which contains regular data rows;
\lii~the \textbf{C}ontrol group, which consists of two rows (\texttt{C0} and \texttt{C1})  with all-0 and all-1 values; and 
\liii~the \textbf{B}itwise group, which contains six rows designated for computation (four regular rows, \texttt{T0}, \texttt{T1}, \texttt{T2}, \texttt{T3}; and two rows, \texttt{DCC0} and \texttt{DCC1},  of dual-contact cells {for NOT)}.
%The D-group contains regular rows that store program or system data. The C-group consists of two constant rows, called C0 and C1, that contain all-0 and all-1 values, respectively, used to control the execution of AND (C0 row) and OR (C1 row) operations. The D-group and the C-group are connected to the regular local row decoder, which selects a single row at a time. The B-group contains six regular rows %, called T0, T1, T2, and T3; 
%and two rows of dual-contact cells%, 
%whose d-wordlines are called DCC0 and DCC1, and whose n-wordlines are called $\overline{\mbox{DCC0}}$ and $\overline{\mbox{DCC1}}$, respectively
\revdel{The B-group rows are designated to perform bitwise operations. They are all connected to a special local row decoder that can simultaneously activate three rows using a single address 
(i.e., perform a \gfi{\gls{TRA}}).}
%Using a typical subarray size of 1024 rows~\cite{chang2014improving, kim2012case, kim2018solar,Tiered-Latency_LEE,kim2019d}, Ambit splits the row addressing into 1006 D-group row addresses, 2 C-group row addresses, and 16 B-group rows addresses.
}

% \begin{figure}[ht]
%     %\vspace{-5pt}
%     \centering
%     \includegraphics[width=\linewidth]{figures/ambit_subarray-crop.pdf}
%     \caption{Ambit's subarray organization.}
%     \label{fig_subarray_ambit}
%    % \vspace{-10pt}
% \end{figure}

\paratitle{Generalizing In-DRAM Majority} SIMDRAM~\cite{hajinazarsimdram} \omiii{proposes} a three-step framework to implement \gfi{\gls{PuD}} operations. In the first step, SIMDRAM converts an AND/OR/NOT-based representation of the desired operation into an equivalent optimized MAJ/NOT-based representation. \rA{\changerA{\rA{\#A1}}By doing so, SIMDRAM reduces the number of TRA operations required to implement the operation.} 
In the second step, SIMDRAM generates the required sequence of DRAM commands to execute the desired operation. \rA{\changerA{\rA{\#A1}}Specifically, this step translates the MAJ/NOT-based implementation of the operation into \texttt{AAPs}/\texttt{APs}.
This step involves \li~allocating the designated compute rows in DRAM to the operands and \lii~determining the optimized sequence of \texttt{AAPs}/\texttt{APs} that are required to perform the operation.} 
This step's output is a \uprog, i.e., the optimized sequence of \texttt{AAPs}/\texttt{APs} that will be used to execute the operation at runtime. \gfcrii{Each \uprog corresponds to a different \emph{bbop} instruction, which is one of the CPU ISA extensions to allow programs to interact with the SIMDRAM framework (see \cref{sec:design:isa}).} 
In the third step, SIMDRAM uses a control unit in the memory controller to \omi{execute the \emph{bbop} instruction using the corresponding} \uprog. 
%SIMDRAM uses a control unit in the memory controller that transparently issues the sequence of \texttt{AAPs}/\texttt{APs} to DRAM, as dictated by a \emph{bbop} instructions. %Once a \emph{bbop} instruction is complete, the result of the in-DRAM computation is held in DRAM. 
\rA{{SIMDRAM implements} 16 \emph{bbop} instructions, including abs, add, bitcount, div, max, min, mult, ReLU, sub, and-/or-/xor-reduction, equal, greater, greater equal, and if-else.} 

\begin{figure}[ht]
    \centering
    \includegraphics[width=\linewidth]{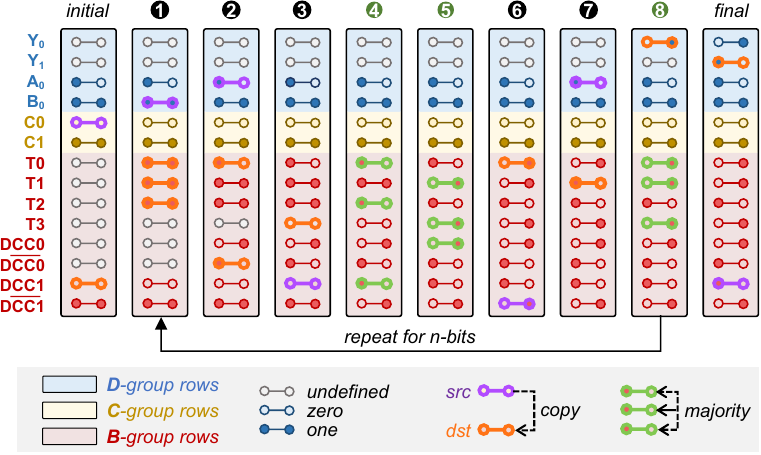}%
    %\vspace{-8pt}
    \caption{\gfcriii{Full adder operation in SIMDRAM.}}
    \label{fig:simdram:example}
\end{figure}

\rA{\changerA{\rA{\#A1}}Fig.~\ref{fig:simdram:example} illustrates how SIMDRAM executes a \gfcrii{one}-bit full addition operation using the sequence of row copy (\texttt{AAP}) and majority (\texttt{AP}) operations in DRAM. The figure shows one iteration of the full adder computation that computes \texttt{Y$_{0}$ = A$_{0}$ + B$_{0}$ + C$_{in}$}. 
First, }\gfi{SIMDRAM uses a \emph{vertical} data layout, where all bits of a \gfcrii{data element} are placed in a single DRAM column\revdel{ (i.e., in a single bitline),} when performing \gfi{PuD} computation.} 
%By doing, SIMDRAM can execute bit-shift operations, which is essential \gfi{in} complex computations without extra area cost. 
%For example, SIMDRAM can perform a left-shift-by-one operation by copying (using RowClone) the data in DRAM row $j$ to DRAM row $j+1$. 
Consequently, SIMDRAM employs a \emph{bulk bit-serial \gls{SIMD}} execution model, where each \gfcrii{data element} is mapped to a column of a DRAM row. This allows a DRAM subarray to operate as a \emph{\gfi{PuD} SIMD engine}, where a single bit-serial operation is performed over a large number of independent \gfcrii{data elements} \omii{(i.e., as many \gfcrii{data elements} as the size of a logical DRAM row, for example, 65,536)} at once. \rA{Second, as shown in the figure, each iteration of the full adder requires five \texttt{AAP}s \gfcrii{(\circled{1}, \circled{2}, \circled{3}, \circled{6}, \circled{7}) and three \texttt{AP}s (\circledgreen{4}, \circledgreen{5}, \circledgreen{8})}. 
A bit-serial addition of $n$-bit operands needs $n$ iterations, thus $(8 × n + 2)$ \texttt{AP}s \omii{and} \texttt{AAP}s~\omii{\cite{hajinazarsimdram}}.
}

% To enable a fast conversion between horizontally-layout data (used by the CPU) and vertically-layout data (used by \emph{bbop} instructions), SIMDRAM implements a data \emph{transposition unit} placed between the last-level cache and the memory controller. The transposition unit tracks data structures that are consumed/produced by a \emph{bbop} instruction (manually identified by the programmer). It automatically transposes such data structures' cache lines from a horizontal to vertical data layout (during cache evictions) and from a vertical to horizontal data layout (during cache line reads from main memory).     

%\paratitle{In-DRAM Look-Up Tables} 

%\paratitle{Alternative Processing-using-DRAM Architectures}

%% file: sections/03_motivation.tex
\section{Motivation}
\label{sec:motivation}

\gf{The efficiency of \juani{state-of-the-art} \gfi{\gls{PuD}} \juani{substrates} can be subpar \juani{when the\revdel{amount of} \gls{SIMD} parallelism that exists in an application\revdel{'s code} is smaller than} or not a multiple of the size of a DRAM row.} 
To quantify the\revdel{ amount of} \gf{SIMD} parallelism \omi{some} real-world applications inherently \gfcrii{possess}, we \omi{profile} the \emph{maximum vectorization factor} of \gfcrii{twelve real-world} applications. The vectorization factor\revdel{ (or vectorization width)} of a\revdel{ given} loop is the number of scalar operands that fit into a \gls{SIMD} register~\cite{pohl2018cost,trifunovic2009polyhedral}. % (e.g., a common vectorization factor for a loop that operates over \SI{4}{\byte} integers in a system with Intel AVX-512~\cite{avx512doc} \gls{SIMD} support is 16, or $\frac{512}{4\times8}$). 
%
% Since the loops we are interested in for processing-using-DRAM offloading are embarrassingly parallel, there is no data dependency across loop iterations.
%
We calculate the maximum vectorization factor by multiplying the vectorization factor of a single loop iteration and the loop's trip count~\cite{sokulski2022spec}.
%Since manually extracting \gls{SIMD} parallelism from an application can be daunting, 
We leverage modern compilers' loop auto-vectorization engines, which allows us to have an initial understanding of the \gls{SIMD} parallelism that a large number of applications \gfcrii{possesses}. 

\gfcrii{For our analysis, 
we use 
\li~the LLVM compiler toolchain~\cite{lattner2008llvm} (version 12.0.0) to \emph{automatically} vectorize loops in the application, and 
\lii~an LLVM pass~\cite{sarda2015llvm, lopes2014getting,writingpass} that instruments each application's loop to, during execution, gather  \emph{dynamic} information \omii{about} each vectorized loop, i.e., the loop trip count, execution count, execution time, and instruction breakdown~\cite{llvmprofiler}.
We compile each application using the clang compiler~\cite{lattner2008llvm}, using the appropriate flags to enable the loop auto-vectorization engine and its loop vectorization report (i.e., \texttt{-O3 -Rpass-analysis=loop-vectorize -Rpass=loop-vectorize}).\footnote{\gfcrii{See \cref{sec:methodology} for the description of our applications and their dataset.}}} 
\gfi{We assume SIMDRAM~\cite{hajinazarsimdram} as the target \gls{PuD} architecture.}

%To understand the in-DRAM \gls{SIMD} utilization, performance, and energy consumption when offloading such \gls{SIMD}-prone operations to the processing-using-DRAM architecture, we perform a \emph{loop offloading analysis}, where we offload the automatically vectorized loops in an application to the SIMDRAM architecture. In this analysis, the vectorization factor we use for all vectorized loops is equal to the size of a logical DRAM row (i.e., 65'536 operands). In case the maximum vectorization factor of a loop is smaller than 65'536, we pad the remaining columns in the target DRAM row with zeros (which represents \underutilization). 

\begin{figure}[ht]
    \centering
    \includegraphics[width=\linewidth]{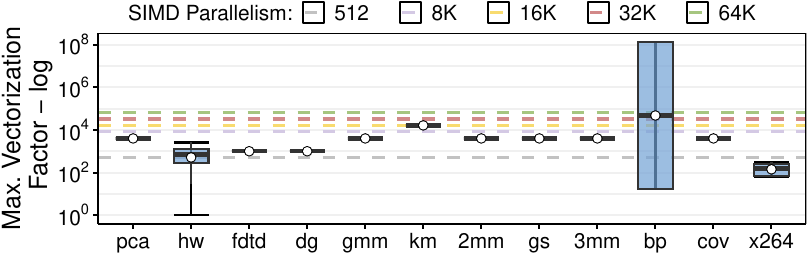}
    \caption{\gfi{Distribution of} maximum vectorization factor \omi{across all vectorized loops}. Whiskers extend to the minimum and maximum data points on either side of the box. \omi{Bubbles depict} average values.}
    \label{fig_max_utilization}
\end{figure}

%\paratitle{Vectorization Factor Analysis} 
\gfi{Fig.}~\ref{fig_max_utilization} shows the \gfi{distribution of} maximum vectorization factors (y-axis) \omi{of} all the vectorizable loops in an application (x-axis). \changeB{\#B3}\revB{We indicate different amounts of SIMD parallelism with horizontal dashed lines for reference.} We make two observations. 
First, the maximum vectorization factor varies \omi{both} within \omi{an} application and across different applications. Our analysis shows maximum vectorization factors as low as \gf{8} and as high as \gf{134,217,729}. 
Second, \omi{only} a small \omi{fraction} of vectorized loops have enough maximum vectorization factor \changeB{\#B3}\revB{(i.e., values above the green horizontal dashed line)} to fully exploit the \gls{SIMD} parallelism of SIMDRAM. On average, only \gf{0.11}\% of all vectorized loops have a maximum vectorization factor equal to or \juani{greater} than a DRAM row \gfcrii{(i.e., \juani{greater} than 65,536 data elements)}. 
We conclude that 
\li~real-world applications have \juani{varying} degrees of \gls{SIMD} parallelism\revdel{ that can be exploited for \gfi{PuD} computation}; and
\lii~\juani{these varying degrees of \gls{SIMD} parallelism rarely take \omi{full} advantage of the \gfcrii{very}-wide \gls{SIMD} width of state-of-the-art \gls{PuD} substrates.}

\paratitle{Problem \& Goal} \gf{We observe that the rigid granularity of \gfi{\gls{PuD}} architectures limits their efficiency \omi{(and thus their \emph{effective} applicability)} for many applications.} Such applications \juani{would benefit from} a variable-size \gls{SIMD} substrate that can \gfcriii{dynamically} adapt to the \omii{varying levels of \gls{SIMD} parallelism (i.e., different vectorization factors) an application exhibits during its execution}. 
Therefore, our \textit{goal} is to design a flexible \gfi{\gls{PuD}} substrate that
\li~adapts to the \omi{varying} \gfcriii{levels of} \gls{SIMD} parallelism \omiii{present in} an application, and 
\lii~maximizes the utilization of the \omi{very-wide} \gfi{\gls{PuD}} engine \changeD{\#D1}\revD{by \omi{concurrently \omii{exploiting}} parallelism across \omi{\emph{different}} \gls{PuD} \gfcrii{operations} \omii{(potentially from different applications)}.}

%
%
% We draw three conclusions from our analysis. 
% First, the rigid granularity of processing-using-DRAM architectures limits their applicability and efficiency for many applications.
% Such applications require a variable-size \gls{SIMD} substrate that can adapt to the various-size vectorization factor the application presents.
% Second, naively offloading all vector operations to be executed by the processing-using-DRAM architecture leads to subpar performance or even slowdowns.
% Third, enabling communication across DRAM bitlines can allow the execution of a broader set of vector operations in DRAM (i.e., vector reduction operations).

% overcomes the two limitations caused by the large and rigid granularity of in-DRAM execution. in this work is to enhance state-of-the-art processing-using-DRAM substrates with solutions that can
% \li~flexibility and efficiently extract vectorization opportunities from general-purpose application targeting in-DRAM computing while
% \lii~maximizing \gls{SIMD} utilization of the underlying processing-using-DRAM substrate. 

%% file: sections/04_key_ideas.tex
\section{\prop: A MIMD \gfi{\gls{PuD}} Architecture}
\label{sec:idea}

\prop is a hardware/software co-designed \gfi{\gls{PuD} system} that enables fine-grained \gfi{\gls{PuD}} computation at low cost and low programming effort. The \emph{key idea} of \prop is to leverage fine-grained DRAM activation for \gfi{\gls{PuD}}, \gfi{which} provides three benefits. First, it enables \prop to allocate \emph{only} the appropriate computation resources (based on the maximum vectorization factor \juani{of a} loop) for a target loop, \omi{thereby} reducing \underutilization and energy waste. 
Second, \prop can currently execute \omi{multiple} independent operations inside a single DRAM subarray \omi{\emph{independently} in separate} \gfi{DRAM mats}. This allows \prop to \juani{operate} as a \gls{MIMD} \gfi{\gls{PuD}} substrate, increasing overall throughput. Third, \prop implements low-cost \gfcri{interconnects} that enable moving data across DRAM columns \gfcrii{\emph{across} and \emph{within} DRAM mats} by combining fine-grained DRAM activation with simple modifications to the DRAM I/O circuitry. This enables \prop to implement reduction operations in DRAM without \juani{any intervention of the host CPU cores.}

\subsection{\prop: Hardware Overview}
%\label{sec:idea:hardware}

%\subsection{Fine-Grained Processing-using-DRAM Execution}

\begin{figure}[ht]
    \centering
    \includegraphics[width=\linewidth]{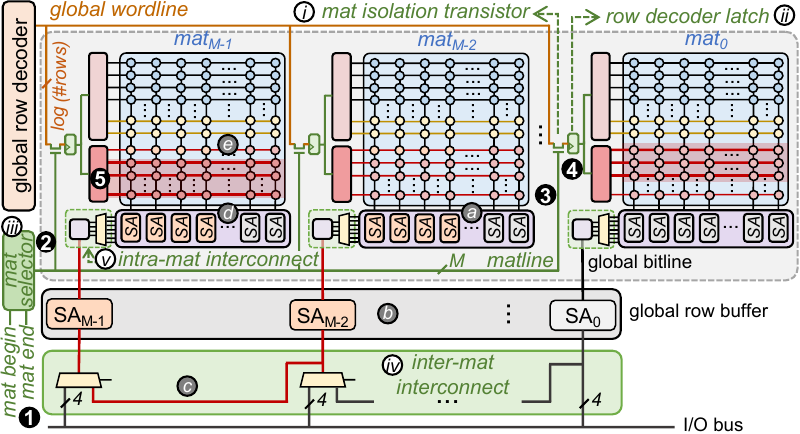}
    \caption{\omiii{\prop subarray and bank organization. {Green-colored boxes represent newly added hardware components.}}}
    \label{fig_subarray_matdram}
\end{figure}

\gfi{Fig.}~\ref{fig_subarray_matdram} shows an overview of the DRAM organization of \prop. Compared to the baseline Ambit subarray organization, \prop adds four new components (\juani{colored} in green) to a DRAM subarray and DRAM bank, which \juani{enable} 
\li~fine-grained \gfi{\gls{PuD}} execution;
\lii~global I/O data movement; and
\liii~local  I/O data movement. 
%In this section, we first introduce each component at a high-level, highlighting their goal and operation.
% In the following sections, we describe their detailed hardware implementation and timing analysis. 

\paratitle{Fine-Grained \gfi{\gls{PuD}} Execution} To enable fine-grained \gfi{\gls{PuD}} execution, \prop modifies Ambit's subarray and the DRAM bank with three new hardware structures: the \emph{mat isolation transistor}, the \emph{row decoder latch}, and the \emph{mat selector}. 
\gfcrii{At a high level, 
the \emph{mat isolation transistor} allows for the  independent access and operation of DRAM mats within a subarray while 
the \emph{row decoder latch} enables the execution of a \gls{PuD} operation in a range of DRAM mats that the \emph{mat selector} defines.}

First, the \emph{mat isolation transistor} \gfcrii{(\circlediii{i} in Fig.~\ref{fig_subarray_matdram})} segments the global wordline connected to the local row decoder in \emph{each} DRAM mat \gfi{in} a subarray. 
Second, the \emph{row decoder latch} \gfcrii{(\circlediii{ii})} stores the bits from the global wordline used to address the local row decoder. 
Third, the \emph{mat selector} \gfcrii{(\circlediii{iii})}, shared across all DRAM mats \gfi{in} a subarray, asserts one or more mat isolation transistors. \gfcrii{The \emph{mat selector} enables the connection between the global wordline and  the \emph{row decoder latches} belonging to a range of DRAM mats. When issuing \gls{PuD} operations, the memory controller specifies the \emph{logical} address of the \emph{first} and \emph{last} DRAM mats that the \gls{PuD} operation targets (called \emph{logical mat range}). 
Internally, each DRAM chip \li~identifies whether \emph{any} of its DRAM mats belong to the logical mat range and 
\lii~translates the logical mat range into the appropriate \emph{physical mat range}, which is used as input for the \emph{mat selector}. 
With these structures, \omi{\emph{different} \gls{PuD}  operations can execute in \emph{different} ranges of DRAM mats.}}  

For example, to execute a \gfi{\gls{TRA}} in \juani{only} $mat_0$, \prop performs \gfi{four} steps:
\li~when issuing a \gfi{\gls{TRA}}, the memory controller sends, alongside the row address information, the \emph{logical mat range} \matrange = \texttt{[\#0,\#0]} to address  $mat_0$ (\circled{1} in \gfi{Fig.}~\ref{fig_subarray_matdram});  
\lii~the \emph{mat selector} (\circled{2}) receives the logical mat range, translates it to the appropriate \emph{physical mat range}, and raises the \emph{matline} \juani{corresponding} to $mat_0$, which asserts $mat_0$'s \emph{mat isolation transistor} (\circled{3}) and connects the global wordline to $mat_0$'s row decoder latch;
\liii~the bits of the global wordline used to drive $mat_0$'s local row decoder are stored \juani{in} $mat_0$'s \emph{row decoder latch} (\circled{4});
\liv~finally, $mat_0$'s local row decoder drives the appropriate rows in $mat_0$'s DRAM array based on the value stored \omi{in} the row decoder latch. From here, the DRAM row activation (and thus, \gfi{\gls{PuD}} computation) proceeds as described in \gfi{\cref{sec:background:dram}}, only \juani{involving} the DRAM rows in $mat_0$. Since the \omi{per-mat} row decoder latch stores the local row address for a given row activation \omi{in a mat}, the memory controller can issue a \gfi{\gls{TRA}} to another DRAM mat while $mat_0$ is being activated (\circled{5}).
%Therefore, enabling concurrent execution of different triple-row activation commands across different DRAM mats in a \gls{MIMD} fashion.  

\paratitle{Global I/O Data Movement} To enable data movement across different mats, \prop implements an  \emph{inter-mat \gfcri{interconnect}} by slightly modifying the connection between the I/O bus and the global row buffer \gfcrii{(\circlediii{iv} in Fig.~\ref{fig_subarray_matdram})}. The inter-mat \gfcri{interconnect} relies on the observation that the sense amplifiers in the global row buffer \juani{have \emph{higher} drive} than the sense amplifiers in the local row buffer~\cite{keeth2007dram, wang2020figaro}, allowing to directly drive data from the global row buffer into the local row buffer.\footnote{Prior work~\cite{wang2020figaro} leverages the same observation to \omi{copy} DRAM columns from one subarray to another.} \juani{To leverage this observation}, \prop adds a 2:1 multiplexer to the input/output port of each \gfcrii{set of \emph{four} 1-bit  sense amplifiers} in the global row buffer. The multiplexer selects whether the data that is written to the sense amplifier \omi{set} $SA_i$  comes from the I/O bus or from the neighbor sense amplifier \omi{set} $SA_{i-1}$. 
%Since each sense amplifier $SA_M$ is connected to $mat_M$, the inter-mat network creates a path from $mat_{i-1}$ $\rightarrow$ $SA_{i-1}$ $\rightarrow$  $SA_{i}$ $\rightarrow$ $mat_i$. 

To manage inter-mat data movement, \prop exposes a new DRAM command to the memory controller called \texttt{GB-MOV} (\underline{g}lo\underline{b}al I/O \underline{mo}ve). 
%
%\paratitle{The \underline{G}lo\underline{b}al I/O \underline{Mo}ve Command} 
The \texttt{GB-MOV} command takes as input: 
\li~the logical mat range \matrange, row address, and column address of the \emph{source} DRAM row \gfcrii{and column}; and
\lii~the logical mat range \matrange, row address, and column address of the \emph{destination} DRAM row \gfcrii{and column}. 
With the inter-mat \gfcri{interconnect} and new DRAM command, \prop can move \emph{four} bits\footnote{\gfi{The number of bits the inter-mat \gfcri{interconnect} can move at once depends on the number of \glspl{HFF} \gfcrii{\emph{already present}} in a DRAM mat. We assume that each mat has four \glspl{HFF}, as prior works suggest~\cite{zhang2014half,ha2016improving,oconnor2017fine}.}} of data from a source row \gfcriii{and column (\texttt{row$_{src}$}, \texttt{column$_{src}$}) in $mat_{M-2}$ to a destination row \gfcrii{and column (}\texttt{row$_{dst}$}, \texttt{column$_{dst}$}) in $mat_{M-1}$\gfcrii{, in a DRAM subarray with $M$ DRAM mats}.} Once the memory controller receives a \texttt{GB-MOV} command, it \juani{performs} three steps. 
First, the memory controller issues an \texttt{ACT} to \gfcrii{the source} row in $mat_{M-2}$, which loads the target DRAM \gfcriii{\texttt{row}$_{src}$} to $mat_{M-2}$'s local sense \omi{amplifiers} (\circledii{a} in \gfi{Fig.}~\ref{fig_subarray_matdram}). Concurrently, the memory controller issues an \texttt{ACT} to \gfcrii{the destination} row  in $mat_{M-1}$, which connects \gfcriii{\texttt{row}$_{dst}$} to $mat_{M-1}$'s local sense \omi{amplifiers}. 
Second, the memory controller issues a \texttt{RD} with the address of the four-bit \emph{source} column to $mat_{M-2}$. The column select command loads the four-bit \gfcriii{\texttt{column}$_{src}$} from $mat_{M-2}$'s local sense \omi{amplifiers} to its \glspl{HFF}, and $mat_{M-2}$'s \gfcrii{\glspl{HFF} drive} the corresponding \gfcrii{set of four one-bit sense amplifiers} $SA_{M-2}$ \omii{in} the global row buffer (\circledii{b}).
Third, the memory controller issues a \texttt{WR} \gfcrii{with the address of the four-bit \emph{destination} column to $mat_{M-1}$}. Since the \texttt{WR} corresponds to a \texttt{GB-MOV} command, the multiplexer that connects $mat_{M-1}$'s \glspl{HFF} to the global row buffer takes as input the added datapath coming from $SA_{M-2}$ instead of the conventional datapath coming from the I/O bus  (\circledii{c}). \juani{As a result}, the data stored in $SA_{M-2}$ \omii{is} loaded into $SA_{M-1}$, which in turn drives $mat_{M-1}$'s \glspl{HFF} and local sense amplifiers (\circledii{d}). 
\gfcrii{Once the four-bit column coming from \gfcrii{\texttt{row}$_{src}$} is written into $mat_{M-1}$'s local sense amplifiers, the local sense amplifiers finish the \texttt{WR}  by restoring the local bitlines in $mat_{M-1}$ to \texttt{VDD} or \texttt{GND}, thereby storing the four-bit column coming from \gfcriii{\texttt{column}$_{src}$} as a column of \gfcriii{\texttt{row}$_{dst}$} (\circledii{e}).}

\gfcrii{The \emph{conservative worst-case latency} of a \texttt{GB-MOV} command (i.e., where the addresses of the source and the destination rows differ) is equal to $t_{RAS} + t_{RELOC} + t_{WR} + t_{RP}$; where 
$t_{RAS}$ is latency from the start of row activation until the completion of the DRAM cell's charge restoration,
$t_{RELOC}$~\cite{wang2020figaro} is the latency of turning on the connection between the source and destination local sense amplifiers; 
$t_{WR}$ is the minimum time interval between a \texttt{WR} and a \texttt{PRE} command, which allows the sense amplifiers to restore the data to the DRAM cells;
$t_{RP}$ is the latency between issuing a \texttt{PRE} and when the DRAM bank is ready for a new row activation.}

 %In Section~\ref{sec:design:indramexec}, we discuss the execution latency of a \texttt{GB-MOV} command and how \prop employs it to implement in-DRAM vector reduction. 

\begin{figure}[ht]
    \centering
    \includegraphics[width=\linewidth]{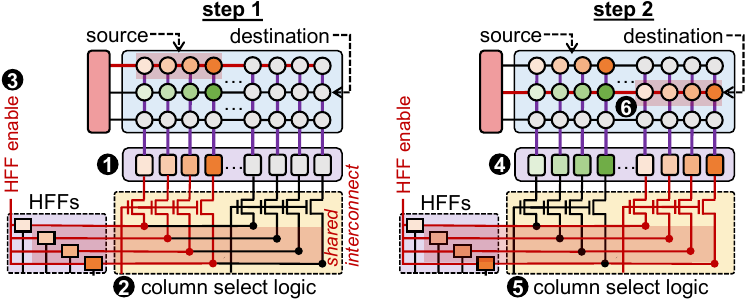}
    \caption{\prop intra-mat \gfcri{interconnect}.}
    \label{fig_intra_mat}
\end{figure}

\paratitle{Local I/O Data Movement} To enable data movement \omi{across columns} \emph{within} a DRAM mat, \prop implements an  \emph{intra-mat \gfcri{interconnect}} \gfcrii{(\circlediii{v} in Fig.~\ref{fig_subarray_matdram})}, which does \emph{not} require any hardware modifications. 
Instead, it modifies the sequence of steps DRAM executes during a column access operation. There are two \emph{key observations} \juani{that enable} the intra-mat \gfcri{interconnect}. 
First, we observe that the local bitlines \gfcrii{of a DRAM mat} \emph{already} share an interconnection path via the \glspl{HFF} and column select logic \gfcrii{(as Fig.~\ref{fig_intra_mat} illustrates)}. 
Second, the \glspl{HFF} in a DRAM mat can latch and \emph{amplify} the local row buffer's data~\cite{keeth2007dram,o2021energy}. 
%This happens because the local sense amplifiers have limited drive capability and cannot quickly drive the global bitlines. %Therefore, modern DRAM architectures employ several layers of data amplification to improve signal integrity and DRAM performance. \Based on the two key observations, the intra-mat network operates by latching and amplifying the source data into the mat's \glspl{HFF} and then allowing the \glspl{HFF} to drive the latched data into the destination column. 

To manage intra-mat data movement, \prop exposes a new DRAM command to the memory controller called \texttt{LC-MOV} (\underline{l}o\underline{c}al I/O \underline{mo}ve). 
The \texttt{LC-MOV} command takes as input: 
\li~the logical mat range \matrange of the target row, 
\lii~the row and column \gfcrii{addresses} of the \emph{source} DRAM row and column; and
\liii~the row and column \gfcrii{addresses} of the \emph{destination} DRAM row and column. 
With the intra-mat \gfcri{interconnect} and new DRAM command, \prop can move \emph{four} bits of data from a source row \gfcrii{and column (\texttt{row$_{src}$}, \texttt{column$_{src}$}) to a destination row \gfcrii{and column (}\texttt{row$_{dst}$}, \texttt{column$_{dst}$}) in $mat_{M}$.} 
Once the memory controller receives \omi{an} \texttt{LC-MOV} command, it \omi{performs} two steps, which \gfi{Fig.}~\ref{fig_intra_mat} illustrates. 
In the \emph{first step}, the memory controller performs an \texttt{ACT}--\texttt{RD}--\texttt{PRE}\revdel{ command sequence} targeting \gfcrii{\texttt{row$_{src}$}, \texttt{column$_{src}$}} in $mat_{M}$. The \texttt{ACT} loads \gfcrii{\texttt{row$_{src}$}} to $mat_{M}$'s local sense amplifier (\circled{1} in \gfi{Fig.}~\ref{fig_intra_mat}).
The \texttt{RD} moves \gfcrii{four bits from  \texttt{row$_{src}$}, as indexed by \texttt{column$_{src}$},} into the mat's \glspl{HFF} by \omi{enabling} the appropriate transistors in the column select logic~(\circled{2}). 
The \glspl{HFF} are then enabled by transitioning the \emph{\gls{HFF} enable} signal from low to high. This allows the \glspl{HFF} to \emph{latch} and \emph{amplify} the selected \gfcrii{four-bit} data column from the local sense amplifier~(\circled{3}).
The \texttt{PRE} closes \gfcrii{\texttt{row$_{src}$}}. 
Until here, the \texttt{LC-MOV} command operates exactly as a regular  \texttt{ACT}--\texttt{RD}--\texttt{PRE} command sequence. 
However, differently from a regular \texttt{ACT}--\texttt{RD}--\texttt{PRE}, the \texttt{LC-MOV} command does \emph{not} lower the \emph{\gls{HFF} enable} signal when the \texttt{RD} finishes. This allows the \gfcrii{four-bit} data \gfcrii{from \texttt{column$_{src}$}} to reside in the mat's \glspl{HFF}. 
In the \emph{second step}, the memory controller performs an \texttt{ACT}--\texttt{WR}--\texttt{PRE} targeting \gfcrii{\texttt{row$_{dst}$}, \texttt{column$_{dst}$}} in $mat_{M}$. 
The \texttt{ACT} loads \texttt{row$_{dst}$} into the mat's local row buffer (\circled{4}), and the \texttt{WR} asserts the column select logic to \gfcrii{\texttt{column$_{dst}$}}, creating a path between the \glspl{HFF} and the local row buffer (\circled{5}). Since the \emph{\gls{HFF} enable} signal is kept high, the \glspl{HFF} will \emph{not} sense and latch the data from \gfcrii{\texttt{column$_{dst}$}}. Instead, \gfcrii{the \glspl{HFF} overwrite} the data stored in the local sense amplifier with the previously \gfcrii{four-bit}  data latched from \gfcrii{\gfcrii{\texttt{column$_{src}$}}}. The new data stored in the mat's local sense amplifier propagates through the local bitlines and is written to the destination DRAM cells (\circled{6}). 

\gfcrii{The \emph{conservative worst-case latency} of an \texttt{LC-MOV} command (i.e., where the addresses of the source and the destination rows differ) is equal to $2 \times (t_{RAS} + t_{RP}) + t_{RELOC} + t_{WR}$.}
% Finally, the memory controller issues a \texttt{PRECHARGE} command to prepare the subarray to subsequent requests. 
%In case the \texttt{LC-MOV} command targets columns in the \emph{same} DRAM row \juani{of the same mat}, the memory controller does \emph{not} need to issue the first \texttt{PRE}, reducing the data movement's latency. 

%Note that the \texttt{LC-MOV} command operates by following regular DRAM \texttt{RD} and \texttt{WR} commands. The primary difference is that the \texttt{LC-MOV} command manipulates the behavior of the \glspl{HFF} to enable data to be moved across DRAM columns. 

\subsubsection{\gfi{\gls{PuD}} Vector Reduction} 
\begin{figure}[ht]
     \centering
     \includegraphics[width=\linewidth]{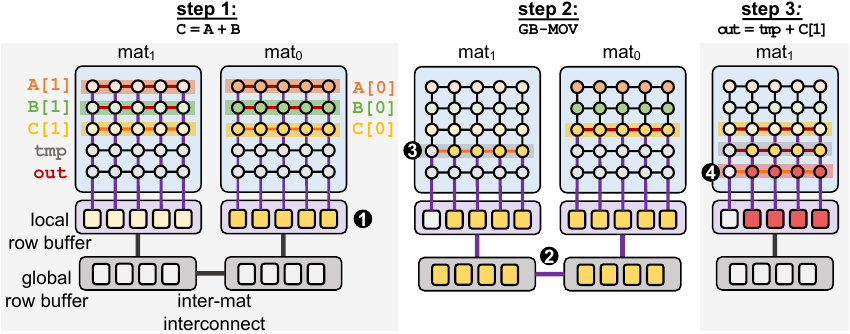}
    \caption{\gfcriii{An example of a \gls{PuD} vector reduction in \prop.}}
     \label{fig_vector_reduction}
 \end{figure}

We describe how \prop uses the inter-\omi{mat} and intra-mat \gfcri{interconnects} to implement \gfi{\gls{PuD}} vector reduction. To do so, we use a simple example, where \prop executes a vector addition followed by a vector reduction, i.e., \texttt{out+=(A[i]+B[i])}. We assume that DRAM has \omi{only} \gfcrii{two} mats, and the \gfcrii{data elements of the} input arrays \texttt{A} and \texttt{B} \omi{are evenly distributed} across \gfcrii{the two DRAM mats, as Fig.~\ref{fig_vector_reduction} illustrates. \prop executes a vector reduction in three steps.}
%
% \prop executes the \gfi{\gls{PuD}} vector reduction operation in three main steps: 
% \li~map operation (i.e., \texttt{C[i] = A[i] + B[i]});
% \lii~512-element vector reduction operation (i.e., \texttt{tmp[j] += C[i]}); and
% \liii~4-element vector reduction operation (i.e., \texttt{out[3:0] += tmp[j]}). To execute the vector reduction operations, \prop uses the  \texttt{GB-MOV} and \texttt{LC-MOV} to implement an adder tree in DRAM in seven main steps. 
%
\gfcrii{In the first step,} \prop executes a \gls{PuD} addition operation over the data in  \gfcrii{the two DRAM mats (\circled{1})}, storing the temporary output data \texttt{C} into the same mats \omi{where} the computation takes place (i.e., \texttt{C} = \texttt{\{}\texttt{C[0]}$_{mat0}$, \texttt{C[1]}$_{mat1}$\texttt{\}}). 
\gfcrii{In the second step}, \prop issues a \texttt{GB-MOV} to move \omii{part} of the temporary output \texttt{C[0]} stored in \gfcrii{$mat_{0}$ to a temporary row \texttt{tmp} in $mat_{1}$} (\texttt{tmp}$_{mat1}$  $\leftarrow$ \texttt{C[0]}$_{mat0}$) \gfcrii{via the inter-mat interconnect (\circled{2}--\circled{3}})\gfcrii{,  \omii{four bits} (i.e., four data elements) \omii{at a} time. \prop \emph{iteratively} executes step 2 until \emph{all} data elements of \texttt{C[0]} are copied to $mat_{1}$}.   
% Third, %once the \texttt{GB-MOV} command finishes executing, 
% \prop issues a second \texttt{GB-MOV} to move the portion of the temporary output stored in $mat_{0}$ to $mat_{1}$ ($mat_{1}$  $\leftarrow$ \texttt{C[0]}$_{mat0}$). 
% Fourth, \prop concurrently executes addition operations in $mat_3$ and $mat_1$, which computes \texttt{C[3] + C[2]} in $mat_3$ and \texttt{C[1] + C[0]} in  $mat_1$.   
% Fifth, \prop issues a \texttt{GB-MOV} to move the produced temporary output of \texttt{C[1] + C[0]}, from $mat_1$ to $mat_3$.
\gfcrii{In the third step, once the \texttt{GB-MOV} finishes, \prop executes the final addition operation, i.e. \texttt{tmp} + \texttt{C[1]}, in $mat_1$. The \gfcrii{final} output of the \gfcrii{vector reduction operation} is stored in the destination row \texttt{out} in $mat_1$ (\circled{4}).} 
Once the vector reduction operation finishes, the temporary output array stored in \gfcrii{$mat_1$} holds as many \gfcrii{data} elements as the number of DRAM columns in a mat (\gfcrii{e.g.}, 512 \gfcrii{data} elements). 
\prop allows reducing the temporary output vector further to an output vector with \gfcrii{\emph{four} \gfcrii{data} elements} using the intra-mat \gfcri{interconnect}. The process is analogous to that employed during the 512\omii{-element} vector reduction: \prop uses the intra-mat \gfcri{interconnect} and the \texttt{LC-MOV} command to implement an adder tree inside a single DRAM mat.\footnote{\gfi{The number of \texttt{GB-MOV} and \texttt{LC-MOV} commands \omii{issued depends on} the bit-precision of the input operands~\omii{\cite{hajinazarsimdram}}.}}  

%\subsubsection{Timing Analysis}

%\subsection{Intra-MAT Communication}

%\subsubsection{Timing Analysis}

% \textcolor{red}{\noindent\rule{8.5cm}{2pt}}
% \begin{center}
%     \vspace{-12pt}
%     \textcolor{red}{\textbf{STALE TEXT BEGIN.}}
% %    \vspace{-15pt}
% \end{center}

\subsection{\prop: \gfcrii{Control \& Execution}}

\gfhpca{To \gfcrii{enable} \prop to execute in a \gls{MIMD} fashion, we need to efficiently 
\li~\gfcrii{\emph{encode} and \omi{\emph{communicate}} information regarding the target DRAM mats (i.e., the target mat range) in a \emph{timely} manner (i.e., respecting DRAM timing parameters) while
\lii~\emph{orchestrating} the execution of independent \gls{PuD} operations across the DRAM mats of a DRAM subarray.
To do so, we take a \emph{conservative} design approach: we aim to integrate \prop in commodity DRAM chips by providing an implementation 
\li~\emph{compatible} with existing DRAM standards and 
\lii~\gfcriii{that does \emph{not} add new pins to a DRAM chips.}}}

\paratitle{Encoding MAT Information} \gfhpca{\prop needs a compact way to encode the target mat information, since a DRAM module often contains many DRAM mats. To solve this issue, \prop \omi{only allows} a \gls{PuD} \gfcrii{operation} \omi{to be executed in a \emph{physically contiguous} set of DRAM mats}.\footnote{In \cref{sec:operating:system}, we describe how we enforce \gfcrii{physically contiguous} mat allocation.} 
In this way, \gfcrii{when executing the DRAM commands (i.e., \texttt{ACT}s and \texttt{PRE}s) that realize a \gls{PuD} operation,} the memory controller only needs to \omi{provide} the \emph{first} and \emph{last} (\emph{logical}) mats an \texttt{ACT} target. 
Then, \prop internally decides which (\emph{physical}) mats fit into the \omi{provided} mat range. 
To do so, \prop implements a simple \emph{chip select logic} and \emph{mat identifier logic} inside the I/O circuitry of each DRAM chip. The \emph{chip select logic} and \emph{mat identifier logic} take as input the \emph{logical mat range} and \omi{output} 
\li~if DRAM mats placed in a chip belong to the mat range, and 
\lii~the physical mat range. 
\gfcrii{\omiii{In case a DRAM mat placed in a chip belongs to the mat range,} the DRAM chip queues the physical mat range in the \emph{mat queue} (which we describe later in this section). 
The \emph{physical mat range} is used as input for the \emph{mat selector} (see Fig.~\ref{fig_subarray_matdram}).}
Since there are up to 128 DRAM mats in a DDR4 module~\cite{lee2021greendimm}, \prop uses \omii{14~bits} to encode the logical mat range \gfcrii{(\omii{7~bits each} for \emph{mat begin} and \emph{mat end}, each)}\gfcrii{, from which 
\li~the three most significant bits are used to identify the target DRAM chip and 
\lii~the four \omii{least} significant bits are used to identify individual mats.} The \emph{chip select logic} and \emph{mat identifier logic} comprise simple hardware elements: four \gf{comparators}, two \omi{2-input} AND gates, two 2:1 multiplexers, and a \omii{3-bit} \emph{chip id register} in each DRAM chip. }

\paratitle{\omi{Communicating} MAT Information} 
\prop needs to \omi{communicate} to the DRAM \gfcrii{chip} information regarding the target mats during a \gfi{\gls{PuD}} operation. 
However, it is challenging to \omi{communicate} the mat information alongside an \texttt{ACT} due to the narrow DRAM command/address (C/A) bus interface, since the memory controller uses most of the available pins during a row activation for row address and command \omi{communication}.\footnote{There are 27 C/A pins in a DDR4 chip~\cite{jedec2017jedec}, from which only three pins are \emph{not} used during an \texttt{ACT} command.} 
Our \emph{key idea} to solve this issue is to overlap the latency of \omi{communicating} the mat information to DRAM with the latency of DRAM commands in a \uprog in two ways: 
\li~\texttt{ACT}--\texttt{ACT} overlap, and 
\lii~\texttt{PRE}--\texttt{ACT} overlap.
The first case (\texttt{ACT}--\texttt{ACT} overlap) happens when issuing a row copy operation (\texttt{AAP}). 
In this case, the mat information required by the second \texttt{ACT} is transmitted immediately \emph{after} issuing the first \texttt{ACT}, exploiting the delay between \omi{two} activations. The mat information is buffered once it reaches DRAM.
The second case (\texttt{PRE}--\texttt{ACT} overlap) happens when issuing the first \texttt{ACT} in a row copy operation or the \texttt{ACT} in a \gls{TRA}. 
We notice that
\li~\gfcrii{the first \texttt{ACT}  command in an \texttt{AAP}/\texttt{AP}} is \emph{always} preceded by a \texttt{PRE} (due to a previous \texttt{AAP}/\texttt{AP}, or due to a previous DRAM request), and 
\lii~a \texttt{PRE} does \emph{not} use the row address pins, since it targets a DRAM bank (not a DRAM row). \gfcrii{Thus, \prop uses the row address pins during a \texttt{PRE} \gfcrii{that immediately precedes the first \texttt{ACT} in an \texttt{AAP}/\texttt{AP} command sequence to communicate the mat information}.\footnote{\rA{\changerA{\rA{\#A2}}If there are insufficient pins in the DDRx interface to communicate mat information (e.g., as in DDR5~\cite{jedec2020ddr5}), \prop utilizes multiple DRAM C/A cycles to propagate the mat information. For example, in DDR5, \prop still performs \texttt{PRE-ACT} overlap, communicating the mat information in two cycles. Note that \omi{an} extra cycle does \emph{not} impact \prop's performance, since in a \texttt{PRE-ACT} command sequence, the \texttt{PRE} still needs to wait for the completion of the \texttt{ACT} for more than two DRAM C/A cycles.}}} 

\paratitle{Timing \omi{of} MAT Information} \gfhpca{\prop needs to \gfcrii{communicate} the mat information \emph{before} a respective \texttt{ACT} in a \uprog{}. \gfcrii{Communicating} the mat information immediately \emph{after} the memory controller issues the \texttt{ACT} would open the \emph{entire} DRAM row (instead of only the relevant portion of the DRAM row). To solve this issue, we devise a simple \gfi{queuing}-based mechanism for partial row activation. 
Our mechanism relies on the fact that the \gfcrii{execution} order \gfcrii{of} \texttt{ACT}s and \texttt{PRE}s \gfcrii{in} a \uprog is \emph{deterministic}.\footnote{To realize a \gls{PuD} \omii{operation}, the memory controller \emph{must} respect the order in which \texttt{ACT} and \texttt{PRE} commands are specified in the \uprog{}. Therefore, during \gls{PuD} execution, \texttt{ACT}s and \texttt{PRE}s in a \uprog{} cannot be reordered, and the behavior of the \uprog{} is thus deterministic. If the memory controller is \omi{performing} maintenance operation to a DRAM bank, the \texttt{AAP}/\texttt{AP} commands of a \gls{PuD} operation wait until the maintenance operation finishes.} Thus, we can add to each DRAM command in \gfcrii{an \texttt{AAP}/\texttt{AP}} the information about when the  DRAM circuitry should propagate the mat information. 
\revdel{The memory controller enforces that \glspl{TRA} from different \uprogs \emph{are} overlapped, but \glspl{TRA} from the \emph{same} \uprog \emph{are serialized}. }\prop leverages \gfcrii{this} key idea by \gfcrii{adding} a \emph{mat \gfi{queue}} to the I/O logic of each DRAM chip and adding extra \omi{functionality} to the existing \texttt{ACT} and \texttt{PRE} commands to control the mat queue:
\li~\texttt{ACT-enqueue} issues an \texttt{ACT} to \texttt{row\_addr} in the first DRAM clock cycle and enqueues [\texttt{mat\_begin,mat\_end}] in the second DRAM clock cycle; 
\lii~\texttt{PRE-enqueue} issues a \texttt{PRE} to \texttt{bank\_\omi{id}} and enqueues [\texttt{mat\_begin,mat\_end}];
\liii~\texttt{ACT-dequeue} issues an \texttt{ACT} to \texttt{row\_addr} and dequeues from the mat queue.  }

\begin{figure}[!ht]
    \centering
    \includegraphics[width=0.6\linewidth]{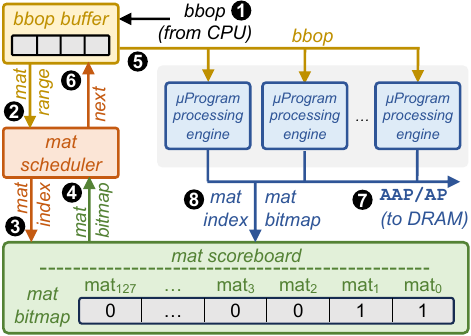}
    \caption{\prop control unit \omi{in the memory controller}.}
    \label{fig_control_unit}
\end{figure}

%\vspace{-15pt}
\paratitle{Orchestrating MAT Information} \prop needs to execute different \gfi{\gls{PuD}} \gfcrii{operations} concurrently. To this end, we implement a control unit inside the memory controller \gfcrii{on the CPU die}, which Fig.~\ref{fig_control_unit} illustrates. 
\prop leverages SIMDRAM control unit to translate \omi{each} \emph{bbop} \omi{instruction} into \omi{its corresponding} \uprog{} and adds extra circuitry to 
\li~schedule \omi{each \uprog{}} based on \omi{its} target \omi{mats} and
\lii~maintain multiple \uprog contexts. 
\gfcrii{\prop control unit consists of \gf{four} main components.
First, \emph{bbop buffer}, which stores \emph{bbops} dispatched by the host CPU.
Second,  \emph{mat scheduler}, which schedules the most appropriate \emph{bbop} to execute depending on the \emph{bbop}'s mat range and current mat utilization. 
Third, \emph{mat scoreboard}, which tracks whether a given mat is being \omii{used} by a \emph{bbop} instruction. The \emph{mat scoreboard} stores \omii{an} $M$-bit \emph{mat bitmap} \omii{that keeps track of which mats are \omiii{currently} in use}, where $M$ is the number of mats in the DRAM module. 
The \emph{mat scoreboard} can index a range of positions in the \emph{mat bitmap} using a \emph{mat index}.
Fourth, several \gfi{(\omii{e.g.}, eight)} \emph{\gf{\uprog} processing engines}, \omii{each of} which \omii{translates} a \emph{bbop} into its respective \gf{\uprog} and \omiii{controls} \omii{the \emph{bbop}'s} execution. \revdel{A \gf{\uprog} processing engine is equivalent to \omii{a} single SIMDRAM control unit.}
}

\gfcrii{\prop control unit works in four steps.} In the first step, \prop control unit enqueues an incoming \emph{bbop} instruction dispatched by the host CPU (\circled{1} in Fig.~\ref{fig_control_unit}) in the \emph{bbop} buffer. 
In the second step, the mat scheduler scans the \emph{bbop} buffer from the oldest to the newest element. Then, the mat scheduler employs an online first fit algorithm~\cite{garey1972worst} to select a \emph{bbop} to be executed. For each \emph{bbop} in the \emph{bbop} buffer, the algorithm:
\li~extracts the mat range information encoded in the \emph{bbop} (\circled{2}), which is used to index the \emph{mat scoreboard} (\circled{3});
\lii~reads the mat bitmap to identify whether the mats belonging to the \emph{bbop}'s mat range are currently free or busy (\circled{4});
\liii~in case the mats are free, the mat scheduler writes a new mat bitmap to the mat scoreboard, indicating that the given mat range is now busy, selects the current \emph{bbop} to be executed by allocating \gfcrii{and copying the \bbop to} a free \uprog processing engine \gfcrii{(\circled{5})}, and \omi{removes} the current \emph{bbop} from the \emph{bbop} buffer (\circled{6});
\liv~in case the mats \gfcrii{belonging to the \bbop's range} are busy, the mat scheduler reads the next available \emph{bbop} from the \emph{bbop} buffer and repeats \li--\liii.
In the third step, one or multiple \uprog processing engines execute their allocated \emph{bbop}, issuing \texttt{AAP}s/\texttt{AP}s to \gfcrii{the DRAM chips (\circled{7})}\revdel{, one \uprog processing engine per cycle}. \rA{The \uprog processing engine is \changerA{\rA{\#A4}}responsible for maintaining the timing of \texttt{AAP}/\texttt{AP}  commands\revdel{(i.e., the timing delay between \texttt{ACT} and \texttt{PRE})}. In our design, we avoid the need to maintain state for \emph{all} DRAM mats in a DRAM module \emph{individually} by: 
\li~only allowing a \gls{PuD} \gfcrii{operation} to address a contiguous range of DRAM mats, \omi{which share state as} they execute the same sequence of \texttt{ACT-PRE} \omii{commands} and 
\lii~limiting the number of concurrent \gls{PuD} \gfcrii{operations} to the number of \uprog processing engines available in the control unit.} In the fourth step, when a \uprog processing engine finishes executing, it frees \omi{its allocated} mats by \omi{correspondingly updating the mat bitmap in the} mat scoreboard (\circled{8}) \omi{and notifies the CPU that the execution of the \bbop instruction is done}.

%% file: sections/05_design.tex
\section{\prop: Software Support}
\label{sec:idea:software}

\begin{figure}[ht]
    \centering
    \includegraphics[width=\linewidth]{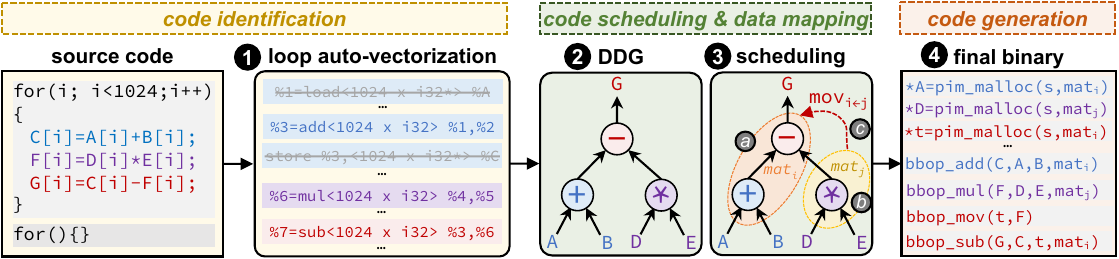}
    \caption{\revA{\prop's compilation flow.}}
    \label{fig_compilation_flow}
\end{figure}

%\prop provides a hardware substrate that exploits data-level and instruction-level parallelism from applications. However, manually identifying, transforming, and scheduling computation to \prop can be challenging. Thereby, 
To ease \prop's programmability, we provide compiler support to transparently map \gls{SIMD} operations to \prop. 
\gfi{Fig.}~\ref{fig_compilation_flow} illustrates \gf{\prop's} compilation flow, which we implement using LLVM~\cite{lattner2008llvm}\omi{: we take} a C/C++ application's source code as input, 
\omi{perform} \gf{three} transformations passes, and  \omi{output} a binary with a mix of CPU and \gfcrii{\gfi{\gls{PuD}} instructions.} \revdel{\rC{\changerC{\rC{\#C2}}We implement {two} transformations and analysis passes at the middle-end, and one at the back-end of the LLVM's compiler toolchain.}}  

\paratitle{Pass 1: Code Identification} The first pass is responsible for \emph{code identification}. Its goal is to identify 
\li~loops that can be successfully auto-vectorized and 
\lii~the appropriate vectorization factor of a given vectorized loop. The code identification pass takes as input the application's LLVM intermediate representation (IR) generated by the compiler's front-end. It produces as output an optimized IR containing \gls{SIMD} instructions that will be translated to \emph{bbop} instructions. 
We leverage the native LLVM's loop auto-vectorization pass~\cite{AutoVect65:online} to identify and transform loops into their vectorized form \revA{(\circled{1} in \gfi{Fig.}~\ref{fig_compilation_flow})}.\footnote{Prior works~\cite{ahmed2019compiler,devic2022pim} also \omi{leverage} modern \omi{compilers'} loop auto-vectorization engines to generate instructions to \gfcrii{processing-near-memory (\gls{PnM}) architectures equipped with \gls{SIMD} engines.}} We apply two modifications to LLVM's loop auto-vectorization pass. 
First, \revdel{we modify how the loop auto-vectorization pass selects the best-performing vectorization factor for a  loop. I}instead of using a cost model to choose the vectorization factor that leads to the highest performance improvement compared to a scalar version of the same loop, we \emph{always} select the \emph{maximum} vectorization factor for the loop. \rC{\changerC{\rC{\#C2}}This is important because the native cost model takes into account the hardware characteristics of \omi{a} target CPU \gls{SIMD} engine (i.e., number of available vector registers, SIMD width of the target execution engine, the latency of different \gls{SIMD} instructions), which are not representative of \omi{our \prop} engine with a variable \gls{SIMD} width.}
Second, \rC{we modify the code generation routine for a given vectorized loop. Concretely, for a given vectorized loop, }we identify and remove memory instructions related to {each} arithmetic \gls{SIMD} operation (i.e., load/store instructions that manipulate vector registers)\revdel{. This optimization is \juani{necessary} } since \gfi{\gls{PuD}} operations directly manipulate the data stored in DRAM\rC{; \changerC{\rC{\#C2}}thus, there is no need to \emph{explicitly} move data into/out \gls{SIMD} registers}. %

\paratitle{Pass 2: Code Scheduling \& Data Mapping} The \gf{second} pass is responsible for \emph{code scheduling and data mapping}. Its goal is to improve overall SIMD utilization by allowing the distribution of independent \gfi{\gls{PuD}} instructions across DRAM mats. \revdel{The key idea behind our code scheduling pass is to distribute \emph{bbop} instructions across DRAM mats based on the \emph{data dependency} of operands.}
Since \gls{PuD} instructions operate \changerC{\rC{\#C2}}directly on the data stored in DRAM, the DRAM mat where the data is allocated determines the efficiency and utilization of the \gls{PuD} SIMD engine. If operands of \omi{independent} instructions are distributed across different DRAM mats, such instructions can be executed concurrently. Likewise, operands of dependent instructions are mapped to the same DRAM mat. 
In that case, intermediate data that one instruction produces and the next instruction consumes do \emph{not} need to be moved across different DRAM mats, improving energy efficiency. \omi{Leveraging these observations,} the code scheduling pass takes as input all \emph{bbop} instructions the code identification pass generates and outputs new \emph{bbop} instructions containing metadata regarding \gf{their} mat location (i.e., \emph{mat label}). The code scheduling pass works in two steps.

In the first step, the code scheduling pass creates a data-dependency graph (DDG) \revA{of the vectorized instructions} \revA{(\circled{2})}. \rC{\changerC{\rC{\#C2}}Each node represents a \emph{bbop} instruction, incoming edges represent input, and outgoing edges represent output of the \emph{bbop}.} In the second step, the code scheduling pass
takes as input the DDG and employs a data scheduling algorithm to distribute \emph{bbop} instructions across DRAM mats \revA{(\circled{3})}. The data scheduling algorithm traverses the DDG \gfcrii{in \emph{topological order} to respect dependencies between \emph{bbop} instructions} using a depth-first search (DFS) kernel\gfcrii{, which is a common algorithm for topological ordering~\cite{cormen2022introduction,tarjan1976edge},} and performs \gf{three} operations. 
First, the algorithm traverses the \emph{left} nodes in the DDG, assigning a single \emph{mat label i} to nodes in the \emph{left} path \revA{(\circled{3}-\circledii{a})}. \revdel{This guarantees that instructions with data dependency reside in the same DRAM mat.}
Second, when the algorithm reaches a leaf node, it traverses the \emph{right} sub-tree in the DDG. In this case, the algorithm assigns a new \emph{mat label j} to the nodes in the \emph{right} path in the sub-tree \revA{(\circled{3}-\circledii{b})}.
Third, once the algorithm visits all the nodes in the \emph{right} sub-tree, it returns to the parent node of the sub-tree. Since the parent node has already been visited when descending into the left path, the left and right sub-tree nodes will be assigned to different mats while having data dependencies across them (as indicated by the parent node). In this case, the algorithm creates a data movement  \emph{bbop} instruction (see \gfi{\cref{sec:design:isa}}) to move the output produced by the right sub-tree from \emph{mat label j} to \emph{mat label i} \revA{(\circled{3}-\circledii{c})}. 
\rC{This process \changerC{\rC{\#C2}}repeats until the algorithm visits all nodes in the DDG.}

% backtrand
% \lii~different mat IDs to all edges and nodes of different connect components, which guarantees that independent in-DRAM \emph{bbop} vector instructions are executed concurrently in different DRAM mats.  
% -> malloc with virtual mat id malloc (size, virtual\_mat\_id, vectorization\_size).
% -> size comes from the data structure 
% -> vectorization\_size = total number of mats that will be required, comes from step 1.
% -> virtual\_mat\_id = indicates grouping. Produced by the data scheduling algorithms.
% -> algorithm has three steps.
% 1: identify connected components 
% 2: nodes within the same connect component gets assign the same  virtual mat id 
% 3: nodes across different connect components get assigned different virtual mat ids. 
% * Pim malloc informs the operating system about the allocation need. The OS then:
% 1. allocates the data accordling (we discuss how this can be implemnted later)
% 2. register the virtual mat id and the assgined mat range to the memory controller (see x). 

\paratitle{Pass \gf{3}: Code Generation}  The \gf{third} pass is responsible for 
\li~\emph{data allocation} and \lii~\emph{code generation}. It takes as input the LLVM IR containing both CPU and \emph{bbop} instructions \omi{(with metadata)} and produces a binary to the target ISA \revA{(\circled{4})}. 
To implement data allocation, the code generation pass first identifies calls for memory allocation routines (e.g., \texttt{malloc}) associated with  operands of \emph{bbop}s and replaces such memory allocation routines with a specialized \gls{PIM} memory allocation routine (i.e., \texttt{pim\_malloc}, see \gfi{\cref{sec:operating:system}}). \texttt{pim\_malloc} receives as input the \emph{mat label} assigned to its associated \emph{bbop} instruction. 
Second, the pass inserts \omi{a} \texttt{bbop\_trsp\_init} \omi{instruction} right after \omi{each} \texttt{pim\_malloc} \omi{call} for each memory object that is an input/output of \gfcrii{a} \emph{bbop} instruction. This instruction registers the memory object in \prop's transposition unit (\gfi{\cref{sec:exc:control}}). Similar to the \texttt{pim\_malloc} call, the \texttt{bbop\_trsp\_init} instruction receives as input the \emph{mat label} assigned to its associated \emph{bbop} instruction. To implement code generation, we modify LLVM's X86 back-end to identify \emph{bbop} instructions and generate the appropriate \gfcrii{assembly code}. \gfhpca{In case the application uses parallel primitives (e.g., OpenMP pragmas~\cite{dagum1998openmp}) to parallelize outermost loops, the code generation pass interacts with the underlying runtime system to statically distribute \emph{bbop} instructions from innermost loops across the  available DRAM mats  in a subarray, i.e., mats \gfcrii{with unassigned \emph{mat labels}}. 
\rC{\changerC{\rC{\#C2}}This allows \prop to execute in a \gfcrii{\gls{SIMT}}~\omi{\cite{lindholm2008nvidia,nvidia2009nvidia}} fashion for manually parallelized applications.} }

% \subsection{Putting All Together: MAT-Based Computing}
% \label{sec:idea:hardware}

% - MAT-based computing enables a fine-grained multiple-instruction multiple-data processing-using-DRAM substrate
% - Each MAT becomes an independent SIMD engine 
% -> Enables different operations to be mapped to a single DRAM subarray 
% -> Enables fine-grained SIMD parallelism inside DRAM
% - Enables new operations: vector reduction 

\section{System Support \omi{for \prop}}
\label{sec:system:integration}

We envision \prop as a tightly-coupled accelerator for the host processor. As such, \prop relies on the host processor for \omi{its system integration}, \omi{which includes} ISA support \gfcrii{(\cref{sec:design:isa})}, 
\gfcrii{instruction execution \& data transposition} \gfcrii{(\cref{sec:exc:control})}, and 
\gfcrii{operating system support for address translation and data allocation \& alignment} \gfcrii{(\cref{sec:operating:system}).}

\subsection{Instruction-Set Architecture}
\label{sec:design:isa}

Table~\ref{table_isa_format} shows the CPU ISA extensions that \prop exposes to the compiler.\footnote{\changeE{\#E3}\revE{\prop ISA extensions are vector-oriented by design. We did \emph{not} use \omii{an} \omi{existing} ISA because we needed to define new fields for \prop that do \emph{not} exist in current vector ISAs (e.g., mat label information). \rD{\changerD{\rD{\#D6}}Instead, we propose to extend the baseline CPU ISA with \prop instructions since there is \omi{usually} more than enough unused opcode space to support the extra opcodes that \prop requires~\cite{lopes2013isa, lopes2015shrink}. Extending the CPU ISA to interface with accelerators is a common approach~\gfcrii{\cite{hajinazarsimdram, PEI, seshadri2017ambit, doblas2023gmx,razdan1994high}}}.}} There are five types of instructions: 
\li~object initialization instructions, 
\lii~1-input arithmetic instructions, 
\liii~2-input arithmetic instructions, 
\liv~predication instructions, and
\lv~data movement instructions. 
The first three types of \prop instructions are inherited from \omi{the} SIMDRAM ISA~\omi{\cite{hajinazarsimdram}}. 
\gfcrii{These instructions can be further divided into two categories:
\li~operations with one input operand (e.g., bitcount, ReLU), 
and 
\lii~operations with two input operands (e.g., addition, division, equal, maximum). 
To enable predication, \prop uses the \texttt{bbop\_if\_else} instruction \gfcriii{that SIMDRAM introduces}, \omii{which takes as input three operands: two input arrays (\texttt{src$_1$} and \texttt{src$_2$}) and \omiii{one} predicate array (\texttt{select}).}} We modify such instructions by including two new fields:
\li~\emph{mat label} (ML), which identifies groups of instructions that must execute inside the same DRAM mat, and
\lii~\emph{vectorization factor} (VF), which dictates how many scalar operands are packed within the vector instruction. These two new fields are automatically generated by \prop's compiler passes (\gfi{\cref{sec:idea:software}}).

\begin{table}[ht]
\tempcommand{0.8}
\vspace{5pt}
\caption{\prop ISA extensions.}
\vspace{-5pt}
\label{table_isa_format}
\resizebox{\linewidth}{!}{
    \begin{tabular}{@{}ll@{}}
    \toprule
    \textbf{Type}                & \multicolumn{1}{c}{\textbf{ISA Format}}    \\ \midrule
    Initialization & \texttt{bbop\_trsp\_init addr, size, n, ML}                         \\
    1-Input Arith.           & \texttt{bbop\_op dst, src, size, n, ML, VF}                                \\
    2-Input Arith.           & \texttt{bbop\_op dst, src$_1$, src$_2$ size, n, ML, VF}               \\
    Predication                  & \texttt{bbop\_if\_else dst, src$_1$, src$_2$, sel, size, n, ML, VF} \\ 
    Data Move & \texttt{bbop\_mov dst, dst\_idx, src, src\_idx, size, n} \\ 
    \bottomrule
    \end{tabular}
}
%\vspace{-5pt}
\end{table}

% In such instructions, \texttt{bbop\_op} represents the opcode of the \prop operation,
% \texttt{src} and \texttt{dst} represent source and destination \emph{arrays}; \texttt{size} represents the number of elements in the source and destination arrays; \texttt{n} represents the number of bits in each array element; and \texttt{sel} represents the
% predicate array.

%\prop supports all 16 operations SIMDRAM proposes.
Data movement instructions allow the compiler to trigger inter-\omi{mat} and intra-mat data movement operations. In a data movement instruction, \texttt{dst} and \texttt{src} represent the source and destination \emph{arrays}; \texttt{dst\_idx} and \texttt{src\_idx} represent the first position of the first element inside the source and destination arrays to be moved; \texttt{size} represents the number of elements to move from source to the destination array; \texttt{n} represents the number of bits in each array element. \prop control unit automatically identifies the \emph{mat range} the data movement instruction targets by calculating the distance between the source and destination arrays, taking into account the indexes and number of elements to move. In case the source and destination mats are the same, \prop control unit translates the data movement instruction into \omi{an} \texttt{LC-MOV} command\omi{:} otherwise, \omi{a \texttt{GB-MOV}} command.

\subsection{Execution \& Data Transposition}
\label{sec:exc:control}

\paratitle{Instruction Fetch and Dispatch}
\prop relies on the host CPU to offload \emph{bbop} instructions to DRAM since they are part of the CPU ISA. Assuming that the host CPU consists of one or more out-of-order cores, \prop leverages the host processor's front-end to 
\li~identify and 
\lii~dispatch to \prop control unit \emph{only} \omi{independent} \emph{bbop}s. This simplifies the design of \prop control unit since no in-flight \emph{bbop} instructions will have data dependencies. As a result, \prop control unit can freely schedule \uprogs to the \gfi{\gls{PuD}} \gls{SIMD} engine as they arrive.  

\paratitle{\gfcrii{Data Coherence}} \gfcrii{Input arrays to \prop may be generated or modified by the CPU, and the data updates may reside only in the cache (e.g., because the updates have not yet been written back to DRAM). To ensure that \prop does not operate on stale data, programmers are responsible for flushing cache lines~\cite{guide2016intel, manual2010arm} modified by the CPU. \prop can leverage coherence optimizations tailored to \gls{PIM} to improve overall performance~\cite{lazypim,boroumand2019conda}.}

\paratitle{\prop Transposition Unit}
\prop transposition unit shares the same hardware components and functionalities as \omi{the} SIMDRAM transposition unit~\omi{\cite{hajinazarsimdram}}, which includes:
\li~\emph{object tracker}, a small cache that keeps track of the memory objects used by \emph{bbop} instructions;
\lii~an \emph{horizontal to vertical transpose} unit, which converts cache lines of memory objects stored in the object tracker from a horizontal to vertical data layout during \omi{a} \gls{LLC} writeback; 
\liii~a \emph{vertical to horizontal transpose} unit, which converts cache lines of memory objects stored in the object tracker from a vertical to horizontal data layout  during an LLC read request;
\liv~\emph{store} and \emph{fetch} units, which generate memory read/write requests using the transpose units' output data. 
One main limitation of \omi{the} SIMDRAM transposition unit is that it needs to fill \emph{at least} an entire DRAM row with vertically-\omi{laid out} data before the execution of a \emph{bbop}. \revdel{In the case number of operands the \emph{bbop} uses is lower than the DRAM row size, SIMDRAM transposition unit fills the remaining space in the DRAM row with `0's. This issue worsens when considering that a vertically laid-out $n$-bit operand spans $n$ different cache lines in DRAM (with each cache line in a different DRAM row). However, since \prop enables fine-grained DRAM activation, \prop transposition unit does \emph{not} need to fill the entire DRAM row with vertically-layout data.} Instead, \prop transposes \omi{only} as much data as required to fill the segment of the DRAM row that the \emph{bbop} instruction operates over. To do so, \omi{the} \prop transposition unit adds information regarding the mat range a memory object operates to the object tracker. 

\subsection{Operating System Support}
\label{sec:operating:system}

\paratitle{Address Translation} 
 As SIMDRAM, \prop operates directly on physical addresses. When the CPU issues a \emph{bbop} instruction, the instruction's virtual memory addresses are translated into their corresponding physical addresses using the same translation lookaside buffer (TLB) lookup mechanisms used by regular load/store operations. 
 
\paratitle{Data Allocation \& Alignment}
\revCommon{\changeCM{\#CQ1}\prop (as other \omi{\gls{PuD}} architectures~\gfcrii{\cite{seshadri2013rowclone, seshadri2018rowclone,
ferreira2022pluto,
seshadri2017ambit,seshadri2019dram,seshadri2015fast,seshadri.bookchapter17,seshadri2016buddy,seshadri2016processing,olgun2022pidram}}) requires OS support to guarantee that data is properly mapped and aligned within the boundaries of the \omi{bank/subarray/mat} that will perform computation. Particularly,} since \gfi{\gls{PuD}} \gfcrii{operations} are executed in-situ, it is essential to enforce that memory objects belonging to the same \emph{bbop} (and their \omi{dependent instructions}) are placed together in the same DRAM mats.
%Otherwise, \prop must execute costly data copy operations to store all memory objects of a given \emph{bbop} inside the same set of DRAM mats before execution. 
%
To achieve this functionality, we propose the implementation of a new data allocation API called \texttt{pim\_malloc}. The main idea is to allow the compiler to inform the OS memory allocator about the memory objects that must be allocated inside the same set of DRAM mats. The \texttt{pim\_malloc} API takes as input the \emph{size} of the memory region to allocate (as a regular \texttt{malloc} instruction) and the \emph{mat label} that the compiler generates (\gfi{\cref{sec:idea:software}}). Then, it  ensures that \emph{all} memory objects with the same \emph{mat label} are \omi{placed together} within \gfcrii{a set of} DRAM mats \gfcrii{that satisfies the target memory allocation size}. \revdel{In case there is no available space within the target mat due to previously allocated data that is not associated with a \emph{bbop} instruction, the \texttt{pim\_malloc} triggers a page fault operation to move one or more pages \revCommon{last-recently used} out of the target mat, freeing space in the target mat for the \texttt{pim\_malloc} to complete.\footnote{\revCommon{\changeCM{\#CQ1}Our analysis assumes that the mat is free whenever a mat is allocated.}} }
%The \texttt{pim\_malloc} also guarantees that memory objects are contiguous in physical memory.

\gfi{To allow the \texttt{pim\_malloc} API to influence the OS memory allocator and ensure that memory objects are placed within specific DRAM mats,\revdel{However, this is challenging due to two main reasons. Firstly, most systems have no knowledge about how the memory controller scrambles data across DRAM module/chip/banks/subarray/mats. Secondly, the \texttt{pim\_malloc} API requires modifications to the OS memory allocation policies to guarantee the allocation of contiguous physical frames for a given memory object. To address these challenges, the OS can obtain information about the DRAM interleaving scheme (e.g., by reverse engineering the bit locations of memory addresses~\cite{kim2020revisiting, orosa2021deeper,yauglikcci2022understanding}), and the \texttt{pim\_malloc} API can create pools of \revCommon{contiguously} allocated pages using huge pages~\cite{santos2022aggressive}.}}
\revCommon{\changeCM{\#CQ1}we propose a new \emph{lazy data allocation routine} (in the kernel) for \texttt{pim\_malloc} objects. This routine has three main components: 
\li~information \omi{about} the DRAM organization (e.g., row, column, and mat sizes), 
\lii~the DRAM interleaving scheme, which the memory controller provides via an open firmware device tree~\cite{devicethree};\footnote{The DRAM interleaving scheme can be obtained by reverse engineering the bit locations of memory addresses~\gfcrii{\cite{kim2020revisiting, orosa2021deeper,yauglikcci2022understanding,loughlin2023siloz}}. \rA{\changerA{\rA{\#A3}}Even though typical DRAM interleaving does \emph{not} take mats into account, it is relatively straightforward to reverse-engineer how a memory address is distributed across the DRAM mats in a DRAM module\gfcrii{, since the mat interleaving is a function of the DRAM chip's organization}. \omi{For example, in} a DDR4 module with 8 chips, 16 mats per chip, and 4 HFFs per mat, a 64~B cache line is evenly distributed across all 128 total mats; i.e., the four least-significant bits of the cache line are placed in mat 0, chip 0, and the four most-significant bits of the cache line are placed in mat 15, chip 7. Our \texttt{pim\_malloc} API takes into account such mat interleaving.}} and 
\liii~a huge \omi{page} pool for \texttt{pim\_malloc} objects (configured during boot time), which guarantees that virtual addresses assigned to a \texttt{pim\_malloc} object are contiguous in the physical address space and \gfhpca{that DRAM mats are free whenever a \texttt{pim\_malloc} object is allocated.}  
The allocation routine uses the DRAM address mapping knowledge to split the huge pages into different memory regions. Then, when an application calls the \texttt{pim\_malloc} API, the allocation routine selects the appropriate memory region that satisfies \texttt{pim\_malloc}.  \rD{\changerD{\rD{\#D7}}Internally, the \texttt{pim\_malloc} API operates using three main sub-tasks, depending on the order of the data allocation: 
\li~\texttt{pim\_preallocate}, for data pre-allocation;
\lii~\texttt{pim\_alloc}, for the first data allocation; and
\liii~\texttt{pim\_alloc\_align}, for subsequent aligned allocations.}}

\rD{\changerD{\rD{\#D7}}\li~\underline{Pre-Allocation.} The first sub-task's role is to indicate the number of huge pages available for \gls{PuD} allocations. We \omii{leave it to the user} to provide the number of huge pages used for \gls{PuD} \gfcrii{operations} because huge pages are scarce in the system.}

\rD{\changerD{\rD{\#D7}}\lii~\underline{First Allocation.} The second sub-task uses the \emph{worst-fit allocation scheme}~\cite{johnson1973near} to manage the allocation of memory regions in the huge page pool. The \emph{key idea} behind this placement strategy is to optimize the remaining \omi{memory} space \omi{after} allocations to increase the chances of accommodating another process in the remaining space. For the first \gls{PuD} memory allocation, the \texttt{pim\_alloc} sub-task simply scans an \emph{ordered array} data structure (similar to the one used in the Linux Kernel buddy allocator algorithm~\cite{knowlton1966programmer}, where each entry represents the number of memory regions in a single subarray) to select the subarray with the \emph{largest} amount of memory regions available. If the requested memory allocation requires more than one memory region, \prop \gfcrii{iteratively} scans the  \emph{ordered array}, searching for the next largest memory region until the memory allocation is fully satisfied. Once enough space is allocated, \texttt{pim\_alloc} sub-task creates a new allocation object and inserts it in an \emph{allocation hashmap}, which is indexed by the allocation's virtual address. The sub-task needs to keep track of allocations since it might need to find a memory region from the \emph{same} subarray/mat when performing future \omi{\emph{aligned allocations}}.} 

\rD{\changerD{\rD{\#D7}}\liii~\underline{Aligned Allocation.} After allocating \gfcrii{a} memory \gfcrii{region} for the first operand in a \gls{PuD} \gfcrii{operation}, the user can use this memory region as a regular memory object. However, when allocating the remaining operands for a \gls{PuD} \gfcrii{operation}, the \texttt{pim\_malloc} API needs to guarantee data alignment for all memory objects within the same DRAM subarray/mat. To this end, the third sub-task (\texttt{pim\_malloc\_align}) identifies \gfcrii{a} previously allocated memory region to which the current memory allocation must be aligned (based on the compiler-generated \emph{mat labels}). The \texttt{pim\_malloc\_align} sub-task works in five main steps. 
First, it searches the \emph{allocation hashmap} for a match with previously allocated memory regions. If a match is not found, the allocation fails.
Second, if a match is found, the \texttt{pim\_malloc\_align} sub-task iterates through the identified previously-allocated memory regions. 
Third, for each memory region, the sub-task identifies its source subarray/mat address and tries to allocate another memory region \omi{in} the same subarray/mat for the new allocation.
Fourth, if the subarray/mat of a given memory region has no free region, the sub-task allocates a new memory region from another subarray/mat following the worst-fit allocation scheme. Since we use a worst-fit allocation scheme \gfcrii{that always selects the \emph{largest} \omii{number} of memory regions available during memory allocation for the \emph{first} operand of a \gls{PuD} operation}, we have a good chance of having a single subarray/mat holding memory regions for \gfcrii{the remaining operands of a \gls{PuD} operation}.
Fifth, since memory regions might come from different huge pages, we must perform \texttt{re-mmap} to map such memory regions into contiguous virtual addresses.}

% This work assumes that the \texttt{pim\_malloc} API can influence the OS memory allocator and the system's \gls{MMU} to ensure that memory objects are continuous in physical space and placed within particular DRAM mats. However, enabling such behavior in practice is challenging for two main reasons. 
% First, in most systems, the OS and the \gls{MMU} have \emph{no} knowledge about how the memory controller scrambles data across DRAM module/chip/banks/subarray/mats (i.e., the DRAM interleaving scheme the memory controller employs is oblivious to the OS). This means that even if the OS tries to allocate continuous physical frames for a given memory object, there are no guarantees that such physical frames will be stored continuously within a DRAM module. 
% Second, the \texttt{pim\_malloc} API requires non-trivial modifications to the OS memory allocation policies to ensure the allocation of continuous physical frames for a given memory object.  
% These challenges can be addressed as follows. 
% First, the OS can obtain information regarding the DRAM interleaving scheme either by cooperating with memory manufacturers or by reverse engineering the bit locations of memory addresses, as done by prior works~\cite{kim2020revisiting, orosa2021deeper,yauglikcci2022understanding}. 
% Second, the \texttt{pim\_malloc} API can create pools of continuously allocated pages using huge pages~\cite{santos2022aggressive}. We leave the concrete implementation of both solutions to future work.

\paratitle{Mat Label Translation} To keep track of the \omi{mapping} between \emph{mat label}s and allocated \emph{mat ranges}, \prop adds a small \emph{mat translation table} alongside the page table. The table is indexed by hashing the \emph{mat label} with the \emph{process ID}. It stores in each entry the associated \emph{mat range} that the memory allocator assigned to that particular  \emph{mat label}. When the CPU dispatches a \emph{bbop}, the CPU  \li~accesses the \emph{mat translation table} to obtain the \emph{mat range} assigned to the given \emph{bbop}, and
\lii~replaces the \emph{mat label} with the \emph{mat range}.   
% \subsection{Programming \& Execution Model}
% \label{sec:design:programming}

%\paratitle{Vertical Data Layout for Bit-Serial Computation.}

%\paratitle{SIMD + VLIW}

% \subsection{Compiler Design}
% \label{sec:design:compiler}

% \subsubsection{Identifying Loop Candidates}

% \subsubsection{Offload Cost Model}

% \subsubsection{Static Instruction Scheduling}

% \subsubsection{Binary Generation}

% \subsection{In-DRAM Execution}
% \label{sec:design:indramexec}

%\subsection{Limitations \& Next Steps}
%\label{sec:design:limitations}

%% file: sections/06_evaluation.tex
\section{Methodology}
\label{sec:methodology}

 We implement \prop using the gem5 simulator~\cite{gem5} and compare it to a real multicore CPU (Intel Skylake~\cite{intelskylake})\revC{, a real high-end GPU (NVIDIA A100~\cite{a100}),} and a state-of-the-art \gfi{\gls{PuD} framework}  (SIMDRAM~\cite{hajinazarsimdram}). \rD{\changerD{\rD{\#D1}}In all our evaluations, the CPU code is optimized to leverage AVX-512 instructions~\cite{firasta2008intel}.} Table~\ref{table_parameters} shows the system parameters we use\revdel{ in our evaluations}.\revdel{ To measure CPU performance, we implement a set of timers in \texttt{sys/time.h}~\cite{systime}.} To measure CPU energy consumption, we use Intel RAPL~\cite{hahnel2012measuring}. \revC{We capture GPU kernel execution time that excludes data initialization/transfer time. To measure GPU energy consumption, we use the \texttt{nvml} API~\cite{NVIDIAMa14}.}
We implement SIMDRAM on gem5\gfcrii{, taking into account that the latency of executing the back-to-back \texttt{ACT}s is only \omii{1.1}$\times$ the latency of $t_{RAS}$~\ambit,} and validate our implementation rigorously with the results reported in \cite{hajinazarsimdram}. \gf{We use CACTI~\cite{cacti} to evaluate \prop and SIMDRAM energy consumption, where we take into account that each additional simultaneous row activation increases energy consumption by 22\%~\cite{seshadri2017ambit, hajinazarsimdram}. 
Our simulation accounts for the additional latency imposes by \prop's mat isolation transistors and row decoder latches\revdel{ upon DRAM operations} (i.e., \gfcrii{measured (using CACTI~\cite{cacti, muralimanohar2007optimizing}) to incur} less than 0.5\% extra latency for an \texttt{ACT}).
\gfcrii{We open-source our simulation infrastructure  at \url{https://github.com/CMUSAFARI/MIMDRAM}}.} 

% We evaluate \emph{two} \prop's implementations:
% \li~\emph{\prop-AOp}, an area-optimized implementation that implements fine-grained DRAM activation \emph{only} to the DRAM rows in the B-group's portion of the DRAM subarray (as described in Section~\ref{}); and 
% \lii~\emph{\prop-TOp}, a throughput-optimized implementation that implements fine-grained DRAM activation  in the \emph{entire} DRAM subarray (i.e., for both D-group and B-group portions of the DRAM subarray). 
% While \emph{\prop-AOp} minimally modifies the design of a DRAM subarray, its peak throughput is limited by the fact that row copy operations must be serialized (Section~\ref{}). In contrast, \emph{\prop-TOp} relaxes this limitation and allows both row copy and triple-row activation operations to be concurrently executed across different DRAM mats, at the cost of a higher area cost. 

%\omii{We \omiii{use} \omiii{the same} vertical data layout in our Ambit \omiii{and SIMDRAM implementations}, \omiii{which} enables us to (1) evaluate all 16 SIMDRAM operations in Ambit using their equivalent AND/OR/NOT-based implementation\omiii{s}, and (2) highlight the benefits of Step 1 in the \mech framework (i.e., using an optimized MAJ/NOT-based implementation of the operations).} %We \omi{modify Ambit to operate on vertically-laid-out data, to illustrate} the benefits of Steps 1 and 2 of our methodology. 
%\omvuii{Our synthetic throughput analysis \omviii{(\cref{sec_performance})} uses 64M-element input arrays.}

\begin{table}[ht]
  \vspace{5pt}
   \caption{Evaluated system configurations.}
   \vspace{-5pt}

   \centering
   \footnotesize
   \tempcommand{1.3}
   \renewcommand{\arraystretch}{0.7}
   \resizebox{\columnwidth}{!}{
   \begin{tabular}{@{} c l @{}}
   \toprule
   \multirow{5}{*}{\shortstack{\textbf{\omi{Real} Intel}\\ \textbf{Skylake CPU~\cite{intelskylake}}}} & x86~\cite{guide2016intel}, 16~cores, 8-wide, out-of-order, \SI{4}{\giga\hertz};  \\
                                                                           & \emph{L1 Data + Inst. Private Cache:} \SI{256}{\kilo\byte}, 8-way, \SI{64}{\byte} line; \\
                                                                           & \emph{L2 Private Cache:} \SI{2}{\kilo\byte}, 4-way, \SI{64}{\byte} line; \\
                                                                           & \emph{L3 Shared Cache:} \SI{16}{\mega\byte}, 16-way, \SI{64}{\byte} line; \\
                                                                           & \emph{Main Memory:} \SI{64}{\giga\byte} DDR4-2133, 4~channels, 4~ranks \\
   \midrule
      \multirow{3}{*}{\shortstack{\textbf{\omi{Real} \revC{NVIDIA}}\\ \textbf{\revC{A100 GPU~\mbox{\cite{a100}}}}}} &  \revC{\SI{7}{\nano\meter} technology node; 6912 CUDA Cores;}\\ 
                                                                            & \revC{108 streaming multiprocessors, \SI{1.4}{\giga\hertz} base clock;} \\
                                                                            & \revC{\emph{L2 Cache:} \SI{40}{\mega\byte} L2 Cache; \emph{Main Memory:} \SI{40}{\giga\byte} HBM2~\mbox{\cite{HBM,lee2016simultaneous}}} \\
   \midrule

   \multirow{8}{*}{\shortstack{\omi{\textbf{Simulated}} \\ \textbf{SIMDRAM~\cite{hajinazarsimdram}}\\ \textbf{\& \prop}}} &  gem5 system emulation;  x86~\cite{guide2016intel}, 1-core, out-of-order, \SI{4}{\giga\hertz};\\
                                                                             & \emph{L1 Data + Inst. Cache:} \SI{32}{\kilo\byte}, 8-way, \SI{64}{\byte} line;\\
                                                                             & \emph{L2 Cache:} \SI{256}{\kilo\byte}, 4-way, \SI{64}{\byte} line; \\
                                                                             & \emph{Memory Controller:}  \SI{8}{\kilo\byte} row size, FR-FCFS~\cite{mutlu2007stall,zuravleff1997controller}\\
                                                                             & \emph{Main Memory:}  DDR4-2400, 1~channel, 8~chips, 4~rank \\ &
                                        16~banks/rank, 16~mats/chip, 1~K rows/mat, 512~columns/mat\\      &
                                        
                                       \emph{\prop's Setup:} 8~entries mat queue, \SI{2}{\kilo\byte}~\emph{bbop} buffer \\ &
                                       8~\textit{\uprog{} processing engines},   \SI{2}{\kilo\byte}~\emph{mat translation table} \\ 
                                      % \midrule
%   \multirow{5}{*}{\textbf{\prop}} &  gem5 system emulation;  x86~\cite{guide2016intel}, 1-core, out-of-order, \SI{4}{\giga\hertz};\\
%                                                                              & \emph{L1 Data + Inst. Cache:} \SI{32}{\kilo\byte}, 8-way, \SI{64}{\byte} line;\\
%                                                                              & \emph{L2 Cache:} \SI{256}{\kilo\byte}, 4-way, \SI{64}{\byte} line; \\
%                                                                              & \emph{Memory Controller:}  \SI{8}{\kilo\byte} row size, FR-FCFS~\cite{mutlu2007stall,zuravleff1997controller} scheduling\\
%                                                                              & \emph{Main Memory:}  DDR4-2400, 1~channel, 1~rank, 16~banks \\
  
   \bottomrule
   \end{tabular}
   }
   \label{table_parameters}
\end{table}

%\gfbox{Copy paste table 2 ``DRAM'' from sectored dram please myself}

\paratitle{Real-World Applications} We analyze \gf{117 applications from \omi{seven} benchmark suites (SPEC 2017~\cite{spec2017}, SPEC 2006~\cite{spec2006}, Parboil~\cite{stratton2012parboil}, Phoenix~\cite{yoo_iiswc2009}, Polybench~\cite{pouchet2012polybench}, Rodinia~\cite{che_iiswc2009}, and SPLASH-2~\cite{woo_isca1995}) to select applications that} 
\li~are memory-bound, and 
\lii~the most \juani{\omi{time}-consuming} loop can be auto-vectorized. 
From this analysis, we collect \gf{12} multi-threaded \revC{CPU} applications \rD{(\changerD{\rD{\#D3}}as Table~\ref{table:workload:properties} describes)} from different domains (i.e., video compression, data mining, pattern recognition, medical imaging, stencil computation)\revC{, and their respective GPU implementations, when available}\omi{. Our evaluated applications are:} 
\li~525.x264\_r (\texttt{x264}) from SPEC 2017;
\lii~heartwall (\texttt{hw}), kmeans (\texttt{km}), and backprop (\texttt{bs}) from Rodinia;
\liii~\texttt{pca} from Phoenix; and
\liv~\texttt{2mm}, \texttt{3mm}, covariance (\texttt{cov}), doitgen (\texttt{dg}), fdtd-apml (\texttt{fdtd}), gemm (\texttt{gmm}), and gramschmidt (\texttt{gm}) from Polybench.\footnote{\rE{\omi{Several} prior works~\cite{damov,devic2022pim,dualitycache,fujiki2018memory,vadivel2020tdo,iskandar2023ndp,pattnaik2016scheduling} \omi{show} \changerE{\rE{\#E1}} that our selected \gfcrii{twelve} \omi{workloads} can benefit from different types of \gls{PIM} architectures. }}
\gf{Since our base \gfi{\gls{PuD}} substrate (SIMDRAM) does \emph{not} support floating-point, we manually modify the selected floating-point-heavy auto-vectorized loops to operate \juani{on} fixed-point data arrays.\footnote{\changerD{\rD{\#D10}}\rD{We only modify the three applications from the Rodinia benchmark suite to use fixed-point operations. \omi{Prior works~\cite{fujiki2018memory,yazdanbakhsh2016axbench,ho2017efficient} also \omi{employ} fixed-point for the same three Rodinia applications.} The applications from Polybench can be configured to use integers; the auto-vectorized loops in 525.x264\_r use \texttt{uint8\_t}; pca uses integers. }We do \emph{not} observe an output quality degradation when employing fixed-point for the selected loops.} We use the largest input dataset available \revE{and execute each application \emph{end-to-end}} in our evaluations.}\changeE{\#E1}

\begin{table}[ht]
   \caption{Evaluated applications and their characteristics.}
   \tempcommand{1}
   \resizebox{\columnwidth}{!}{%
    \begin{tabular}{|c|c||c|c|c|c|}
\hline
\textbf{\begin{tabular}[c]{@{}c@{}}Benchmark\\  Suite\end{tabular}} & \textbf{\begin{tabular}[c]{@{}c@{}}Application\\ (Short Name)\end{tabular}} & \textbf{\begin{tabular}[c]{@{}c@{}}Dataset \\ Size\end{tabular}} & \textbf{\begin{tabular}[c]{@{}c@{}}\# Vector \\ Loops\end{tabular}} & \textbf{\begin{tabular}[c]{@{}c@{}}VF\\  \{min, max\}\end{tabular}} & \textbf{\begin{tabular}[c]{@{}c@{}}PUD \\ Ops.{$^\dag$}\end{tabular}} \\ \hline \hline
\begin{tabular}[c]{@{}c@{}}Phoenix~\cite{yoo_iiswc2009}\end{tabular} & $^\ddag$pca (\texttt{pca}) & reference & 2 & \{4000, 4000\} & D, S, M, R \\ \hline
\multirow{7}{*}{\begin{tabular}[c]{@{}c@{}}Polybench\\ \cite{pouchet2012polybench}\end{tabular}} & 2mm (\texttt{2mm}) & \begin{tabular}[c]{@{}c@{}}NI = NJ = NK = NL = 4000\end{tabular} & 6 & \{4000, 4000\} & M, R \\ \cline{2-6} 
 & $^\ddag$3mm (\texttt{3mm}) & \begin{tabular}[c]{@{}c@{}}NI = NJ = NK = NL = NM = 4000\end{tabular} & 7 & \{4000, 4000\} & M, R \\ \cline{2-6} 
 & covariance (\texttt{cov}) & \begin{tabular}[c]{@{}c@{}}N = M = 4000\end{tabular} & 2 & \{4000, 4000\} & D, S, R \\ \cline{2-6} 
 & doitgen (\texttt{dg}) & \begin{tabular}[c]{@{}c@{}}NQ = NR = NP = 1000\end{tabular} & 5 & \{1000, 1000\} & M, C, R \\ \cline{2-6} 
 & $^\ddag$fdtd-apml (\texttt{fdtd}) & \begin{tabular}[c]{@{}c@{}}CZ = CYM = CXM = 1000\end{tabular} & 3 & \{1000, 1000\} & D, M, S, A  \\ \cline{2-6} 
 & gemm (\texttt{gmm}) & \begin{tabular}[c]{@{}c@{}}NI = NJ = NK = 4000\end{tabular} & 4 & \{4000, 4000\} & M, R  \\ \cline{2-6} 
 & gramschmidt (\texttt{gs}) & \begin{tabular}[c]{@{}c@{}}NI = NJ = 4000\end{tabular} & 5 & \{4000, 4000\} & M, D, R  \\ \hline
\multirow{3}{*}{\begin{tabular}[c]{@{}c@{}}Rodinia\\ \cite{che_iiswc2009}\end{tabular}} & backprop (\texttt{bs}) & 134217729 input elm.  & 1 & \{17, 134217729\} & M, R  \\ \cline{2-6} 
 & heartwall (\texttt{hw}) &  reference & 4 & \{1, 2601\} & M, R  \\ \cline{2-6} 
 & kmeans (\texttt{km}) & 16384 data points & 2 & \{16384, 16384\} & S, M, R  \\ \hline
\begin{tabular}[c]{@{}c@{}}SPEC 2017\\\cite{spec2017}\end{tabular} & 525.x64\_r (\texttt{x264}) & \begin{tabular}[c]{@{}c@{}}reference input\end{tabular} & 2 & \{64, 320\} & A  \\ \hline
\end{tabular}%
}
\scriptsize{$^\dag$: D = division, S = subtraction, M = multiplication, A = addition, R = reduction, C = copy
}
\newline
\scriptsize{$^\ddag$\gfcrii{: application with independent \gls{PuD} operations}
}
\label{table:workload:properties}
\end{table}

\paratitle{\gfcrii{Multi-Programmed Application Mixes}} \gfcrii{To measure system throughput and fairness, we \emph{manually} create 495 application mixes by randomly selecting eight applications (from our group of 12 applications) for execution co-location. We classify each application mix into \agy{one of} three categories: 
\emph{low}, \emph{medium}, and \emph{high} vectorization factor (VF) mixes based on \gfi{Fig.}~\ref{fig_max_utilization}.
In the \omii{\emph{low}} mix, the maximum VF of \emph{all} eight applications is lower than 16K; 
in the \omii{\emph{medium}} mix, \emph{at least} one application has a maximum vectorization factor between 16K (inclusive) and 64K; and
in the \omii{\emph{large}} mix, \emph{at least} one application has a maximum VF larger than 64K (inclusive). }

\paratitle{\gfcrii{\omii{Comparison to State-of-the-Art} \gls{PIM} Architectures}} \gfcrii{We compare \prop to two other state-of-the-art \gls{PIM} architectures: DRISA~\cite{li2017drisa} and Fulcrum~\cite{lenjani2020fulcrum}. 
DRISA is a \omi{combined} \gls{PuM} \omi{and} \gls{PnM} architecture that \emph{significantly} modifies the DRAM \omii{microarchitecture} and organization to enable bulk in-DRAM computation (\omi{e.g.}, by using 3T1C DRAM cells to execute in-situ bitwise NOR operations \omi{and} by adding logic gates \emph{near} the subarray's sense amplifiers). \revdel{DRISA employs a fine-grained interconnection network to shift data across DRAM columns, thus executing operations in a \emph{bit-parallel} mode (in contrast with SIMDRAM and \prop, which execute operations in a \emph{bit-serial} mode). 
In this analysis, we employ DRISA's 3T1C implementation to contrast both bit-serial and bit-parallel \gls{PuD} execution models.}
Fulcrum is a \gls{PnM} architecture that adds \omi{computation} logic \emph{near} subarrays. 
Fulcrum's primary components are a series of shift registers (called walkers) that latch input/output DRAM rows and a narrow scalar ALU \omi{that executes arithmetic and logic operations}. We model \omi{the} DRISA 3T1C implementation and Fulcrum  \li~\omii{using a DRAM module of equal} \gfcrii{dimensions (i.e., number of DRAM ranks, chips, banks, mats, rows, and columns) as the} baseline DDR4 \gfcrii{DRAM} we use for SIMDRAM and \prop (see Table~\ref{table_parameters}) \gfcriii{and} 
\lii~including all the changes that the DRISA 3T1C and Fulcrum architectures propose to the DRAM cell array and DRAM subarray.}

\section{Evaluation}
\label{sec:eval}

We demonstrate the advantages of \prop 
by evaluating 
\li~\gls{SIMD} utilization and energy efficiency (i.e., performance per Watt) for single applications \gfcrii{(\cref{sec:eval:singleapp})}; 
\lii~\gf{system throughput (in terms of weighted speedup~\cite{snavely2000symbiotic, eyerman2008systemlevel, michaud2012demystifying}), job turnaround time (in terms of harmonic speedup~\cite{luo2001balancing,eyerman2008systemlevel}), and fairness (in terms of maximum slowdown~\cite{kim2010thread, kim2010atlas,subramanian2014blacklisting,subramanian2016bliss, subramanian2013mise, mutlu2007stall, subramanian2015application, ebrahimi2010fairness, ebrahimi2011prefetch, das2009application, das2013application})} for multi-\gfcrii{programmed application mixes} in comparison to the baseline CPU, GPU, and a state-of-the-art \gls{PuD} architecture, i.e., SIMDRAM~\cite{hajinazarsimdram} \gfcrii{(\cref{sec:eval:multiapp})}; 
\lii~\gfcrii{area-normalized performance analysis for single applications  and throughput analysis for multi-programmed application mixes in comparison to state-of-the-art \gls{PIM} architectures, i.e., DRISA~\cite{li2017drisa} and Fulcrum~\cite{lenjani2020fulcrum}   (\cref{sec:eval:otherpims}).}
\gfcriii{\omiii{In most of} our analyses (\cref{sec:eval:singleapp}--\cref{sec:eval:multiapp}), \omiii{to keep our analyses pure,} we \omiii{very} \emph{conservatively} allow \prop to use \omiii{only} a \emph{single} DRAM subarray in a \emph{single} DRAM bank for \gls{PuD} computation. 
In \cref{sec:eval:scalability}, we perform a scalability analysis to evaluate \prop's performance when enabling \emph{multiple} DRAM subarrays and banks for \gls{PuD} computation\omiii{, which reflects a more accurate evaluation of the true benefits of \prop and \gls{PuD}}.} 
Finally, we evaluate \prop's DRAM and CPU area cost \gfcrii{(\cref{sec:eval:area})}.

\subsection{Single-Application \omii{Results}}
\label{sec:eval:singleapp}

\begin{figure}[ht]
\begin{subfigure}{\linewidth}
  \centering
  \includegraphics[width=\linewidth]{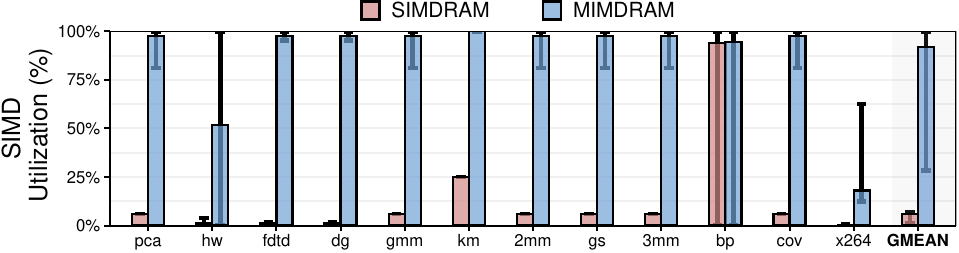}  
  \caption{\omi{\gls{SIMD} utilization. \gf{Whiskers extend to the minimum and maximum \omi{observed} data point values.}}}
  \label{fig:sub-first}
\end{subfigure}
~
\begin{subfigure}{\linewidth}
\centering\includegraphics[width=\linewidth]{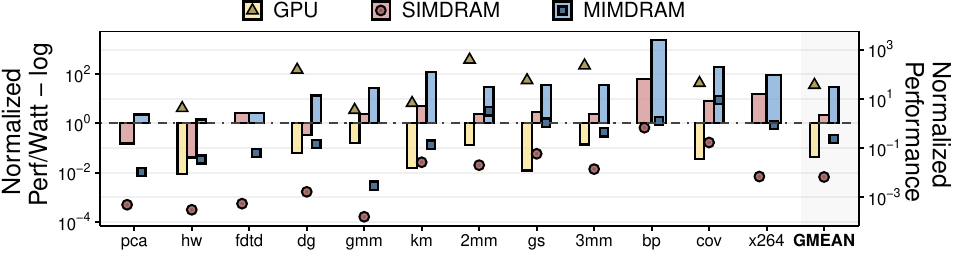}  
  \caption{CPU-normalized performance per Watt (left y-axis; bars) and \revC{performance (right y-axis; dots)}.}
  \label{fig:sub-second}
\end{subfigure}
\caption{Single-application \omii{results} \gfcrii{for processor-centric (i.e., CPU and GPU) and memory-centric (i.e., SIMDRAM and \prop) architectures executing twelve real-world applications}. }
\label{fig_single_app_analysis}
\end{figure}

\gfi{Fig.}~\ref{fig_single_app_analysis} shows \prop's \gls{SIMD} utilization \gf{and normalized energy efficiency (in performance per Watt)} for all \gf{12} applications. Values are normalized to the baseline CPU. 
%\agycomment{Isn't it better to split this figure into two separate figures and have SIMD Utilization and Energy Efficiency as separate subsections of evaluation?}

% \begin{figure}[ht]
%     \vspace{-8pt}
%     \centering
%     \includegraphics[width=\linewidth]{plots/utilization_perf_watt-crop.pdf}
%     \caption{Single-application analysis. \gf{Whiskers extend to the minimum and maximum data point values.}}
%     \label{fig_single_app_analysis}
%     \vspace{-8pt}
% \end{figure}

\paratitle{SIMD Utilization} 
We make two observations from \gfi{Fig.}~\ref{fig_single_app_analysis}a. First, \prop \emph{significantly} improves \gls{SIMD} utilization \omi{over} SIMDRAM. On average across all applications, \prop provides 15.6$\times$ the \gls{SIMD} utilization of SIMDRAM. This is because \prop matches the available \gls{SIMD} parallelism in an application with the underlying \gfi{\gls{PuD}} resources (i.e., \gfi{\gls{PuD}} \gls{SIMD} lanes) by \omi{using} \emph{only} as many DRAM mats as the maximum vectorization factor of a given application's loop. In contrast, SIMDRAM always occupies \omi{\emph{all}} available \gfi{\gls{PuD}} \gls{SIMD} lanes \omi{(i.e., entire subarrays)} for a given operation, resulting in low \gls{SIMD} utilization for applications without a \omi{very}-wide vectorization factor. 
Second, we observe that \gls{SIMD} utilization can vary considerably within an application. For example, \prop's \gls{SIMD} utilization for \texttt{hw} and \texttt{bp} goes from as low as 0.2\% to as high as 100\%. This happens because the \gls{SIMD} parallelism for each vectorized loop in these applications changes at different execution phases. \prop can \gfi{better} adjust to the variation in \gls{SIMD} parallelism \omi{(than SIMDRAM)} due to its flexible design. 
We conclude that \prop \omi{greatly improves} overall \gls{SIMD} utilization for many applications. 

\paratitle{\rC{Performance \&} Energy Efficiency} We make \gfhpca{three} observations from \gfi{Fig.}~\ref{fig_single_app_analysis}b. 
First, \prop \emph{significantly} improves energy efficiency \rC{and performance} \omi{over} SIMDRAM. On average across all applications, \prop provides \efficiencysimdram the energy efficiency \rC{and 34$\times$ the performance} of SIMDRAM. \gf{\prop's higher energy efficiency is due to three main reasons.}
\li~\prop parallelizes the computation of \omi{independent} \emph{bbops} in a single application loop across different mats, improving overall performance. \prop reduces execution time \changerC{\rC{\#C1}}by 2.8$\times$ compared with SIMDRAM, on average across applications with \omi{independent} \emph{bbops} \gfcrii{(i.e., \texttt{pca}, \texttt{3mm}, and \texttt{fdtd}).}
\lii~\prop implements in-situ \gfi{\gls{PuD}} vector reduction operations, while SIMDRAM requires the assistance of the CPU \gfcrii{for} vector reduction, increasing latency and energy consumption. \rC{\changerC{\rC{\#C1}}\prop reduces execution time and energy consumption by 1.6$\times$ and 266$\times$ \omi{over} SIMDRAM, on average across the applications with vector reduction \gfcrii{operations} \gfcrii{(from our twelve applications, only \texttt{fdtd} and \texttt{x264} do \emph{not} require vector reduction operations)}.}
\liii~\prop activates \omi{only} the \omi{necessary} {\gls{PuD}} \gls{SIMD} lanes during an application loop's execution, significantly saving energy  when the application has low \gls{SIMD} utilization. \rC{\changerC{\rC{\#C1}}\prop reduces energy consumption by 325$\times$ \omi{over} SIMDRAM, on average across applications with a maximum vectorization factor lower than 65,536 \gfcrii{(from our twelve applications, only \texttt{bs} exhibits a vectorization factor \emph{higher} than 65,536)}.}
\rD{\changerD{\rD{\#D2}}Second, \prop provides \efficiencycpu/\efficiencygpu the energy efficiency of CPU{/GPU baselines}. \prop's higher energy efficiency is due to its inherent ability to avoid costly data movement operations for memory-bound applications. 
{Third, even though \prop improves performance (by 3.1$\times$, 8.6$\times$, 1.1$\times$, and 1.3$\times$) compared to the baseline CPU for some applications (i.e., \texttt{2mm}, \texttt{cov}, \texttt{gs}, and \texttt{bp}), it \omi{leads to} performance loss compared to the baseline CPU and GPU on average across all applications. This is because, for some applications, the bulk parallelism available inside a \emph{single} DRAM \omiii{subarray and bank} is insufficient to hide the latency of costly bit-serial operations (e.g., multiplication). We observe that enabling \prop in \gfcriii{16} DRAM banks \gfcriii{and 64 subarrays (per bank) allows \prop to provide performance gains compared to} the CPU and the GPU (see \cref{sec:eval:scalability}).}}
%\revC{However, such an increase in energy efficiency comes at the cost of lower performance compared to the baseline CPU/GPU. This is because while the CPU and GPU take advantage of multi-threaded execution, \prop's only source of parallelism comes from the data parallelism available in a vectorized loop, which penalizes performance. This issue can be overcome by allowing \prop to leverage other forms of parallelism for computation (e.g., application-level parallelism, as we evaluate in \cref{sec:eval:multiapp}).} 
We conclude that \prop is an energy-efficient \omiii{and high-performance} \gfi{\gls{PuD}} system.

\subsection{Multi-\gfcrii{Programmed} \omii{Workload Results}}
\label{sec:eval:multiapp} 
\begin{figure}[ht]
    \centering
    \includegraphics[width=\linewidth]{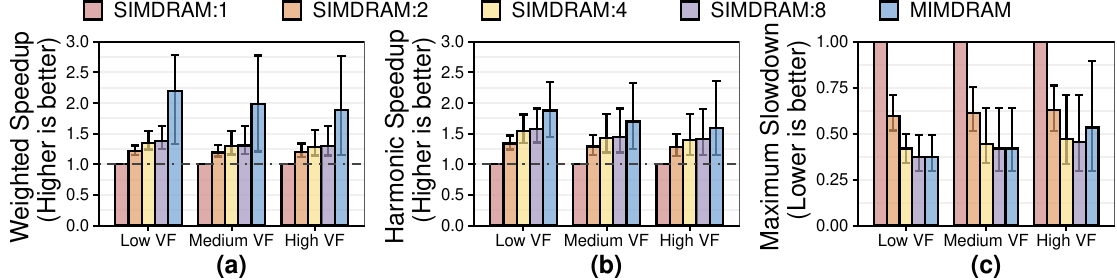}
    \caption{\gfcriii{Multi-\gfcrii{programmed} \omii{workload results} \gf{for three types of application mixes:} 
    \gfcrii{(a) low VF, 
    (b) medium VF, and 
    (c) high VF.
    \emph{VF} stands for vectorization factor.
    \emph{SIMDRAM:X} uses $X$ DRAM banks for computation}. Values are normalized to \emph{SIMDRAM:1}. Whiskers extend to the minimum and maximum observed data point values.}}
    \label{fig_mult_app_analysis}
\end{figure}

\gf{We evaluate SIMDRAM and \prop's impact on system throughput (in terms of weighted speedup~\cite{snavely2000symbiotic, eyerman2008systemlevel, michaud2012demystifying}), job turnaround time (in terms of harmonic speedup~\cite{luo2001balancing,eyerman2008systemlevel}), and fairness (in terms of maximum slowdown~\cite{kim2010thread, kim2010atlas,subramanian2014blacklisting,subramanian2016bliss, subramanian2013mise, mutlu2007stall, subramanian2015application, ebrahimi2010fairness, ebrahimi2011prefetch, das2009application, das2013application}) when executing applications concurrently. 
\revdel{\rD{\changerD{\rD{\#D5}}To do so, we \emph{manually} create 495 application mixes by randomly selecting eight applications (from our group of 12 applications) for execution co-location.} We classify each application mix into \agy{one of} three categories: \emph{low}, \emph{medium}, and \emph{high} vectorization factor (VF) mixes based on \gfi{Fig.}~\ref{fig_max_utilization}.
In the low VF application mix, the maximum VF of \emph{all} eight applications is lower than 16K; in the medium VF application mix, \emph{at least} one application has a maximum vectorization factor between 16K (inclusive) and 64K; and
in the large VF application mix, \emph{at least} one application has a maximum VF larger than 64K (inclusive).} 
To provide a fair comparison, we introduce \gls{MIMD} parallelism in SIMDRAM with bank-level parallelism (BLP)~\cite{mutlu2008parallelism,kim2010thread,kim2012case,kim2016ramulator,lee2009improving}, where each SIMDRAM-capable DRAM bank can independently run an application. We evaluate four configurations of SIMDRAM where 1 (\emph{SIMDRAM:1}), 2 (\emph{SIMDRAM:2}), 4 (\emph{SIMDRAM:4}), and  8 (\emph{SIMDRAM:8}) banks have SIMDRAM computation capability.} \gf{\gfi{Fig.}~\ref{fig_mult_app_analysis} shows the system throughput, job turnaround time (which measures a balance of fairness and throughput), and fairness that SIMDRAM and \prop provide on average across all application mixes. Values are normalized to \emph{SIMDRAM:1}. We make \omiii{three} observations\revdel{ from the figure}.}

\gf{First, \prop \emph{significantly} improves system throughput, job turnaround time, and fairness compared with SIMDRAM. On average, across all application groups, \prop achieves:
\li~1.68$\times$ (min. 1.52$\times$, max. 2.02$\times$) \emph{higher} weighted speedup,
\lii~1.33$\times$ (min. 1.17$\times$, max. 1.72$\times$) \emph{higher} harmonic speedup, and
\liii~1.32$\times$ (min. 0.95$\times$, max. 2.29$\times$) \emph{lower} maximum slowdown than SIMDRAM (averaged across all four configurations). 
\omiii{Second, \prop using a single subarray and single bank for computation, provides 1.68$\times$, 1.54$\times$, and 1.52$\times$ the system throughput of SIMDRAM using 2, 4, and 8 banks for computation, respectively.}
%\agycomment{why not best performing?}
This happens because \prop
\li~utilizes idle resources \agy{\omi{at} DRAM mat granularity} to execute computation as soon as \agy{a mat is}
% they are 
available, thus reducing queuing time \omi{and improving parallelism}; and  
%(in contrast, for SIMDRAM, applications
% need to wait in the execution queue when\agy{are queued as long as} the number of \agy{idle} SIMDRAM-capable DRAM banks is lower than eight);\agycomment{is this true? we schedule resources in a DRAM bank granularity. applications should not wait if there is a bank}
\lii~reduces execution latency of a single application due to its concurrent execution of \omi{independent} \emph{bbop} instructions and support for \gfi{\gls{PuD}} vector reduction.
\omiii{Third}, even though \prop achieves similar fairness compared with \emph{SIMDRAM:4} and \emph{SIMDRAM:8} for application mixes with low and medium VF, \prop's maximum slowdown is 15\% (12\%) higher than \emph{SIMDRAM:8} (\emph{SIMDRAM:4}) for application mixes with high VF. 
This is because in \prop 
\li~applications share the DRAM mats available inside a single DRAM bank and
\lii~\emph{bbops} are dispatched to execution using an online \agy{\emph{first fit}} \agy{algorithm}. In this way, an application in a mix with \emph{high} occupancy and execution latency penalizes an application with \emph{low} occupancy and execution time, \agy{negatively} impacting fairness. In contrast, such interference does \emph{not} happen in \emph{SIMDRAM:8} since each application is assigned \agy{to} a different DRAM bank to execute \agy{at the cost of occupying eight banks instead of one}. \prop's fairness can be further improved by
\li~employing better scheduling algorithms that target quality-of-service~\cite{mutlu2008parallelism, luo2001balancing, xie2014improving,subramanian2013mise,subramanian2014blacklisting,ebrahimi2010fairness} or
\lii~using \omiii{\gls{SLP}~\cite{kim2012case} and} BLP~\cite{mutlu2008parallelism,kim2010thread,kim2012case,kim2016ramulator,lee2009improving}  in MIMDRAM\omiii{, i.e., exploiting multiple subarrays and multiple banks for \prop computation (\cref{sec:eval:scalability}}).\footnote{\omiii{In our extended version~\cite{mimdramextended}, we provide multi-programmed workload results while exploiting \gls{SLP} and BLP for \prop computation.}} We conclude that \prop is an efficient \omi{and high-performance} \gfi{\gls{PuD}} substrate when the system \agy{concurrently} executes several applications. }

\begin{figure}[ht]
    \centering
    \includegraphics[width=0.95\linewidth]{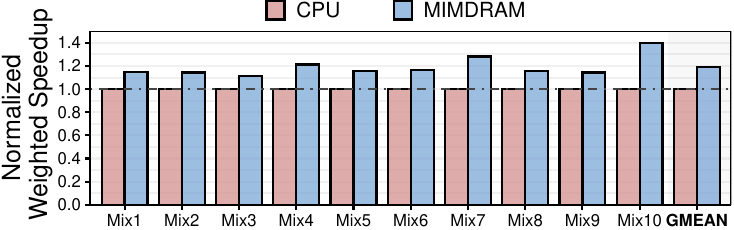}
    \caption{\revC{Multi-\gfcrii{programmed} \omii{workload results} for ten application mixes. Values are normalized to the baseline CPU.}}
    \label{fig_mult_app_analysis_cpu}
\end{figure}

\paratitle{\changeC{\#C1}\revC{CPU Multi-\omii{Programmed Workload Results}}} \revC{We evaluate how \prop performance compares to that of \omi{a state-of-the-art} CPU when executing multiple applications. To do so, we randomly \omi{generate} ten different application mixes, each containing eight applications out of our 12 applications. Then, we run each application mix in our baseline CPU (using multi-threading) and in \prop and compute the achieved system throughput for each system (using weighted speedup). Fig.~\ref{fig_mult_app_analysis_cpu} shows the system throughput \prop achieves compared to the baseline CPU. We observe that \prop improves overall throughput by 19\%. This is because \prop can parallelize the execution of the applications in each application mix across the DRAM mats in a subarray. In contrast, when executing each application mix, the baseline CPU often suffers from contention in its shared resources (e.g., shared cache and DRAM bus). We conclude that \prop is an efficient substrate for highly-parallel environments.}  

\revdel{
\begin{table}[ht]
   \resizebox{0.8\columnwidth}{!}{
\begin{tabular}{|c||c|}
\hline
\textbf{Application Mix} & \textbf{Applications}                 \\ \hline \hline
Mix1                     & 2mm, 3mm, cov, dg, fdtd, gmm, hw, km  \\ \hline
Mix2                     & 2mm, 3mm, cov, fdtd, bp, hw, km, pca  \\ \hline
Mix3                     & 3mm, cov, dg, gmm, bp, hw, km, pca    \\ \hline
Mix4                     & 2mm, 3mm, dg, fdtd, gs, hw, km, pca   \\ \hline
Mix5                     & 2mm, cov, dg, fdtd, gs, bp, hw, pca   \\ \hline
Mix6                     & 2mm, cov, dg, fdtd, gs, hw, km, pca   \\ \hline
Mix7                     & 2mm, 3mm, cov, dg, gmm, bp, hw, pca   \\ \hline
Mix8                     & 2mm, 3mm, cov, dg, gmm, gs, hw, km    \\ \hline
Mix9                     & 2mm, fdtd, gmm, gs, bp, hw, km, pca   \\ \hline
Mix 10                   & 2mm, 3mm, cov, dg, fdtd, gmm, bp, pca \\ \hline
\end{tabular}
}
\end{table}
}

\subsection{\revCommon{Comparison to Other PIM Architectures}}
\label{sec:eval:otherpims}

\revdel{\changeCM{\#CQ2}\revCommon{We compare \prop to two other state-of-the-art \gls{PIM} architectures: DRISA~\cite{li2017drisa} and Fulcrum~\cite{lenjani2020fulcrum}. 
DRISA is a \gls{PuM}/\gls{PnM} architecture that \emph{significantly} modifies the DRAM micro-architecture and organization to enable bulk in-DRAM computation (i.e., by using 3T1C DRAM cells to execute in-situ bitwise NOR operations or by adding logic gates \emph{near} the subarray's sense amplifiers). \revdel{DRISA employs a fine-grained interconnection network to shift data across DRAM columns, thus executing operations in a \emph{bit-parallel} mode (in contrast with SIMDRAM and \prop, which execute operations in a \emph{bit-serial} mode). 
In this analysis, we employ DRISA's 3T1C implementation to contrast both bit-serial and bit-parallel \gls{PuD} execution models.}
Fulcrum is a \gls{PnM} architecture that adds logic \emph{near} subarrays. 
Fulcrum's primary components are a series of shift registers (called walkers) that latch input/output DRAM rows and a narrow scalar ALU.\revdel{By using a single narrow scalar ALU for operations, Fulcrum provides a more flexible execution model than row-wide bitwise \gls{PuD} architectures. 
We compare \prop's and Fulcrum since both works have a similar goal.} We model DRISA 3T1C implementation and Fulcrum using the same baseline DDR4 device we use for SIMDRAM/\prop (see Table~\ref{table_parameters}).}}

\begin{figure}[ht]
    \centering
    \includegraphics[width=\linewidth]{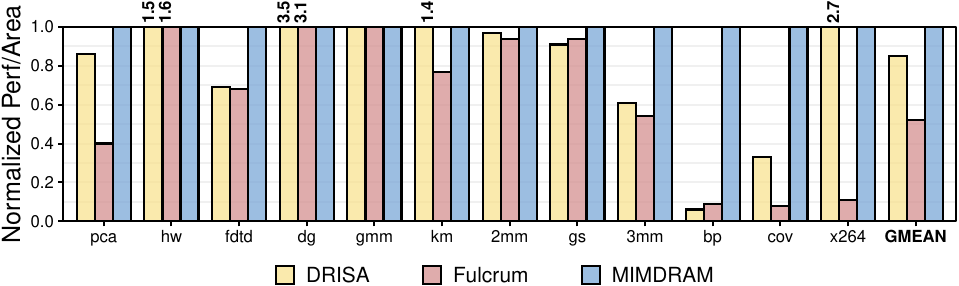}
    \caption{\revCommon{Single-application \omii{results} for different state-of-the-art \gls{PIM} architectures.}}
    \label{fig_single_app_analysis_pim}
\end{figure}

\paratitle{\revCommon{Single-Application \omii{Results}}} \revCommon{We compare the performance of each \gls{PIM} architecture and \prop. Since DRISA and Fulcrum \omi{use large additional} area \omi{(i.e., 21\% and 82\% DRAM area overhead, respectively, over our baseline DDR4 DRAM chip)} to implement \gls{PIM} operations, we report area-normalized results (i.e., performance per area) for a fair comparison. We use the area values reported in both DRISA and Fulcrum's papers, scaled to the baseline DDR4 DRAM device we employ. We allow each mechanism to leverage the data parallelism available in each application by dividing the work evenly across DRISA's \gls{PIM}-capable DRAM banks and Fulcrum's \gls{PIM}-capable subarrays. Fig.~\ref{fig_single_app_analysis_pim} shows the normalized performance per area for all 12 applications. Values are normalized to \prop. We make two observations. 
First, \prop achieves the highest performance per area compared to DRISA and Fulcrum. On average across the 12 applications, \prop performance per area is \mbox{1.18$\times$/1.92$\times$} that of DRISA and Fulcrum. This is because although DRISA and Fulcrum achieve higher absolute performance than \prop (\mbox{7.5$\times$} and \mbox{3.0$\times$}, respectively), such performance benefits come at the expense of \omi{very large} area overheads. \rCommon{\changerCM{\rCommon{\#CQ1}}While MIMDRAM incurs \omi{small} area cost on top of a DRAM array \changerC{\rC{\#C1}}(1.11\% DRAM area overhead, see~\cref{sec:eval:area}), DRISA and Fulcrum incur significantly larger area costs\revdel{ (21\% and 82\% DRAM area overhead, respectively)}.}
Second, for some applications (namely \texttt{hw}, \texttt{dg}, \texttt{km}, and \texttt{x264}), DRISA and Fulcrum achieve higher performance per area than \prop. We observe that such applications are dominated by multiplication operations, which are costly to implement using \prop's bit-serial approach.  
We conclude \prop \gfcrii{is an area-efficient \gls{PIM} architecture, which provides performance benefits compared to state-of-the-art \gls{PIM} architectures for a fixed area budget.}}

\begin{figure}[ht]
    \centering
    \includegraphics[width=\linewidth]{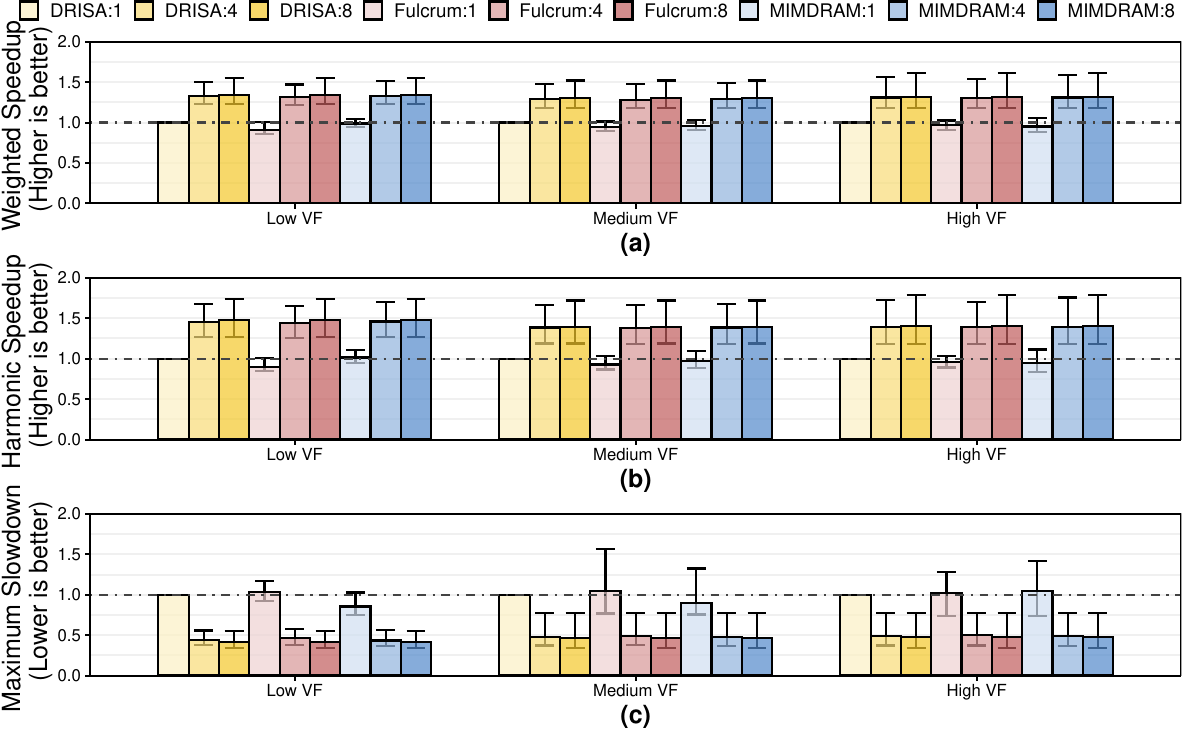}
    \caption{\revCommon{Multi-\omii{programmed workload results} for different \gls{PIM} architectures and three types of application mixes. \emph{VF} stands for vectorization factor. \emph{DRISA:X}\omiii{/\emph{MIMDRAM:X}} (\emph{Fulcrum:X}) uses \emph{X} DRAM banks (subarrays) for computation. \omiii{Values are normalized to \emph{DRISA:1}.} Whiskers extend to the minimum and maximum \gfcrii{observed} data points.}}
    \label{fig_multi_app_analysis_pim}
\end{figure}

\paratitle{\revCommon{Multi-\omii{Programmed Workload Results}}}
\revCommon{Fig.~\ref{fig_multi_app_analysis_pim} shows the system throughput, job turnaround time, and fairness that DRISA, Fulcrum, and \prop provide on average across all application mixes. \revdel{Since each architecture represents a different execution model (i.e., bit-parallel \gls{PuD}, scalar-based \gls{PnM}, and bit-serial \gls{PuD} computing), we consider the performance of \emph{each} architecture when computing the \emph{performance alone} component of the weighted speedup, harmonic speedup, and maximum slowdown metrics.} We employ BLP \omiii{in DRISA and \prop, and} \gls{SLP}~\cite{kim2012case} \omiii{in Fulcrum} to enable \gls{MIMD} execution. \changerCM{\rCommon{\#CQ1}}We make two observations. 
First, \omiii{all three \gls{PIM} architectures achieve similar system throughput. On average across all application mixes and configurations, DRISA, Fulcrum, and \prop achieve 1.20$\times$,  1.17$\times$, and  1.19$\times$ the system throughput of \emph{DRISA:1}, respectively.
Second, when considering a \emph{single} DRAM subarray for computation, \prop achieves 8\%  and 11\% \emph{higher} fairness than DRISA and Fulcrum, respectively. }}

\subsection{\gfcriii{\prop with \gls{SLP} \& BLP}}
\label{sec:eval:scalability}

\begin{figure}[ht]
    \centering
    \includegraphics[width=\linewidth]{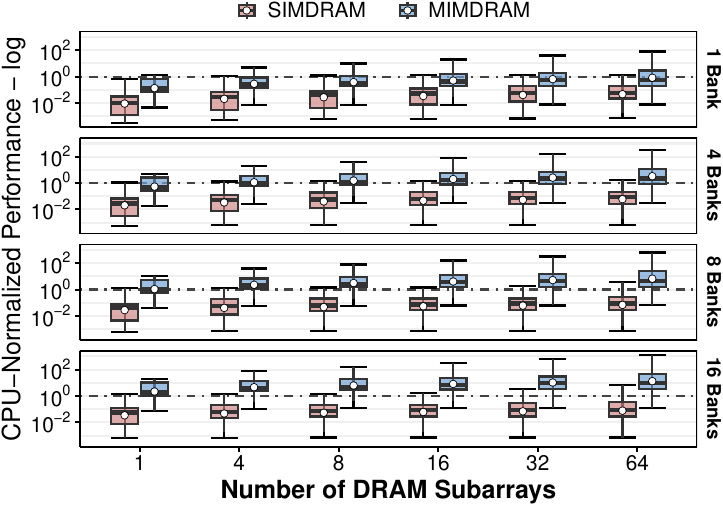}
    \caption{\gfcriii{Distribution of single-application performance across all twelve applications when varying the number of DRAM subarrays and banks for SIMDRAM and \prop. Values are normalized to the baseline CPU. Whiskers extend to the minimum and maximum observed data points on either side of the box. Bubbles depict average values.}}
    \label{fig_scaling}
\end{figure}

\gfcriii{One of the main advantages of \gls{PuD} architectures is the ability to exploit the \omiii{large} internal DRAM parallelism for computation.
A \gls{PuD} substrate can leverage \gls{SLP}~\cite{kim2012case} and BLP~\cite{mutlu2008parallelism,kim2010thread,kim2012case,kim2016ramulator,lee2009improving} techniques to operate \emph{simultaneously} \omiii{exploit} the many DRAM subarrays (e.g., 8--64 \omiii{per bank}) and banks (e.g., 8--16 \omiii{per rank}) in a DRAM chip for \gls{PuD} computation. To this end, we perform a sensitivity analysis of  SIMDRAM and \prop's performance for our twelve applications when using multiple DRAM subarrays (1--64 \omiii{per bank}) and DRAM banks (1--16 \omiii{per rank}) for \gls{PuD} computation, as Fig.~\ref{fig_scaling} depicts.
We make two observations from the figure. 
First, by \emph{fully} leveraging the internal DRAM parallelism in a DRAM chip, \prop can provide \emph{significant} performance gains compared to the baseline CPU. On average across all twelve applications, \prop (using 64 DRAM subarrays \omiii{per bank} and 16 banks for \gls{PuD} computation) achieves 13.2$\times$ the performance of the CPU (and 2$\times$ the performance of the GPU, not shown in the figure). 
Second, in contrast, SIMDRAM \emph{fails} to outperform the baseline CPU, even when fully utilizing the internal DRAM for computation (0.08$\times$ the performance of the CPU when using 64 DRAM subarrays \omiii{per bank} and 16 banks). This is because:
\li~\prop unlocks further parallelism by leveraging idle DRAM mats for computation and
\lii~\prop reduces the latency of costly vector reduction operations.
Third, we observe that \prop \omiii{can lead} to performance loss compared to the baseline CPU \omiii{for some workloads}, even when using all available DRAM subarrays and banks for computation, for two main reasons:
\li~quadratically-scaling \gls{PuD} operations (i.e., multiplication and divisions) or
\lii~\gls{PuD} vector reduction operations dominate \prop's execution time of the application. 
In the first case, \prop's performance could be further improved by leveraging lower-latency algorithms for costly \gls{PuD} operations (e.g., bit-parallel multiplication and division algorithms~\cite{leitersdorf2023aritpim}) \omiii{or performing such complex operations near memory (close to DRAM)~\pnmshort.}
In the second case, \prop would benefit from the assistance of \gls{PnM} architectures to perform faster vector reduction operations in DRAM, at the cost of an increase in area cost.
We conclude that \prop highly benefits from exploiting \gls{SLP} and BLP for \gls{PuD} computation.
}

\subsection{Area Analysis}
\label{sec:eval:area}

\atb{We use CACTI~\cite{cacti,muralimanohar2007optimizing} to model the area of a DRAM chip (Table~\ref{table_parameters}) using a \SI{22}{\nano\meter} technology node. We implement \prop{}'s chip select and mat identifier logic in Verilog HDL and synthesize the HDL using the Synopsys Design Compiler~\cite{synopsysdc} with a \SI{65}{\nano\meter} process technology node.}\footnote{\rB{\changerB{\rB{\#B1}}We use a \SI{65}{\nano\meter} technology node since that is the best CMOS standard cell library we have \omi{access} \omii{to} in \omi{our} environment. We scaled our design to a \SI{22}{\nano\meter} technology node following prior works' methodology~\cite{stillmaker2017Scaling,ghiasi2022genstore,dai2022dimmining,zhang2021sara}.}} 

\paratitle{DRAM Bank Area} 
%\atb{We model the baseline DRAM chip's area as \SI{22.8}{\milli\meter\squared}. 
We evaluate the \agy{area} overhead of
\li~mat isolation transistors, 
\lii~row decoder latches, 
\liii~mat selectors,  
\liv~the wiring to propagate mat selector output to mat isolation transistors (matlines), and
\lv~\omi{multiplexers} and wiring of the inter-mat \gfcri{interconnect}. \prop{} incurs 1.15\% area overhead over the baseline DRAM bank.

% 116.3

\paratitle{DRAM Chip I/O Area} \atb{The total  area overhead for \prop{}'s chip select and mat identifier is \emph{only}
\SI{825.7}{\micro\meter\squared} at a \SI{65}{\nano\meter} technology node. We estimate the equivalent area overhead at a \SI{22}{\nano\meter}  technology node to be \SI{116.3}{\micro\meter\squared}~\cite{stillmaker2017Scaling}.}

% 1.00111210

Overall, \prop{} increases the area of the evaluated DRAM chip (16 banks and I/O) by only 1.11\%.

\paratitle{\omi{\prop} Control Unit \& Transposition Unit Area} \gf{The main components in \omi{the} \prop control unit are the
\li~\emph{bbop} buffer, 
\lii~mat scoreboard, and
\liii~\uprog{} processing engines. We set the size of the \emph{bbop} buffer to \SI{2}{\kilo\byte}, which accommodates up to 1024 \emph{bbop}s. 
%We empirically find that this size is sufficient for our real-world applications.
%\agycomment{how would it scale if real-workload application demands more?} 
\gf{The mat scoreboard requires \omi{128~bits} of storage, one bit per DRAM mat \gfcrii{per subarray}. 
A single \uprog{} processing engine has an area of \SI{0.03}{\milli\meter\squared}. We \gfcrii{empirically} include eight \uprog{} processing engines in our design. \omi{W}e estimate\gfcrii{, using CACTI,} that \prop control unit area is \SI{0.253}{\milli\meter\squared}.} \prop transposition unit has an \omi{area equal to}  \omi{the} SIMDRAM transposition unit (of \SI{0.06}{\milli\meter\squared}).\revdel{\footnote{\gfcrii{We slightly modify the fields of the \emph{object tracker} to fit the \omi{14~bits} mat range information for a memory object in \prop. We  
\li~occupy the unused \omi{7~bits} and 
\lii~reduce the bits used \gfcrii{to store the size information for a memory object from \omi{32~bits} to \omi{25~bits}.}}}} Considering the area of the control and transposition units, \prop has a low area overhead of 0.6\% \omi{over} the die area of a \omi{state-of-the-art} Intel Xeon E5-2697 v3 CPU~\cite{dualitycache}.}

%\agycomment{shouldn't we at least compare its area cost to area cost of SIMDRAM?}

%%% total area is 0.313 mm2 

%% file: sections/07_related_work.tex
\section{Related Work}
\label{sec:related}

To our knowledge, \prop is the first \omi{end-to-end processing-using-DRAM (\gls{PuD})} system \gfcrii{for general-purpose applications that executes operations in a multiple-instruction multiple-data (MIMD) fashion, where \emph{independent} \gls{PuD} operations are executed \emph{concurrently} across the DRAM mats of a DRAM subarray.}  We highlight \prop's key contributions by contrasting it with state-of-the-art processing-in-memory \gfcrii{(PIM)} designs. \omi{We already compared \prop to SIMDRAM~\cite{hajinazarsimdram}, DRISA~\cite{li2017drisa}, and Fulcrum~\cite{lenjani2020fulcrum} both \emph{quantitatively} and \emph{qualitatively} in \cref{sec:eval} and demonstrated \prop's benefits over them. }

\revdel{\paratitle{Fine-Grained DRAM} Several prior works~\cite{lee2017partial,zhang2014half,zhang2017enabling,alawneh2021dynamic,son2014microbank,oconnor2017fine,chatterjee2017architecting,cooper2010fine,ha2016improving,udipi2010rethinking} propose different mechanisms to enable fine-grained DRAM substrates, aiming to alleviate the energy waste caused by coarse-grained DRAM. 
%These works achieve fine-grained DRAM activation and access using varying approaches (e.g., reorganizing the DRAM array and/or modifying the DRAM on-chip interconnect; adding more HFFs~\cite{zhang2017enabling,alawneh2021dynamic,son2014microbank,oconnor2017fine,chatterjee2017architecting}). 
\prop builds on top of such works to enable fine-grained DRAM activation for \gls{PuD} computing. 
Another class of prior work~\cite{yoon2011adaptive,yoon2012dynamic,ahn2009multicore,zheng2008minirank,brewer2010instruction} proposes new DRAM module designs that allow independently operating each DRAM chip in a DRAM module to implement fine-grained DRAM access. The concepts \prop introduces are orthogonal to such works, and can be combined with fine-grained DRAM access.}

\paratitle{Processing-\gfcr{U}sing-DRAM} 
Prior works propose different ways of implementing \gls{PuD} operations, either by 
\li~using the memory arrays themselves to perform  operations in bulk~\cite{seshadri.bookchapter17, seshadri2013rowclone,seshadri2018rowclone,wang2020figaro,seshadri2017ambit, xin2020elp2im, besta2021sisa,deng2018dracc, gao2019computedram,li2017drisa, hajinazarsimdram,li2018scope,ferreira2021pluto,ferreira2022pluto,deng2019lacc,olgun2021quactrng,bostanci2022dr} \omi{or}
\lii~modifying the DRAM sense amplifier design with logic gates for computation~\cite{li2017drisa,zhou2022flexidram}. 
\revdel{\gfcrii{\prop provides two main benefits \omii{over} these \gls{PuD} architectures.}} 
\gfcrii{Since} prior \gls{PuD} architectures execute \gls{PuD} operations at a \omii{coarse} granularity (i.e., at the granularity of a DRAM row access), they can suffer from the \underutilization issue we highlight. 
As in \prop, prior \gls{PuD} architectures can employ fine-grained DRAM for \gls{PuD} operations to mitigate \underutilization.
\gfcriii{We believe that the principles employed in \prop\revdel{, both in hardware and in software,} can benefit other \gls{PuD} architectures, leading to performance\omiii{,} energy-\omiii{efficiency}\omiii{, and programmability} improvements for the underlying \gls{PuD} substrate.}

\paratitle{Programming Support for \gls{PuM}} 
Prior works propose programming models for different types of \gls{PuM} architectures, as
\li~CUDA/OpenAcc~\cite{cheng2014professional, OpenACCA1:online} for in-cache computing~\cite{dualitycache};
\lii~tensor dataflow graphs for \revdel{in-cache~\cite{wang2023infinity} and }in-ReRAM computing~\cite{fujiki2018memory}. By enabling fine-grained DRAM, we believe such programming models can be \omi{now easily} ported to \gfi{\gls{PuD}} computing (for example, by assuming that each DRAM mat executes a different CUDA thread block).  
CHOPPER~\cite{peng2023chopper} improves SIMDRAM's programming model by leveraging bit-slicing compilers and employing optimizations to reduce the latency of a \uprog. Even though CHOPPER simplifies \omiii{programmability} compared to SIMDRAM, it still requires the programmer to re-write applications using the bit-slicing compiler's syntax. Compared to CHOPPER, \prop has two main advantages. First, \prop \emph{automatically} generates code for the \gfi{\gls{PuD}} engine without any code refactoring. Second, since CHOPPER maintains the \gfcrii{very}-wide \gls{SIMD} programming model of SIMDRAM, it \omi{also} suffers from  \gls{SIMD} underutilization.

%% file: sections/08_conclusion.tex
%\glsresetall

\section{Conclusion}
\label{conclusion}

We introduce \prop, a hardware/software co-designed \gfcrii{processing-using-DRAM (\gls{PuD})} substrate that can allocate and control only the needed computing resources inside DRAM for \gfi{\gls{PuD}} computing. 
On the hardware side, \prop introduces simple modifications to the DRAM architecture that enables the execution of
\li~different \gfi{\gls{PuD}} \gfcrii{operations} concurrently inside a single DRAM \gfcrii{subarray} in a \gfcrii{multiple-instruction multiple-data (\gls{MIMD})} fashion, and
\lii~native vector reduction computation.
On the software side, \prop implements a series of compiler passes that automatically identify and map code regions to the underlying \gfi{\gls{PuD}} substrate. 
We experimentally demonstrate that \prop provides significant benefits over state-of-the-art CPU, GPU, and \gfcrii{processing-using-memory (\gls{PuM}) and processing-near-memory (\gls{PnM})} systems. \gfcrii{We hope and believe that our work can inspire more efficient and easy-to-program \gls{PuD} systems. The source code of \prop is freely available at \url{https://github.com/CMU-SAFARI/MIMDRAM}.}
%We hope that future work builds on \prop to further ease the adoption of \gfi{\gls{PuD}} architectures.

% On the hardware side, \prop introduces simple modifications to the DRAM architecture that enables the execution of
% \li~different \gfi{\gls{PuD}} instructions concurrently inside a single DRAM array in a \gls{MIMD}-fashion, and
% \lii~native vector reduction computation.
% On the software side, \prop implements a series of compiler passes that automatically identify and map code regions to the underlying \gfi{\gls{PuD}} substrate. 
% We experimentally demonstrate that \prop provides significant benefits over state-of-the-art CPU, GPU, and \gfi{\gls{PuD}} systems. 
% %We hope that future work builds on \prop to further ease the adoption of \gfi{\gls{PuD}} architectures.